\documentclass{jfm}
\usepackage{graphicx,color}
\usepackage{epstopdf, epsfig}

\shorttitle{Turbulent channel flow of spherical particles with elastic walls}
\shortauthor{M. N. Ardekani, M. E. Rosti and L. Brandt}

\title{Turbulent channel flow of finite-size spherical particles with viscous hyper-elastic walls}

 \author{M. N. Ardekani\aff{1} \corresp{\email{mehd@mech.kth.se}},
 M. E. Rosti\aff{1}
 \and L. Brandt\aff{1}}

\affiliation{\aff{1} Linn\'e Flow Centre and SeRC (Swedish e-Science Research Centre), KTH Mechanics, \\ SE-100 44 Stockholm, Sweden}

\begin{document}

\maketitle

\begin{abstract}
We study single-phase and particulate turbulent channel flows, bounded by two incompressible hyper-elastic walls at bulk Reynolds number $5600$. Different wall elasticities are considered with and without a $10\%$ volume fraction of finite-size neutrally-buoyant rigid spherical particles. A fully Eulerian formulation is employed to account for the fluid-structure interaction at the elastic walls interfaces together with a direct-forcing immersed boundary method (IBM) to model the coupling between the fluid and the particles. The elastic walls are assumed to be made of a neo-Hookean material. We report a significant drag increase and an enhancement of the turbulence activity with growing wall elasticity for both single-phase and particulate cases in comparison with the single phase flow over rigid walls. A drag reduction and a turbulence attenuation is obtained for the particulate cases with highly elastic walls, albeit with respect to the single-phase flow of the same wall elasticity; whereas, an opposite effect of the particles is observed on the flow of the less elastic walls. This is explained by investigating the near-wall turbulence of highly elastic walls, where the strong asymmetry in the magnitude of wall-normal velocity fluctuations (favouring the positive $v^\prime$), is found to push the particles towards the channel centre. The particle layer close to the wall is shown to contribute to the turbulence production by increasing the wall-normal velocity fluctuations, while in the absence of this layer, smaller wall deformation and in turn a turbulence attenuation is observed.
We further address the effect of the volume fraction at a moderate wall elasticity, by increasing the particle volume fraction up to $20\%$. Migration of the particles from the interface region is found to be the cause of a further turbulence attenuation, in comparison to the same volume fraction in the case of rigid walls. However, the particle induced stress compensates for the loss of the Reynolds shear stress, thus, resulting in a higher overall drag for the case with elastic walls. The effect of wall-elasticity on the drag is reported to reduce significantly with increasing volume fraction of particles.  
\end{abstract}


\section{Introduction}\label{sec:Introduction} 
Interaction of elastic structures with multiphase flows is of utmost importance in different fields of science and technology, ranging from biological applications \citep{Freund2014} to energy harvesting \citep{Mckinney1981}. Materials for which the constitutive behaviour is only a function of the current state of deformation are generally known as elastic. In the special case when the work done by the stresses during a deformation process is dependent only on the initial and final configurations, the behaviour of the material is path independent and a stored strain energy function or elastic potential can be defined \citep{Bonet1997}. These so-called hyper-elastic materials show nonlinear stress-strain curves and are generally used to describe rubber-like substances. The aim of this work is to gain an understanding of the interaction between the particulate turbulent flow and hyper-elastic walls.

\subsection{Turbulent channel flow of finite-size particles}\label{subsec:tffsp}
Suspensions of solid particles are relevant in many environmental and industrial processes \citep{Guazzelli2011} such as sediment transport in estuaries \citep{Mehta2014}, blood flow in the human body, pyroclastic flows and pulp fibers in paper making industry \citep{Lundell2011}. 

The first simulations of finite-size particles in a turbulent channel flow, were performed by \cite{Pan1996}. These authors revealed that turbulent fluctuations and stresses increase in the presence of the solid phase. \cite{Matas2003,Loisel2013,Yu2013} reported a decrease of the critical Reynolds number for transition to turbulence in the semi-dilute regime with neutrally-buoyant spherical particles. The simulations by \cite{Shao2012} revealed a decrease of the fluid streamwise velocity fluctuations due to an attenuation of the large-scale streamwise vortices in a turbulent channel flow. Indeed, when the Reynolds number is sufficiently high, the flow becomes turbulent, with chaotic and multi-scale dynamics. In this regime, any solid object larger than the smallest scales of the flow can alter the turbulent structures at or below its size  \citep{Naso2010}, leading to turbulence modulation at large enough volume fractions \citep{Lucci2010,Tanaka2015}. \cite{Lashgari2014, Lashgari2016} documented the existence of three different regimes when changing the volume fraction $\phi$ of neutrally-buoyant spherical particles and the Reynolds number $Re$: a laminar-like regime at low $Re$ and low to intermediate $\phi$ where the viscous stress dominates dissipation, a turbulent-like regime at high Reynolds number and low to intermediate $\phi$ where the turbulent Reynolds stress plays the main role in the momentum transfer across the channel and a third regime at higher $\phi$, denoted as inertial shear-thickening, characterised by a significant enhancement of the wall shear stress due to the particle-induced stresses. \cite{Picano2015} investigated dense suspensions in turbulent channel flow up to volume fraction of $20\%$. Their study revealed that the overall drag increase is due to the enhancement of the turbulence activity up to a certain volume fraction ($\phi \le 10\%$) and to the particle-induced stresses at higher concentrations. \cite{Costa2016} explained that the turbulent drag of sphere suspensions is always higher than what predicted by only accounting for the effective suspension viscosity. They attributed this increase to the formation of a particle wall-layer, a layer of spheres forming near the wall in turbulent suspensions. Based on the thickness of the particle wall-layer, they proposed a relation able to predict the friction Reynolds number as function of the bulk Reynolds number. Indeed, the particle wall-layer was found to have a significant effect on the modulation of the near-wall turbulence, as in the case of  non-spherical particles \citep{Ardekani2017,Eshghinejadfard2017,Ardekani2018AR} where the absence of this layer leads to attenuation of the turbulence activity, resulting in drag reduction. \cite{Picano2015} attribute the formation of the near-wall layer of spherical particles to the strong wall-particle lubrication interaction that stabilizes the particle wall-normal position, forcing it to roll on the wall. In fact, a complex wall might change the particle dynamics in this region, affecting the formation of this layer. Motivated by this, we study here the effect of hyper-elastic walls on the formation of the particle wall-layer and its role in altering the near-wall turbulence.   
 
\subsection{Flow over deformable compliant surfaces}\label{subsec:focw}
Many applications involve flow over complex walls that cannot be assumed as smooth flat surfaces. The study on these complex walls started with the pioneering work of \cite{Nikuradse1933,Nikuradse1950}, who presented a large number of experimental measurements in pipes with walls covered by sand grains. Many studies have been performed since then on the turbulent flow over rough surfaces  \citep{Antonia2001,Cabal2002,Belcher2003,Leonardi2003,Leonardi2004} and porous walls \citep{Beavers1970,Tilton2006,Breugem2006,Suga2010,Rosti2015}. The main results of these studies were to show a destabilizing effect of the complex wall on the flow, the disruption of the the high- and low-speed streaks close to the wall and an increased wall-normal velocity fluctuation at the interface. 

The two-way coupling between the flow and the dynamics of a deformable wall distinguishes these type of walls from the above-mentioned complex rigid surfaces (e.g.\ rough and permeable walls). This coupling allows for non-zero wall-normal velocities at the interface, due to the wall movement. Early experimental studies of \cite{Lahav1973,Krindel1979} showed a significant decrease of the critical Reynolds number for transition to turbulence of the flow in gel-coated tubes. Several studies since then have been devoted to study the linear stability of a fluid flow through flexible channels and pipes with elastic and hyper-elastic walls \citep{Kumaran1995,Srivatsan1997,Kumaran1998a,Kumaran1998b,Kumaran2000}, concluding the possibility of the flow to be unstable even in the absence of fluid inertia. These authors attribute the arising instabilities to the energy transfer from the mean flow to the fluctuations due to the deformation work at the interface \citep{Shankar1999}.   

The effect of elastic surfaces on a fully developed turbulent flow has been rarely studied in the literature. \cite{Luo2003,Luo2005} considered a new class of compliant surfaces, called tensegrity fabrics: a pre-tensioned network of compressive members interconnected by tensile ones. These authors reported a large drag increase and a turbulence activity enhancement, following a resonating condition between the wall deformation and the turbulent flow. Recently, \cite{Rosti20171} performed the first direct numerical simulations of turbulent channel flow, bounded by an incompressible hyper-elastic wall. Their study revealed that the skin friction increases monotonically with the material elastic modulus, while the turbulent flow is affected by the moving wall even at low values of elasticity. These authors reported the elasticity as the key parameter, governing the turbulent flow and and the wall deformation. 

In the case of multiphase flows, a single object interacting with a soft wall has been the subject of several recent works \citep{Skotheim2004,Salez2015,Saintyves2016,Rallabandi2017}, while there is no study in the literature on the suspensions of rigid particles in a channel flow with elastic walls.  

\subsection{Outline}\label{subsec:outline}
In this work, we present the first direct numerical simulations of turbulent channel flow with hyper-elastic walls laden with finite-size rigid spherical particles. The governing equations and the flow geometry are introduced in \S~\ref{sec:Methodology}, followed by the results of the numerical simulations in section \S~\ref{sec:Results}. The main conclusions are finally drawn in \S~\ref{sec:Final_Remarks}. 

\section{Methodology}\label{sec:Methodology}
In this work we study the turbulent flow of an incompressible viscous fluid through a channel with incompressible hyper-elastic walls, laden with non-Brownian neutrally-buoyant finite-size rigid spherical particles. To this purpose, the numerical method in \cite{Rosti20171,Sugiyama2011} with a fully Eulerian formulation is employed to account for the fluid-structure interaction at the interfaces together with a direct-forcing immersed boundary method (IBM) \citep{Uhlmann2005,Breugem2012} to model the coupling between the fluid and the particles. The IBM is combined with the lubrication, friction and collision models for short-range particle-particle interactions \citep{Ardekani2016,Costa2015}, while a sub-grid force \citep{Bolotnov2011,DeVita2019} is included to model the particle-interface interactions. A brief summary to the numerical method is given in \S\ref{subsec:numerics}.

\subsection{Governing equations \& the numerical scheme}\label{subsec:numerics}
For the fluid-structure interaction at the interface, we use the so called one-continuum formulation \citep{Tryggvason2007}, solving only one set of equations for the conservation of momentum and the incompressibility constraint in the fluid phase and the elastic layer:
\begin{eqnarray}
\label{eq:NS1}  
\partial_t u_i  + \partial_j u_i u_j &=&  \frac{1}{\rho} \partial_j \sigma_{ij}  \, ,\\ [8pt]
\partial_i u_i &=& \, 0 \, .
\label{eq:NS2} 
\end{eqnarray}
$\textbf{u}$ is defined as a monolithic velocity vector field, volume averaged \citep{Quintard1994} between the fluid phase and the elastic layer across the interface and $\rho$ is the density, assumed to be the same in both phases. Using a volume of fluid (VoF) approach \citep{Hirt1981,Rosti20171}, the Cauchy stress tensor $\pmb{\sigma}$ can be written as:
\begin{equation}
\label{eq:VoF1}  
\sigma_{ij}  =  \left( 1 - \xi \right) \sigma_{ij} ^f + \xi \sigma_{ij} ^e \, ,
\end{equation}
where the superscripts ${}^f$ and ${}^e$ denote the fluid phase and the elastic phase, respectively. $\xi$ is the local volume fraction, changing smoothly from $0$ in the fluid phase to $1$ in the elastic layer. In particular, the isoline at $\xi=0.5$ represents the interface. The scalar $\xi$ is transported by the local velocity via an advection equation:  
\begin{equation}
\label{eq:V} 
\partial_t \xi \,+\, u_i  \partial_i \xi = 0 \, .\\ [8pt]
\end{equation}

The Cauchy stress tensor $\pmb{\sigma}$ for a Newtonian fluid and an incompressible viscous hyper-elastic material can be written, respectively, as:
\begin{eqnarray}
\label{eq:Sig1}  
\sigma_{ij}^f  &=&  -\, p \delta_{ij} \,+\, \mu^f \left( \partial_j u_i \,+\, \partial_i u_j \right)  \, ,\\ [8pt]
\sigma_{ij}^e  &=&  -\, p \delta_{ij} \,+\, \mu^e \left( \partial_j u_i \,+\, \partial_i u_j \right)  \,+\, G B_{ij} \, ,
\label{eq:Sig2} 
\end{eqnarray}
where $p$ is the pressure, $\delta_{ij}$ the Kronecker delta and $\mu$ the dynamic viscosity. $G \textbf{B}$ is the hyper-elastic contribution for a neo-Hookean material, satisfying the incompressible Mooney-Rivlin law, being $G$ the modulus of transverse elasticity and $\textbf{B}$ the left Cauchy-Green deformation tensor. The tensor $\textbf{B}$ is updated by the following transport equation: 
\begin{equation}
\label{eq:B} 
\partial_t B_{ij} + u_k  \partial_k B_{ij} =  B_{kj} \partial_k u_i +  B_{ik} \partial_k u_j \, ,\\ [8pt]
\end{equation}
where the right hand side of the equation desrcibes the stretching of the elastic material due to the straining action of the flow. 

Following the formulations above and taking into account the IBM force, exerted on the fluid phase by the particles, equation \ref{eq:NS1} can be rewritten in the non-dimensional final form:
\begin{equation}
\label{eq:NonDim} 
\partial_t u_i  \,+\, \partial_j u_i u_j =  -\, \partial_i p_e \,-\, \partial_i p \,+\, \frac{ \left(1-\xi \right) \,+\, \xi  \mu^{e/f}}{Re_b} \, \partial_j \partial_j u_i \, + \, \xi G^* \partial_j B_{ij}  \,+\, f_i \, ,\\ [8pt]
\end{equation}
where $\partial_i p_e$ is the external uniform pressure gradient that drives the flow with a constant bulk velocity $U_b$, $p$ is the modified pressure (the total pressure minus $p_e$), $Re_b$ is the bulk Reynolds number, $\mu^{e/f}$ denotes the dynamic viscosity ratio and $G^*$ is the non-dimensional elastic modulus, normalized by $\rho U_b^2$. The additional term $\textbf{f}$ on the right hand side of equation \ref{eq:NonDim} is the IBM acceleration field, active in the immediate vicinity of a particle surface to enforce the no-slip and no-penetration boundary conditions.

The flow field is resolved numerically on a uniform (cubic), staggered, Cartesian grid while particles are represented by a set of Lagrangian points, uniformly distributed on the surface of each particle.  The governing differential equations are discretized using a second order central finite-difference scheme, except for the advection terms in equations \ref{eq:V} and \ref{eq:B} where
the fifth-order weighted essentially non-oscillatory (WENO) scheme is applied \citep{Sugiyama2011,Rosti20171,Izbassarov2018}. An explicit third-order Runge-Kutta scheme \citep{Breugem2012} is used for the time integration of all terms in equations \ref{eq:V}, \ref{eq:B} and \ref{eq:NonDim} without considering the IBM acceleration field $\textbf{f}$, except for the solid hyper-elastic contribution, which is advanced in time with the Crank-Nicolson method \citep{Min2001}. The first prediction velocity obtained after the time integration step is then used to compute the point forces $\textbf{F}_l$ (normalized by $\rho \Delta V_l$, with $\Delta V_l$ being the volume of each Lagrangian grid point, equal to the volume of an Eulerian grid cell) at each Lagrangian point on the surface of the particles, based on the difference between the particle surface velocity and the interpolated first prediction velocity at that point. The singular forces $\textbf{F}_l$ are then spread into the IBM acceleration field $\textbf{f}$ using the regularized Dirac delta function $\delta_d$ of \cite{Roma1999}. $\textbf{f}$ is then added to the first prediction velocity followed by the pressure-correction scheme used in \cite{Breugem2012} to project the velocity field in the divergence-free space. More details and validations of the numerical scheme can be found in \cite{Breugem2012,Ardekani2016,Rosti20171,Rosti20182,Izbassarov2018}.  

Taking into account the inertia of the fictitious fluid phase, trapped inside the particle volumes, the motion of rigid particles are described by the Newton-Euler Lagrangian equations:
\begin{eqnarray}
\label{eq:NewtonEulerWim1}  
\rho_p V_p \frac{ \mathrm{d} \textbf{u}_{p}}{\mathrm{d} t} &=&  -\rho \sum\limits_{l=1}^{N_L} \textbf{F}_l \Delta V_l + \rho  \frac{ \mathrm{d}}{\mathrm{d} t} \left( \int_{V_p} \textbf{u} \mathrm{d} V  \right)  + \left( \rho_p - \rho \right)V_p\textbf{g} + \textbf{F}_c , \, \\ [8pt] 
\textbf{I}_p \frac{ \mathrm{d} \pmb{\Omega}_{p} }{\mathrm{d} t} &=& -\rho \sum\limits_{l=1}^{N_L} \textbf{r}_l \times \textbf{F}_l \Delta V_l + \rho  \frac{ \mathrm{d}}{\mathrm{d} t} \left( \int_{V_p} \textbf{r} \times \textbf{u} \mathrm{d} V  \right) + \textbf{T}_c  \, ,
\label{eq:NewtonEulerWim2}  
\end{eqnarray}
with $\textbf{u}_p$ and  $\pmb{\Omega}_{p}$ being the translational and the angular velocity of the particle. $\rho_p$, $V_p$ and $\textbf{I}_p$ are the particle mass density, volume and moment-of-inertia tensor and $\textbf{g}$ is the gravity vector. The first term on the right-hand-side of the equations above describes the IBM force and torque as the summation of all the point forces $\textbf{F}_l$ on the surface of the particle, the second term accounts for the inertia of the fictitious fluid phase trapped inside the particle and $\textbf{F}_c$ and $\textbf{T}_c$ are the force and the torque acting on the particles, due to the particle-particle interactions.

When the distance between the particles is smaller than one Eulerian grid cell, the lubrication force is under-predicted by the IBM. To compensate for this inaccuracy and to avoid computationally expensive grid refinements, a lubrication correction model based on the asymptotic analytical expression for the normal lubrication force between spheres \citep{Jeffrey1982} is used. A soft-sphere collision model with Coulomb friction takes over the interaction when the particles touch. The restitution coefficients used for normal and tangential collisions are $0.97$ and $0.1$, with the Coulomb friction coefficient set to $0.15$. More details about these models can be found in \cite{Costa2015,Ardekani2016}.

A sub-grid repellant lubrication-like force $\mathbf{F}_r$ \citep{Clift1978,Bolotnov2011,DeVita2019} is employed to model the particle-interface interactions when the distance between a particle and the interface is less than $1.5$ Eulerian grid size. This sub-grid repellant force is written as:
\begin{equation}
  \mathbf{F}_r = \frac{1}{2} \mu^f U_b D_p \left( \frac{a_1}{d_w}  + \frac{a_2}{d_w^2} \right) \mathbf{n}_w \, ,
\end{equation}
with $a_1$ and $a_2$ being two coefficients, set to $550$ and $55$ in this study. $\mathbf{n}_w$ is the unit vector normal to the wall and $d_w$ is the distance from the interface. 

\subsection{Flow geometry}
We perform direct numerical simulations of pressure-driven, particulate turbulent channel flow, bounded by two incompressible hyper-elastic walls at $y=0$ and $y=2h$, with $h$ being half of the distance between the two interfaces.  A computational domain of size $L_x=6h$, $L_y=2(h+h_e)$ and $L_z=3h$ is considered in the streamwise, wall-normal and spanwise directions, where $h_e=0.25h$ is the height of the elastic walls in the flat unstressed condition. Two impermeable walls are located at $y = -h_e$ and $y=2h + h_e$ with no-slip and no-penetration (ns/np) boundary conditions, while the flow is considered periodic in the two wall-parallel directions. The bulk velocity $U_b$ ($\Sigma \xi u / \Sigma \xi $) is fixed to guarantee a constant bulk Reynolds number $Re_b \equiv 2hU_b/\nu_f = 5600$. This corresponds to an average friction Reynolds number of $Re_\tau \equiv u_{\tau}h/\nu_f \approx180$ in the case of unladen flow over flat rigid walls \citep{Kim1987} with $\nu_f$ and $u_{\tau}$ being the fluid kinematic viscosity and the friction velocity, respectively. The same viscosity is considered in this study for the elastic layer and the fluid phase ($\nu^f = \nu^e$), as it was shown before that the elasticity modulus $G$ is the key parameter affecting the interface and the flow dynamics \cite{Rosti20171}.

Four different modulus of transverse elasticity $G$ are studied here, ranging from an almost rigid interface, $G^*=2$, to a highly elastic case of $G^*=0.25$. Particulate cases are simulated at each $G^*$ with a particle volume fraction $\phi = 10\%$ and the results are compared to the unladen flow with the same wall elasticity. We consider non-Brownian neutrally-buoyant rigid spherical particles with the diameter $D_p=h/9$. This corresponds to $5000$ particles inside the computational domain at $\phi = 10\%$. The effect of volume fraction is further studied at a moderate wall elasticity of $G^*=0.5$, by decreasing and increasing the particle volume fraction to $\phi = 5\%$ and $20\%$. Two extra numerical experiments are also performed here for a better understanding of the flow: i) a simulation, denoted $G1_{FW}$ (fixed wall), where a snapshot of the deformed interface, obtained in the case with the highest wall elasticity, is fixed and frozen in time in order to distinguish the effect of wall elasticity on the near wall turbulence from the modulation caused by their roughness and ii) a case, referred as $G4_{NPWL}$ (no particle wall-layer), where particles bounce back towards the core region of the channel before approaching the interfaces, following a collision with the two virtual walls located at a distance $h/10$ from the two interfaces. A summary to the simulated cases is given in table~\ref{tab:cases}.

The simulations are performed on a uniform Cartesian grid of size $1296 \times 540 \times 648$, corresponding to a resolution of $24$ grid points per particle diameter $D_p$ in the particulate cases. The number of Lagrangian points $N_L$ on the surface of each particle is set to $N_L = 1721$. All cases are started from a fully developed turbulent channel flow in the fluid region ($y = 0$ to $2h$) with a perfectly flat interface (initially) and a random distribution of the particles.  The statistics are collected for about $600 h/U_b$ after the flow has reached a statistical steady state. A CFL number of $0.2$ is used to guarantee the numerical stability of the method. 

\begin{table}
  \begin{center}
\def~{\hphantom{0}}
  \begin{tabular}{lccccccc}
  Case & $Re_b \equiv U_b h/ \nu_f$ & $G^* \equiv G/(\rho U_b^2)$  & $h_e/h$ & $\phi(\%)$  &  $N_p$ & $h/D_p$  & $Re_\tau \equiv u_\tau h/ \nu_f$  \\[5pt]
      $G1$  &  $5600$ &  $0.25$  &  $0.25$ & $0$ & $0$ & --- & $388.4$ \\
      $G2$  &  $5600$ &  $0.5$  &  $0.25$ & $0$ & $0$ & --- & $281.1$ \\
      $G3$  &  $5600$ &  $1$  &  $0.25$ & $0$ & $0$ & --- & $197.2$ \\
      $G4$  &  $5600$ &  $2$  &  $0.25$ & $0$ & $0$ & --- & $184.8$ \\
      $G1_{10\%}$  &  $5600$ &  $0.25$  &  $0.25$ & $10$ & $5000$ & $9$ & $351.4$ \\
      $G2_{10\%}$  &  $5600$ &  $0.5$  &  $0.25$ & $10$ & $5000$ & $9$ & $262.8$ \\
      $G3_{10\%}$  &  $5600$ &  $1$  &  $0.25$ & $10$ & $5000$ & $9$ & $225.1$ \\
      $G4_{10\%}$  &  $5600$ &  $2$  &  $0.25$ & $10$ & $5000$ & $9$ & $212.7$ \\
      $G2_{5\%}$  &  $5600$ &  $0.5$  &  $0.25$ & $5$ & $2500$ & $9$ & $273.1$ \\
      $G2_{20\%}$  &  $5600$ &  $0.5$  &  $0.25$ & $20$ & $10000$ & $9$ & $231.5$ \\
      $G1_{FW}$  &  $5600$ &  ---  &  --- & $0$ & $0$ & --- & $280.2$ \\
      $G4_{NPWL}$  & $5600$ &  $2$  &  $0.25$ & $10$ & $5000$ & $9$ & $166.8$ \\                    
  \end{tabular}
  \caption{Summary of the simulations, performed in this study at a bulk Reynolds number $Re_b=5600$. $h_e$ is the height of the elastic walls in the flat unstressed condition, $G^*$ is the non-dimesionalized modulus of transverse elasticity, $\phi$ denotes the particle volume fraction and $N_p$ the number of the particles inside the domain with diameter $D_p = h/9$. All simulations are performed on a computational domain of size $6h \times 2(h+h_e) \times 3h$ in the streamwise, wall-normal and spanwise directions; the equations are discretized on $1296 \times 540 \times 648$ grid cells. $Re_\tau$ is the obtained mean friction Reynolds number, reported here for all the simulated cases.} 
 \label{tab:cases}
 \end{center}
\end{table}

\section{Results}\label{sec:Results}
\subsection{Drag \& deformation}\label{subsec:drag}
\begin{figure}
  \centering
   \includegraphics[trim={3.9cm 0 6.5cm 0},clip,width=0.496\textwidth]{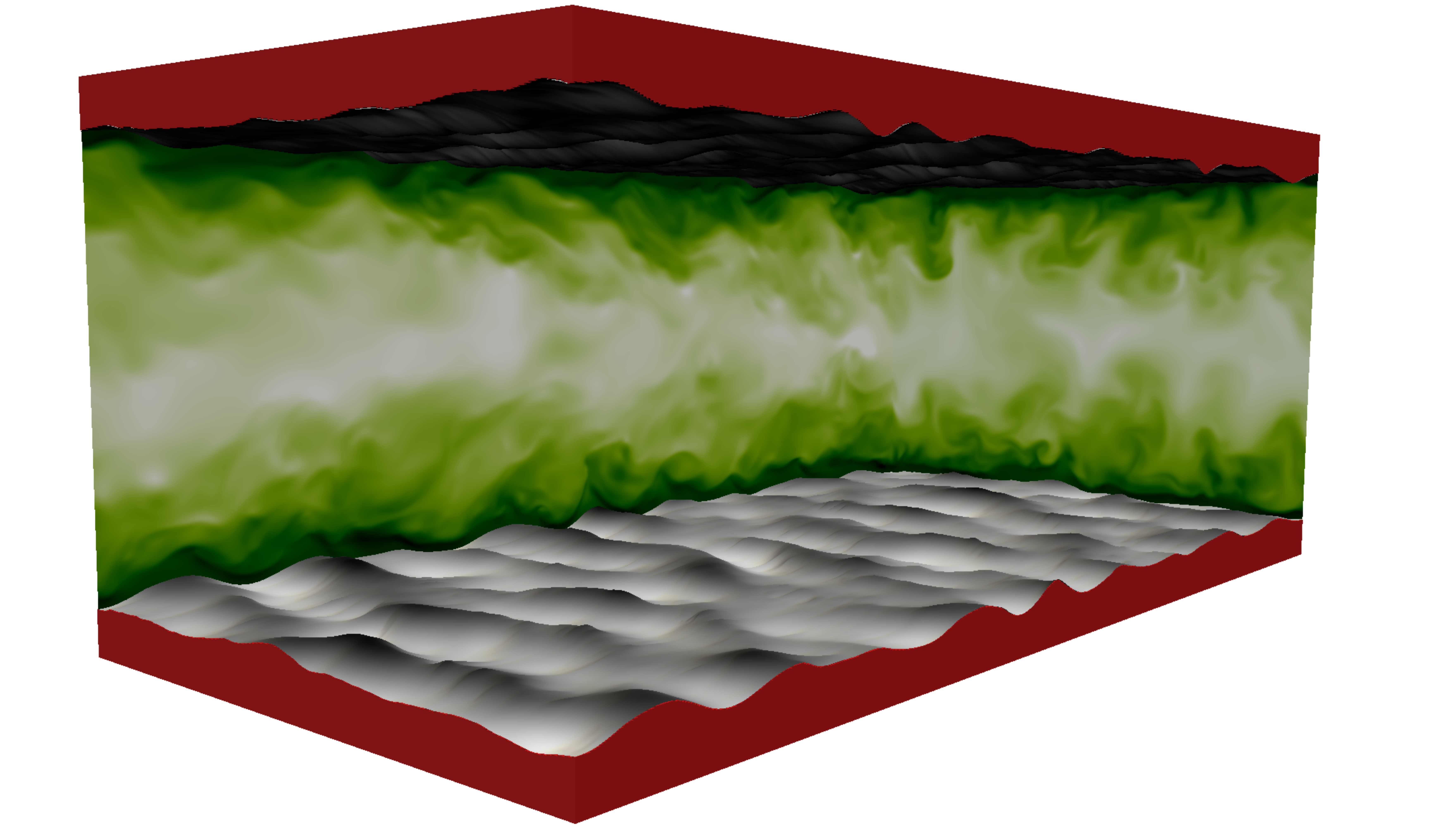}
   \includegraphics[trim={3.9cm 0 6.5cm 0},clip,width=0.496\textwidth]{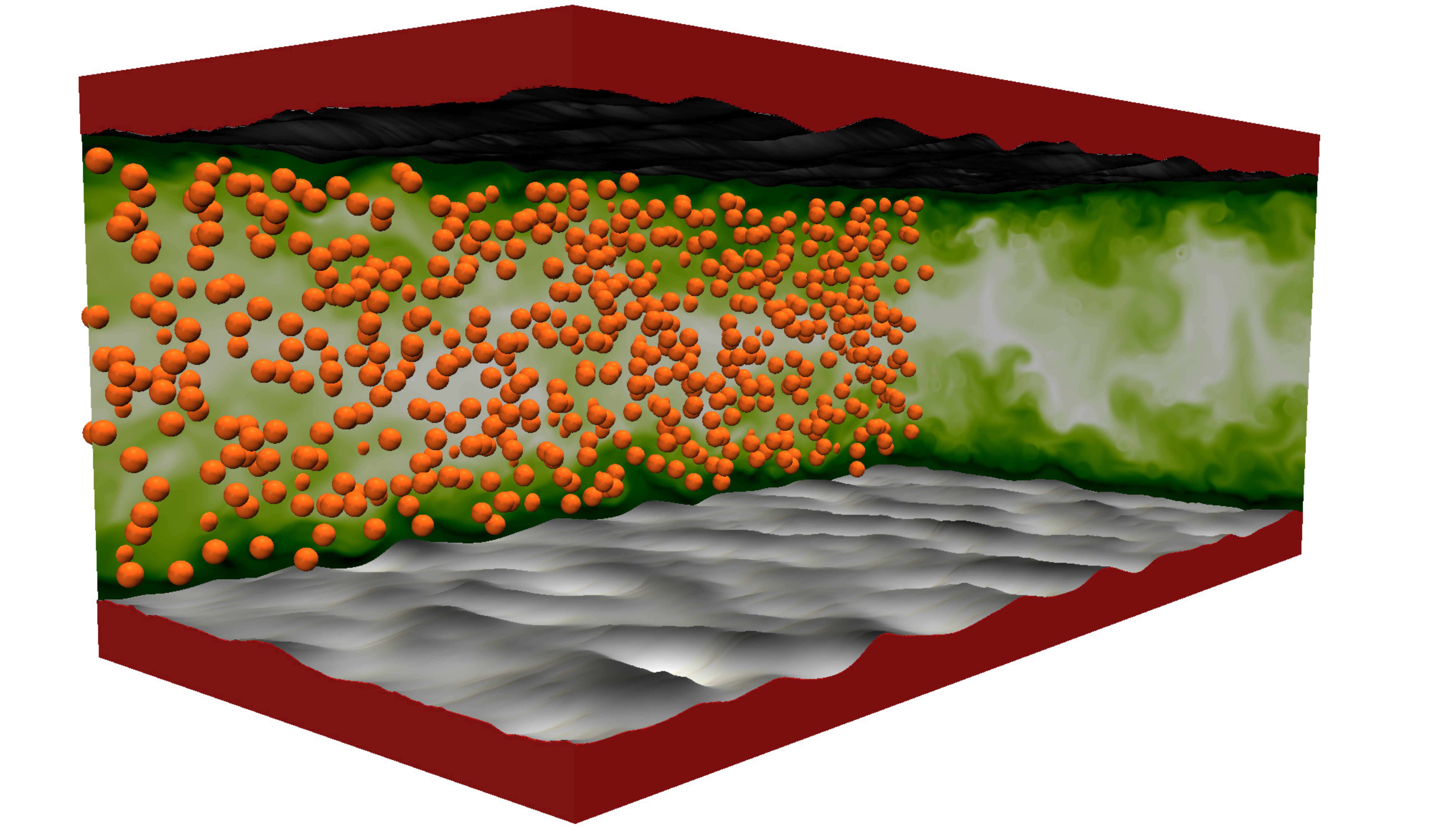} 
   \put(-387,125){\footnotesize $(a)$}
   \put(-194,125){\footnotesize $(b)$} \\ 
   \includegraphics[trim={3.9cm 0 6.5cm 0},clip,width=0.496\textwidth]{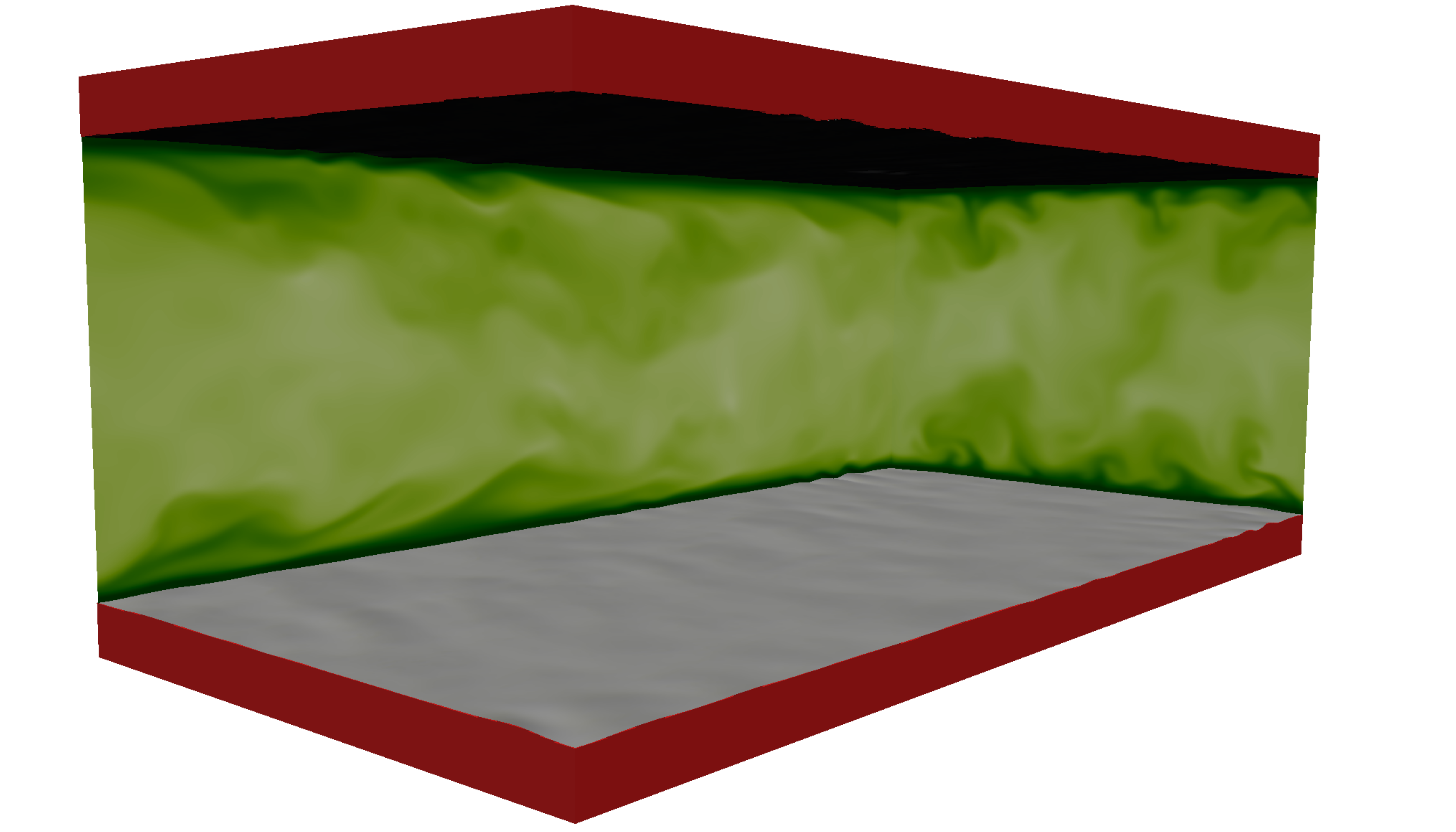}
   \includegraphics[trim={3.9cm 0 6.5cm 0},clip,width=0.496\textwidth]{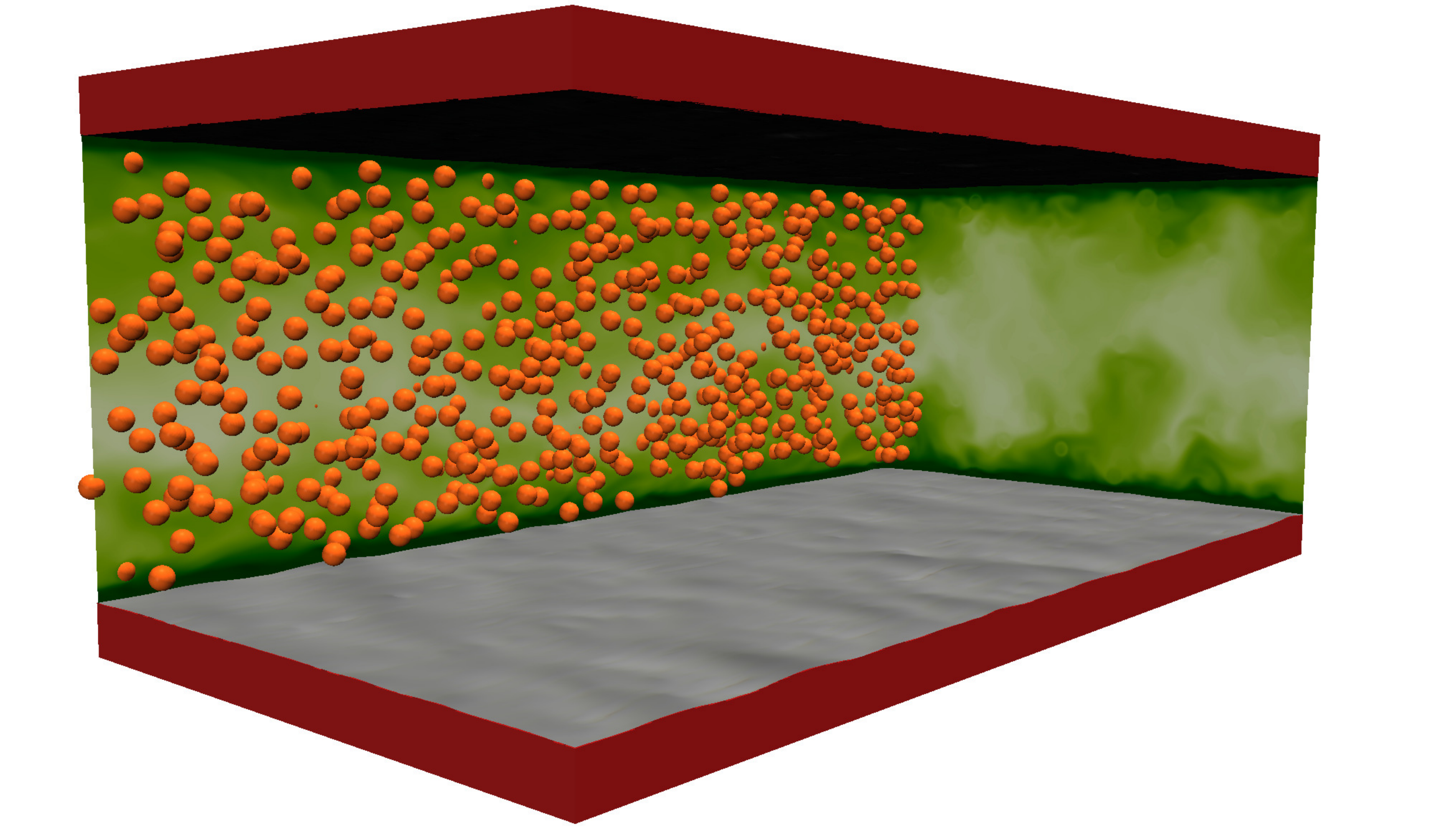}
   \put(-387,125){\footnotesize $(c)$}
   \put(-194,125){\footnotesize $(d)$} \\
  \caption{Instantaneous snapshots of the streamwise velocity $u$ on $x-y$ and $y-z$ planes for $(a)$ $G^*=0.25$ (case $G1$), $(b)$ $G^*=0.25$ at $\phi = 10\%$ (case $G1_{10\%}$), (c) $G^*=1$ (case $G3$)   
and $(d)$ $G^*=1$ at $\phi = 10\%$ (case $G3_{10\%}$). For clarity, only the particles lying within the selected $x-y$ plane are displayed. The colour scale for the streamwise velocity ranges from $0$ (dark green) to $1.5u/U_b$ (white). The elastic walls are represented with the isosurfaces of $\xi=0.5$, coloured by the wall-normal distance ranging from $-0.15h$ (white) to $0.15h$ (black).}
\label{fig:snapshots}
\end{figure}

\begin{figure}
  \centering
   \includegraphics[width=0.496\textwidth]{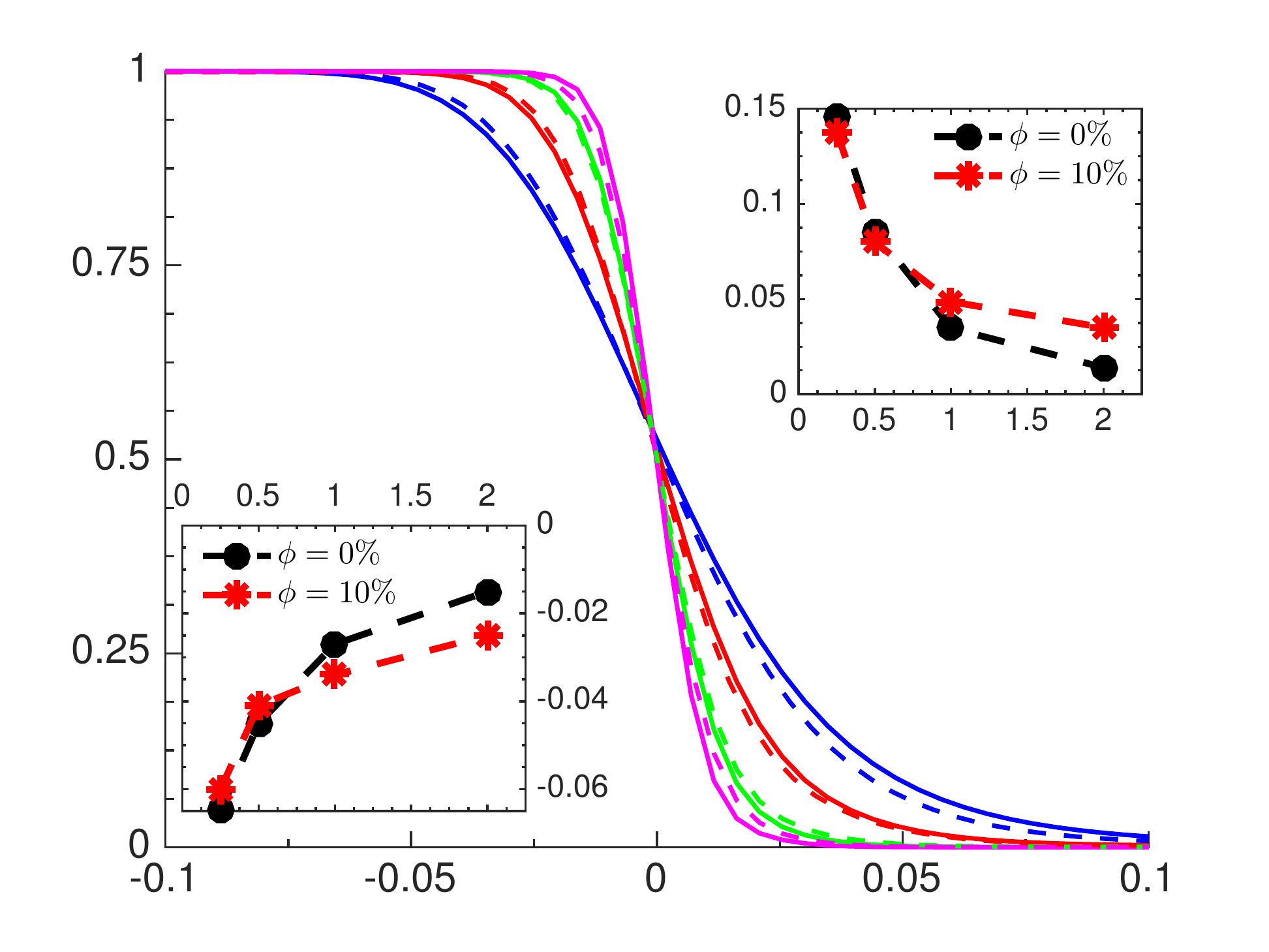}
   \includegraphics[width=0.496\textwidth]{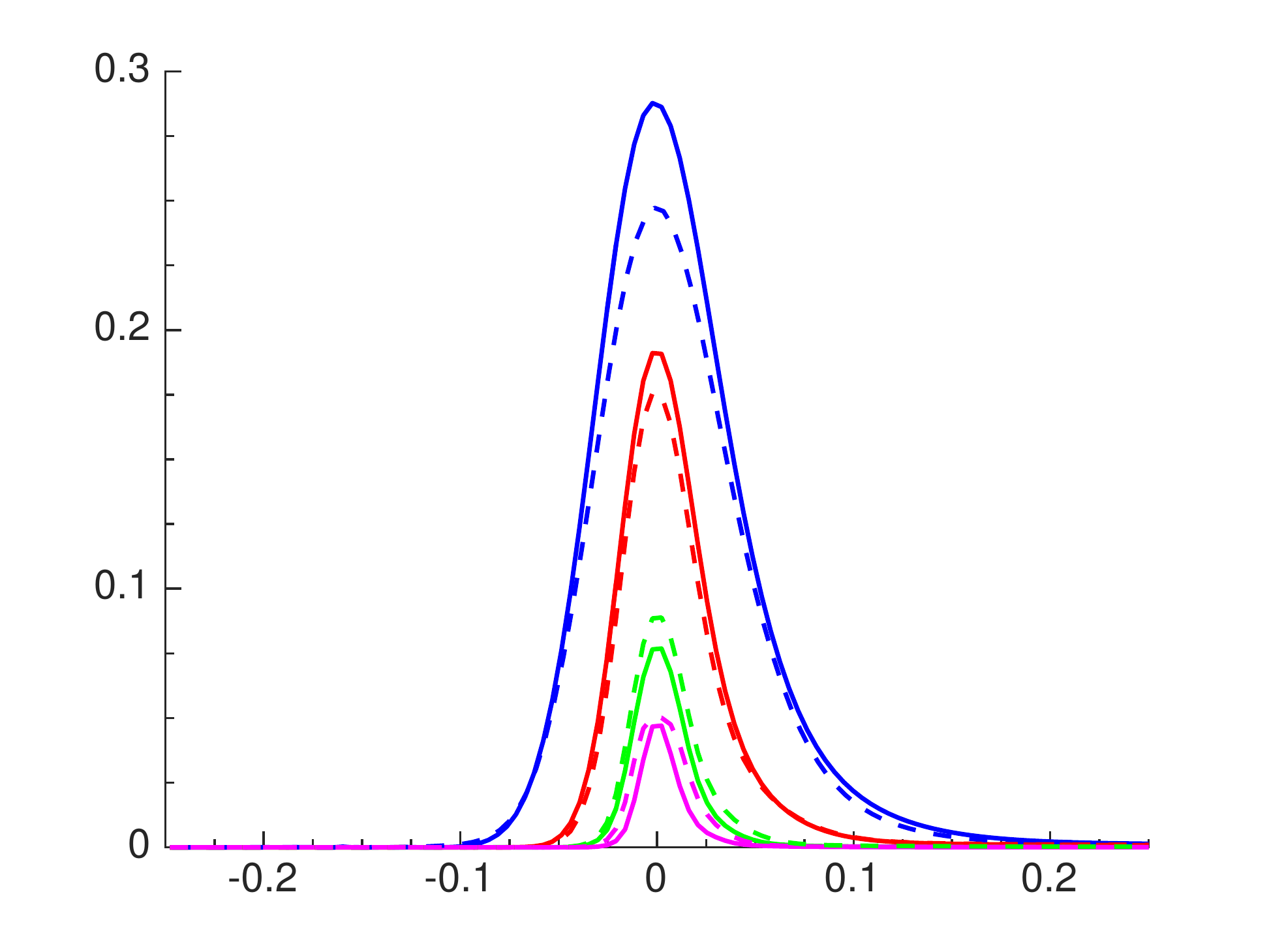} 
   \put(-188,72){\rotatebox{90}{$\xi^{\prime}$}}
   \put(-384,72){\rotatebox{90}{$\overline{\xi}$}}   
   \put(-99,0){{$y / h$}}
   \put(-293,0){{$y / h$}}
   \put(-334,72){{\tiny $G^*$}}
   \put(-242,72){{\tiny $G^*$}}      
   \put(-282,87){\rotatebox{90}{\tiny $y/h {|}_{\overline{\xi}=0.001}$}}    
   \put(-293,63){\rotatebox{-90}{\tiny $y/h {|}_{\overline{\xi}=0.999}$}}       
   \put(-387,130){\footnotesize $(a)$}
   \put(-194,130){\footnotesize $(b)$} \\ 
  \caption{VoF function $\xi$ versus the normalized distance from the interface $y/h$, showing the wall deformation for the different cases under investigation: $(a)$ mean values and $(b)$ root-mean-square (rms) of the fluctuations in $\xi$. The blue, red, green and magenta solid lines are used for the cases $G1$ to $G4$, respectively, while the particulate cases $G1_{10\%}$ to $G4_{10\%}$ are indicated with the dashed lines of the same colour as their single-phase counterparts. The insets in $(a)$ show the maximum and minimum position of the interface for different wall elasticities.}
\label{fig:Deform}
\end{figure}
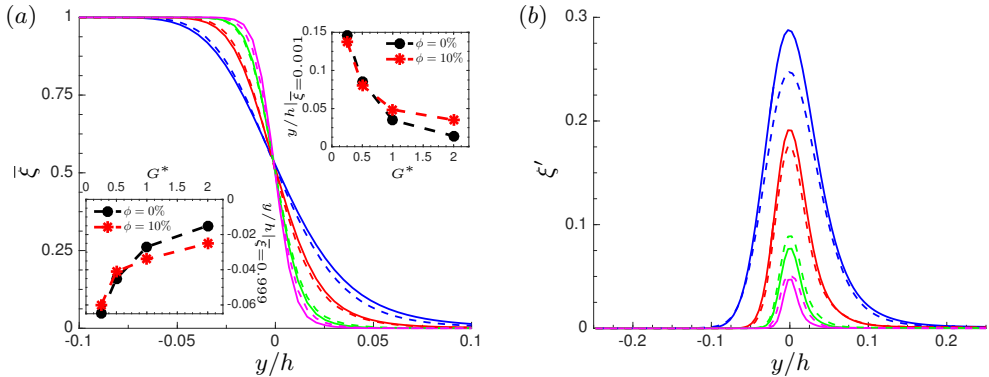

We first display snapshots of the flow and particles in figure~\ref{fig:snapshots}, where the instantaneous streamwise velocity $u$ is depicted on $x-y$ and $y-z$ planes for the cases with $G^*=0.25$ and $1$ at $\phi=0$ and $10\%$. For clarity, just a fraction of the particles (those lying within the visualized $x-y$ plane) are displayed. Particles are observed to decrease the wall deformation and turbulence activity in $G1_{10\%}$ (figure~\ref{fig:snapshots}$(b)$), with respect to the single phase flow at the same wall elasticity (figure~\ref{fig:snapshots}$(a)$). However, an opposite effect is obtained in $G3_{10\%}$ (figures~\ref{fig:snapshots}$(c)$ and $(d)$), where the wall deformation is slightly increased in the presence of the particles. Another observation to note is the spanwise coherency of the wall deformation in figure~\ref{fig:snapshots}$(a)$ that is broken into a less correlated pattern in the particulate case $G1_{10\%}$. The mean ($\overline{\xi}$) and the root-mean-square ($\xi^\prime$) profiles of the VoF function $\xi$ are depicted in figure~\ref{fig:Deform} versus the normalized distance from the interface $y/h$. The maximum and minimum of the interface position, corresponding to the highest crests and the lowest troughs of the wavy interface, are measured where $\overline{\xi}$ reaches $99.9\%$ and $0.1\%$ of its maximum value ($1$) respectively. These are reported in the insets of figure~\ref{fig:Deform}$(a)$ for different wall elasticities with and without particles. The results show an expected decrease of wall deformation with increasing $G^*$ for both single-phase and particulate cases. However, the effect of the particles on the wall deformation is non-monotonic: a slight decrease with respect to the single-phase flow in the cases with highly elastic walls ($G1_{10\%}$ and $G2_{10\%}$), contrary to a larger interface deformation for the cases with less wall elasticity ($G3_{10\%}$ and $G4_{10\%}$). Although the particles do not significantly change the maximum amplitude of deformation, the convergence of $\overline{\xi}$ to zero and one (insets of figure~\ref{fig:Deform}$(a)$) is considerably modified, showing less frequent occurrence of large deformations in the particulate case at low $G^*$ and the opposite behavior at high $G^*$. Profiles of $\xi^\prime$ are depicted in figure~\ref{fig:Deform}$(b)$. The size of the peak at $y=0$ can be used as an indication of the deformation of the interface: $\xi^\prime$ is bounded between $\xi^\prime = 0.5$, when all the grid points in the $y=0$ plane are located inside the fluid region or in the elastic layer and $\xi^\prime = 0$ when the interface is undeformed everywhere. The results show a decrease of the deformation by the presence of the particles in highly elastic walls, while an increase of the deformation is evident in the less deformable cases.    

\begin{figure}
  \centering
   \includegraphics[width=0.496\textwidth]{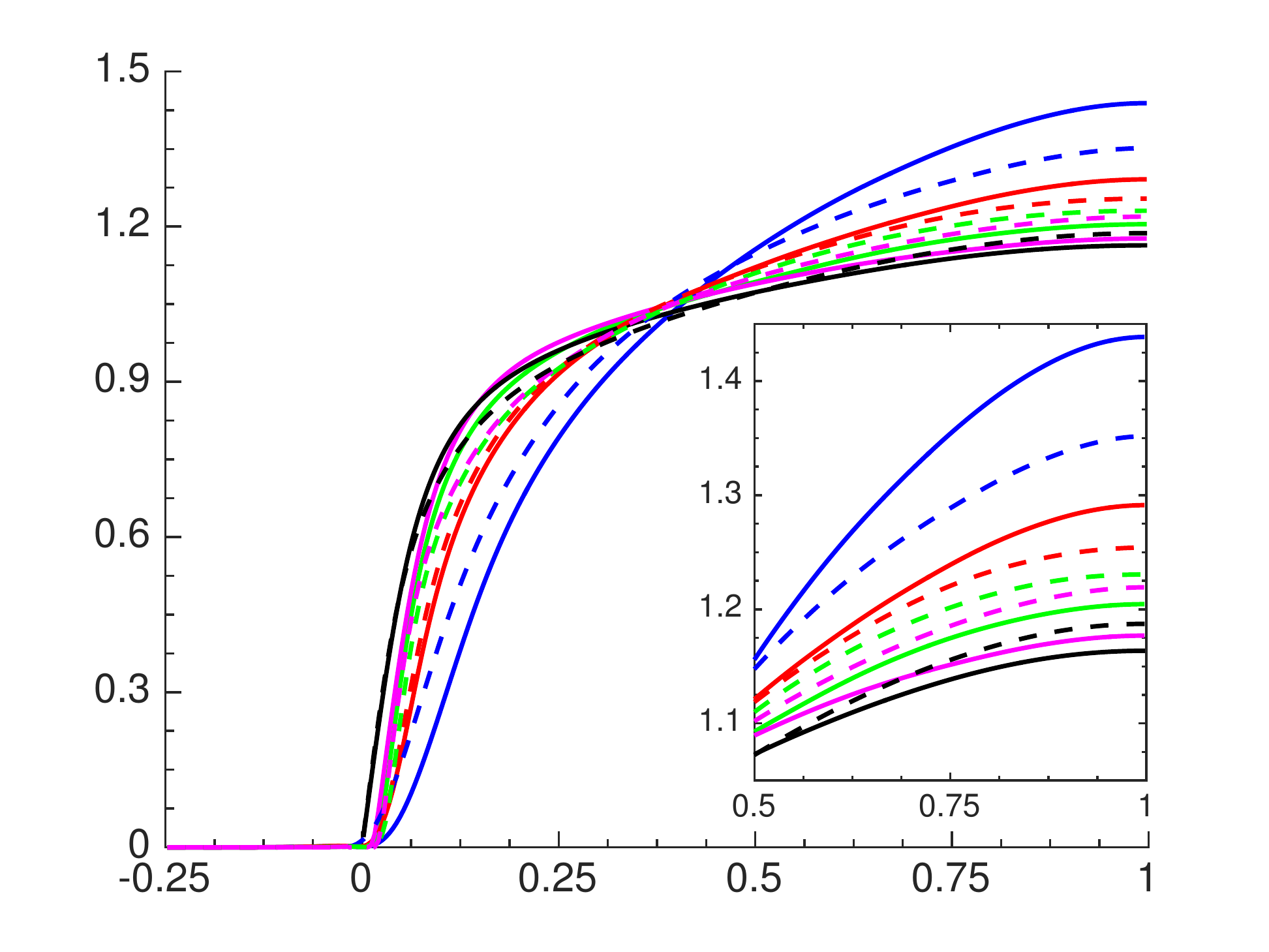}
   \includegraphics[width=0.496\textwidth]{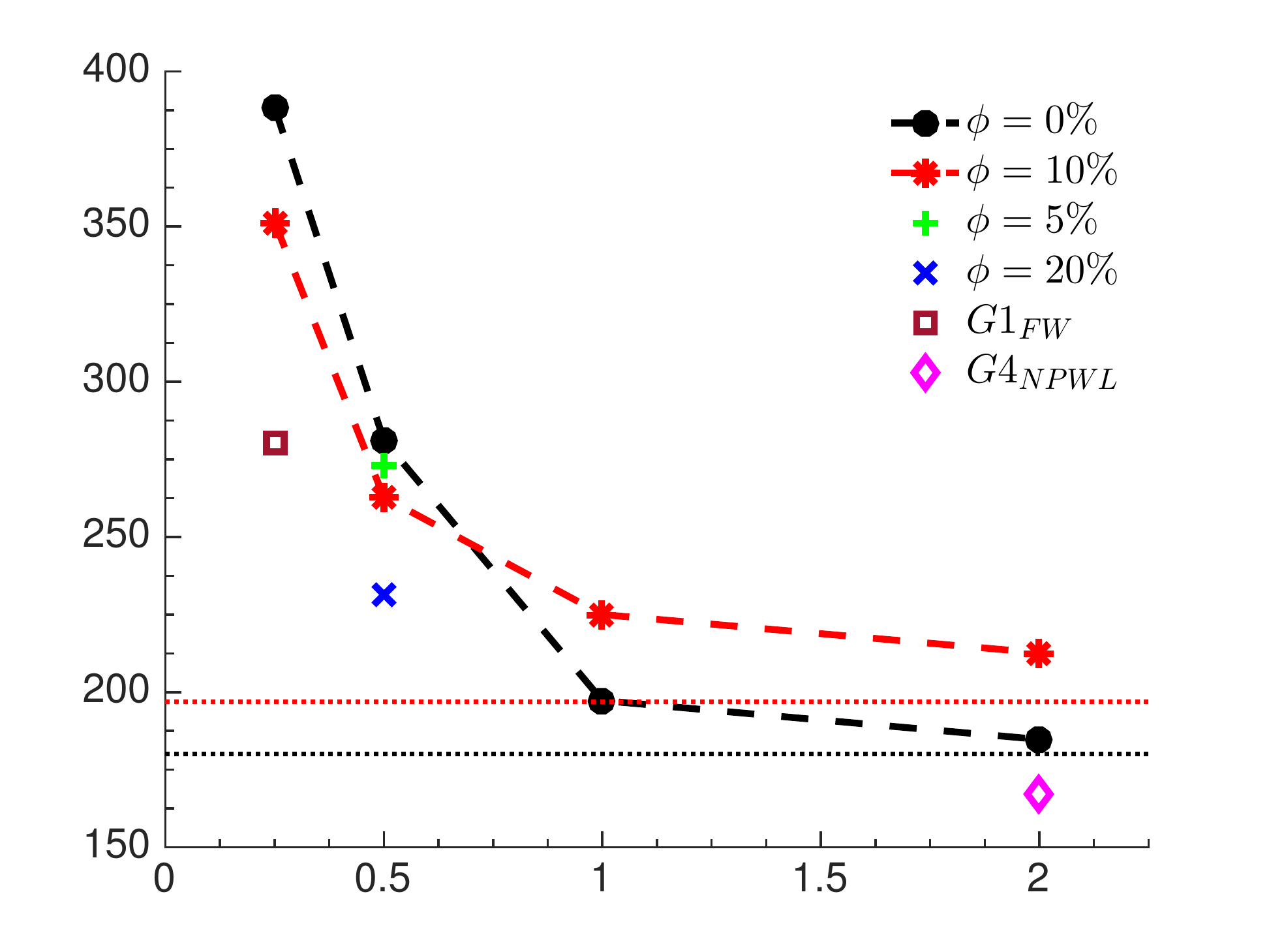} 
   \put(-97,0){{$G^*$}}
   \put(-293,0){{$y / h$}}   
   \put(-188,68){\rotatebox{90}{$Re_{\tau}$}}
   \put(-384,66){\rotatebox{90}{$U / U_b$}}  
   \put(-387,130){\footnotesize $(a)$}
   \put(-194,130){\footnotesize $(b)$} \\ 
  \caption{$(a)$ Mean velocity profiles, normalized by the bulk velocity $U_b$. The colour scheme is the same as figure~\ref{fig:Deform} with two extra black lines, pertaining the results for rigid walls \citep{Ardekani2017}: a solid line for single-phase flow and a dashed line for particulate flow at $\phi=10\%$. The inset displays the increase of the mean velocity at the core of the channel. $(b)$ The friction Reynolds number $Re_{\tau}$, obtained for all the cases considered in this study, versus the modulus of transverse elasticity $G^*$. The black and red dashed horizontal lines in $(b)$ represent $Re_{\tau}$ for the single-phase and the particulate flow over smooth rigid walls, respectively.}
\label{fig:Retau}
\end{figure}

\begin{figure}
  \centering
   \includegraphics[width=0.496\textwidth]{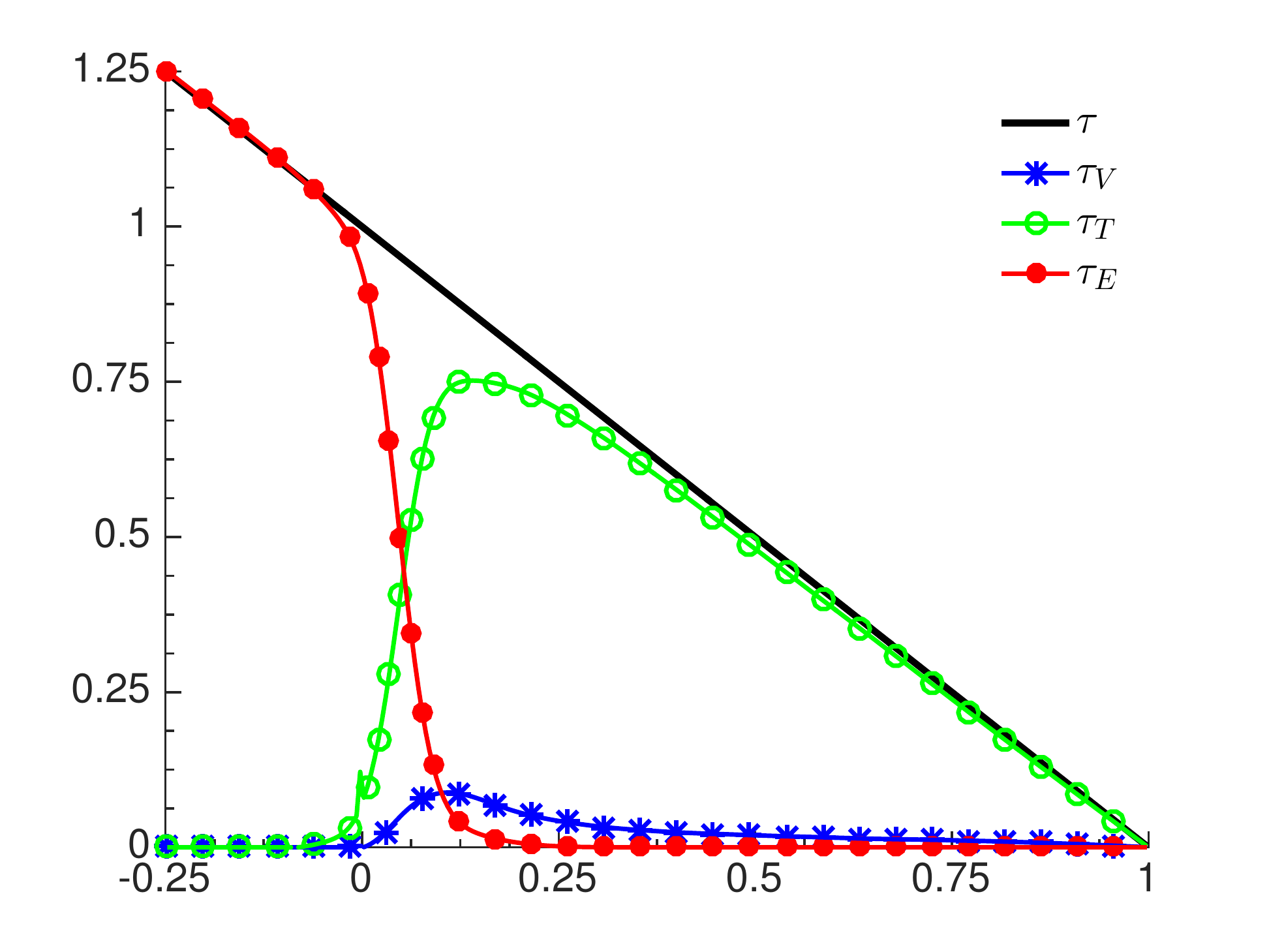}
   \includegraphics[width=0.496\textwidth]{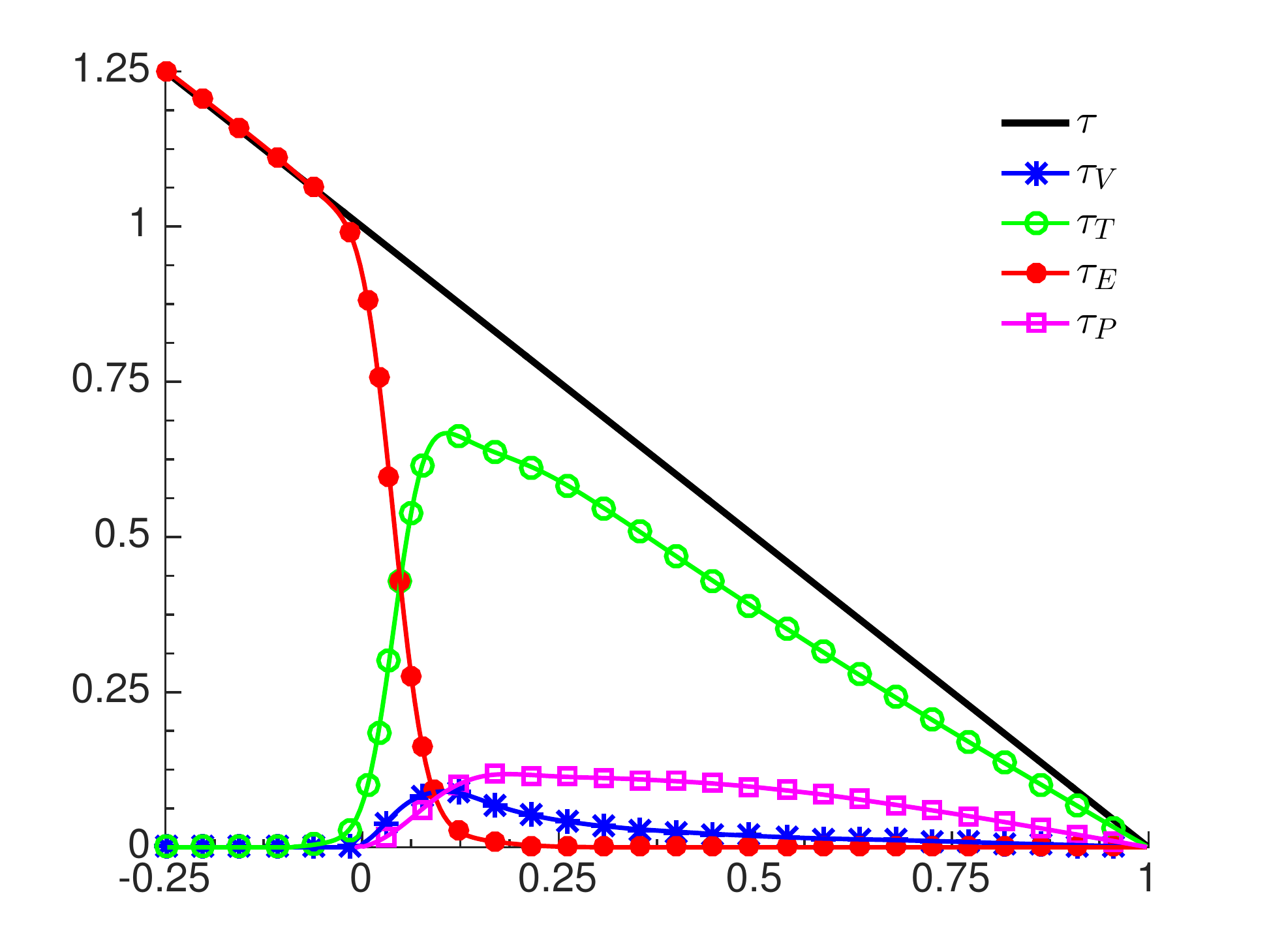} 
   \put(-382,68){\rotatebox{90}{$\tau^+_i$}}
   \put(-189,68){\rotatebox{90}{$\tau^+_i$}}    
   \put(-99,0){{$y / h$}}
   \put(-293,0){{$y / h$}}   
   \put(-387,130){\footnotesize $(a)$}
   \put(-194,130){\footnotesize $(b)$} \\ 
   \includegraphics[width=0.496\textwidth]{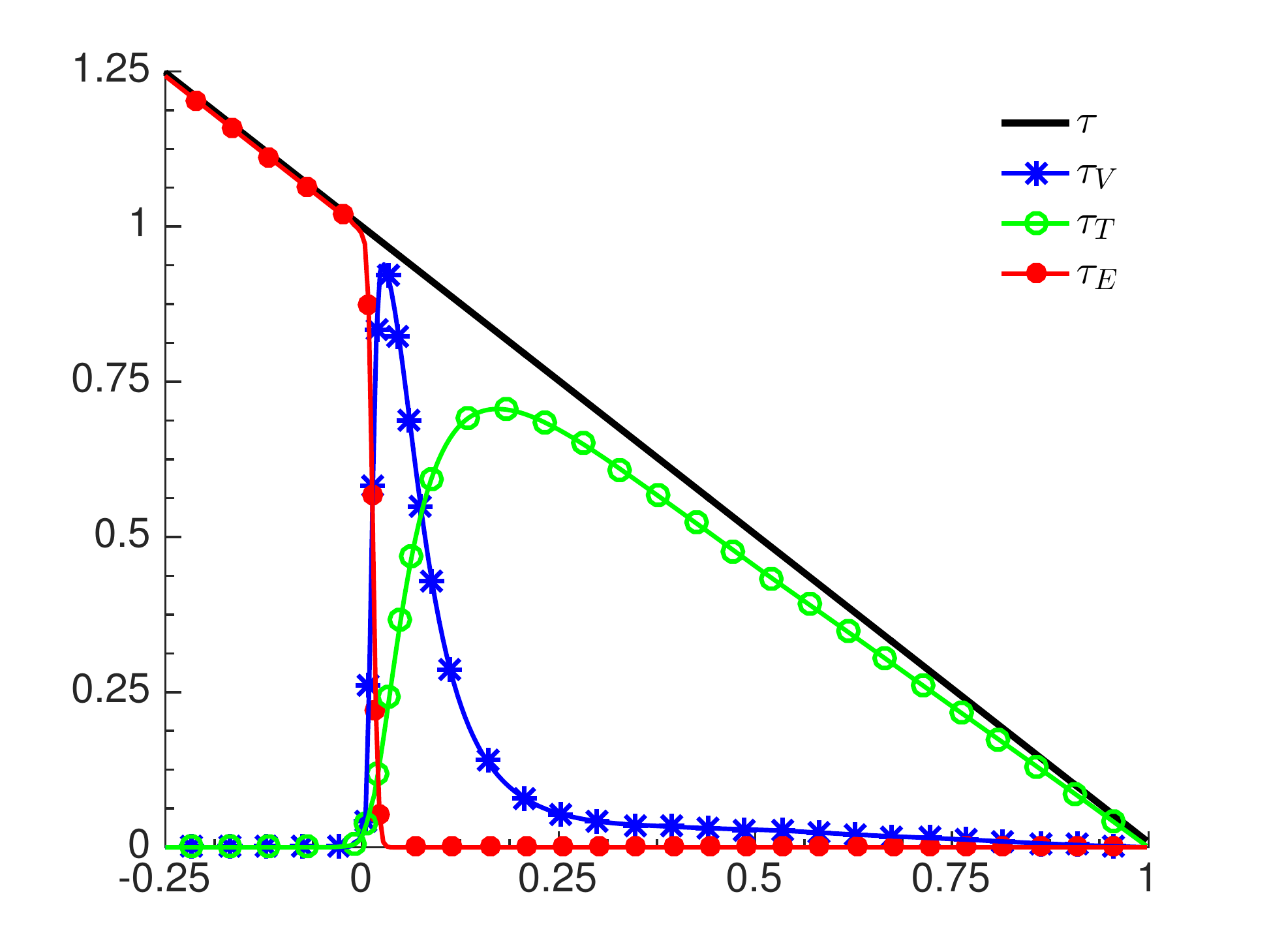}
   \includegraphics[width=0.496\textwidth]{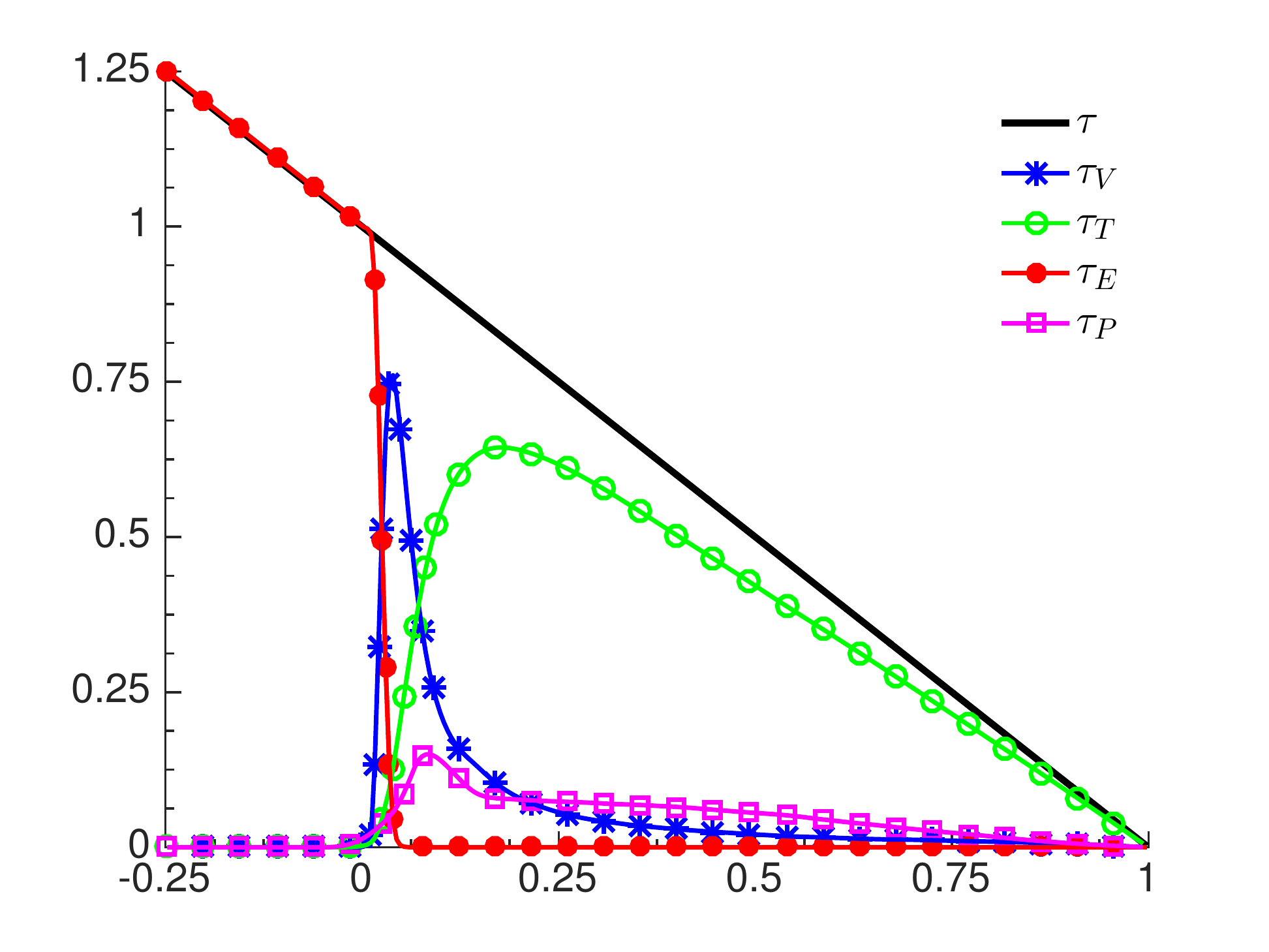}
   \put(-382,68){\rotatebox{90}{$\tau^+_i$}}
   \put(-189,68){\rotatebox{90}{$\tau^+_i$}}    
   \put(-99,0){{$y / h$}}
   \put(-293,0){{$y / h$}}   
   \put(-387,130){\footnotesize $(c)$}
   \put(-194,130){\footnotesize $(d)$} \\
   \includegraphics[width=0.496\textwidth]{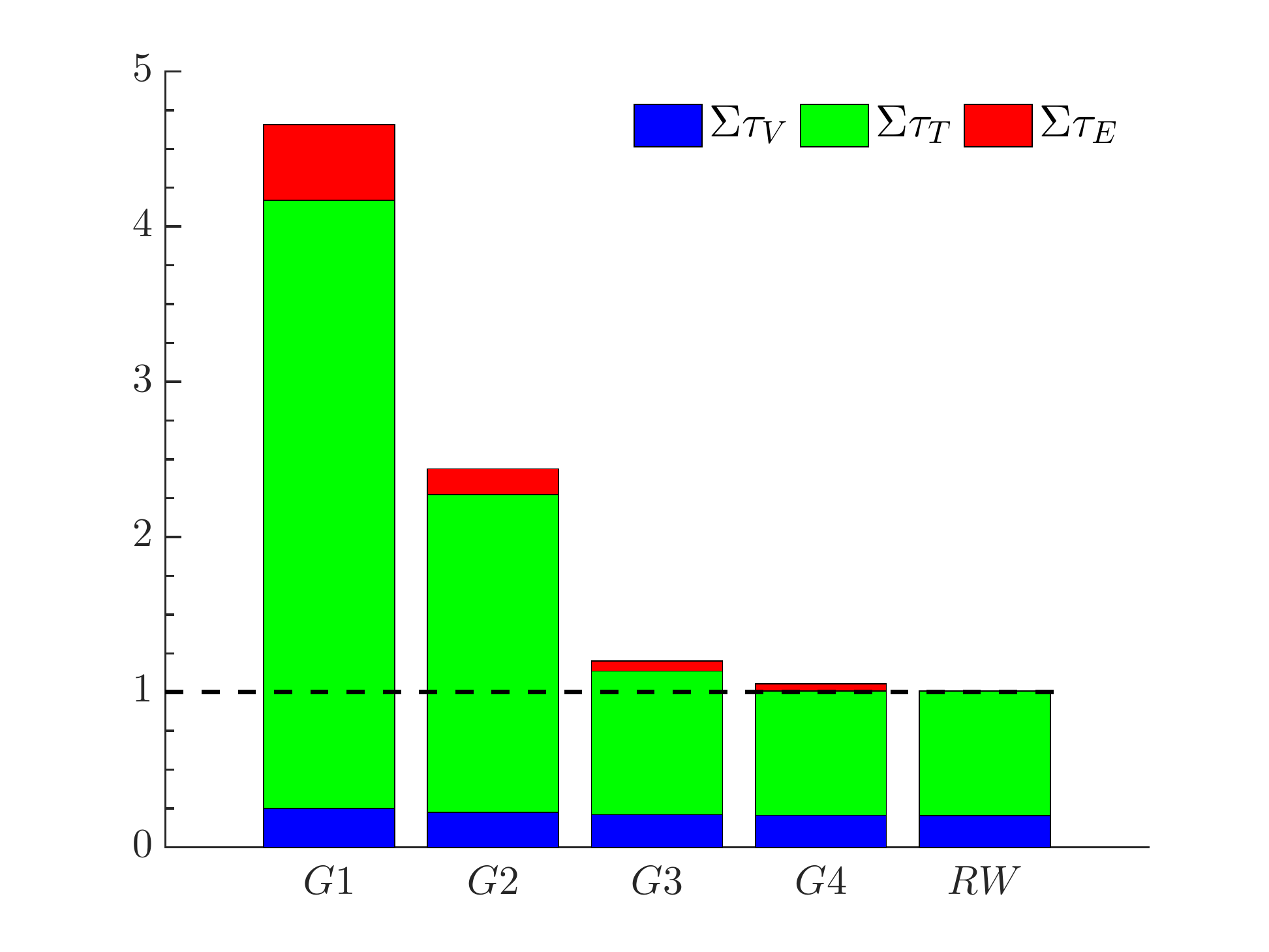}
   \includegraphics[width=0.496\textwidth]{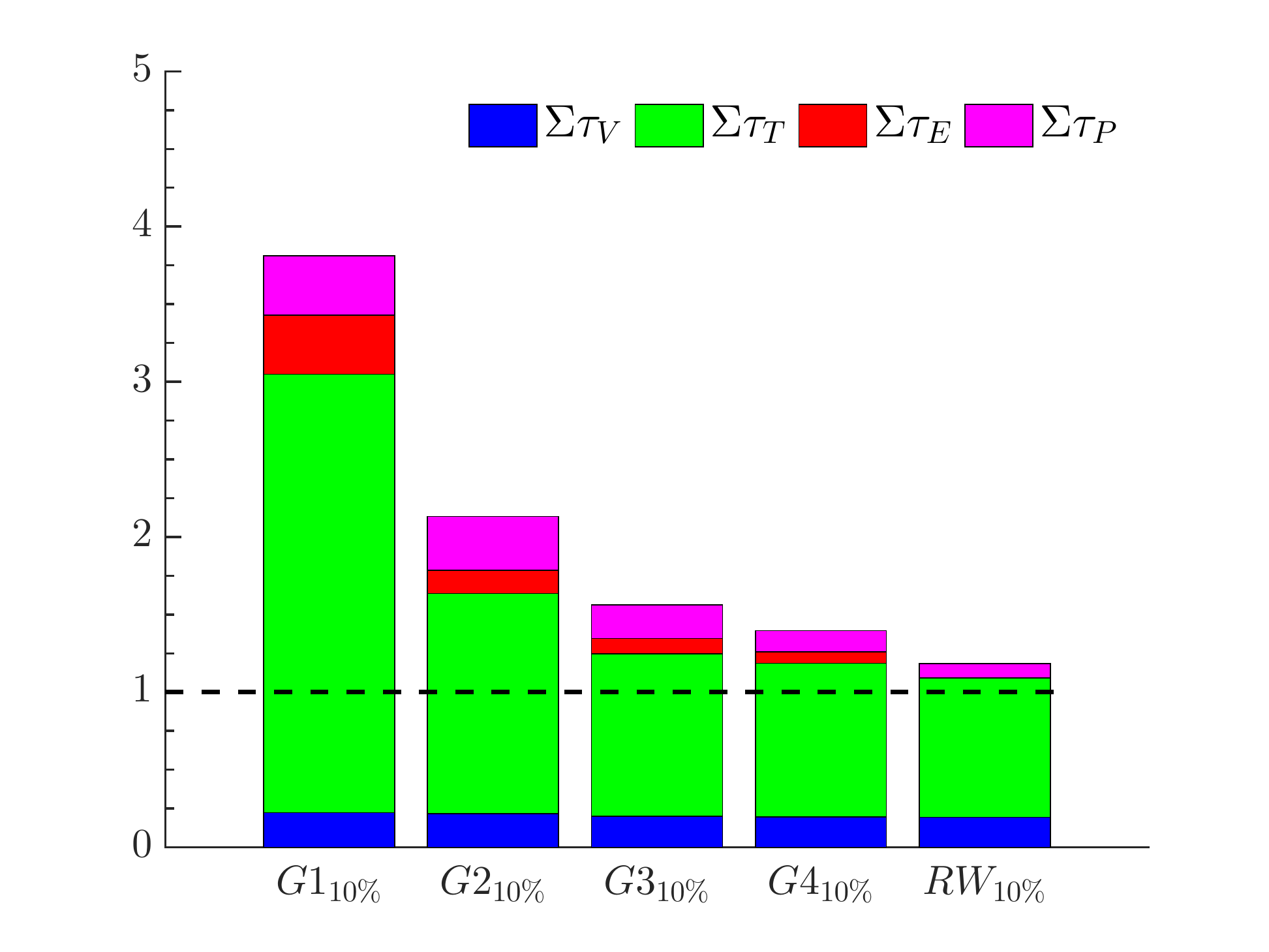} 
   \put(-382,50){\rotatebox{90}{$2 \Sigma \,\tau_{i} \, / \, h \tau_{RW}$}}    
   \put(-189,50){\rotatebox{90}{$2 \Sigma \,\tau_{i} \, / \, h \tau_{RW}$}}           
   \put(-387,130){\footnotesize $(e)$}
   \put(-194,130){\footnotesize $(f)$} \\    
  \caption{Momentum budget, normalized with $\rho u^2_\tau$, for $(a)$  $G1$, $(b)$ $G1_{10\%}$, $(c)$ $G4$ and $(d)$ $G4_{10\%}$. Here, $\tau$ is the total stress, $\tau_V$ the viscous shear stress, $\tau_T$ the turbulent Reynolds shear stress of the combined phases, $\tau_E$ the hyper-elastic contribution and $\tau_P$ the particle induced stress. Panels $(e)$ and $(f)$ show the total contribution of each stress to the drag, normalized by the wall shear stress of the single-phase flow with rigid walls, $\tau_{RW}$ (the dashed line in the figure). $RW$ and $RW_{10\%}$ are the cases with rigid walls taken by \cite{Ardekani2017}.}
\label{fig:budgetp}
\end{figure}
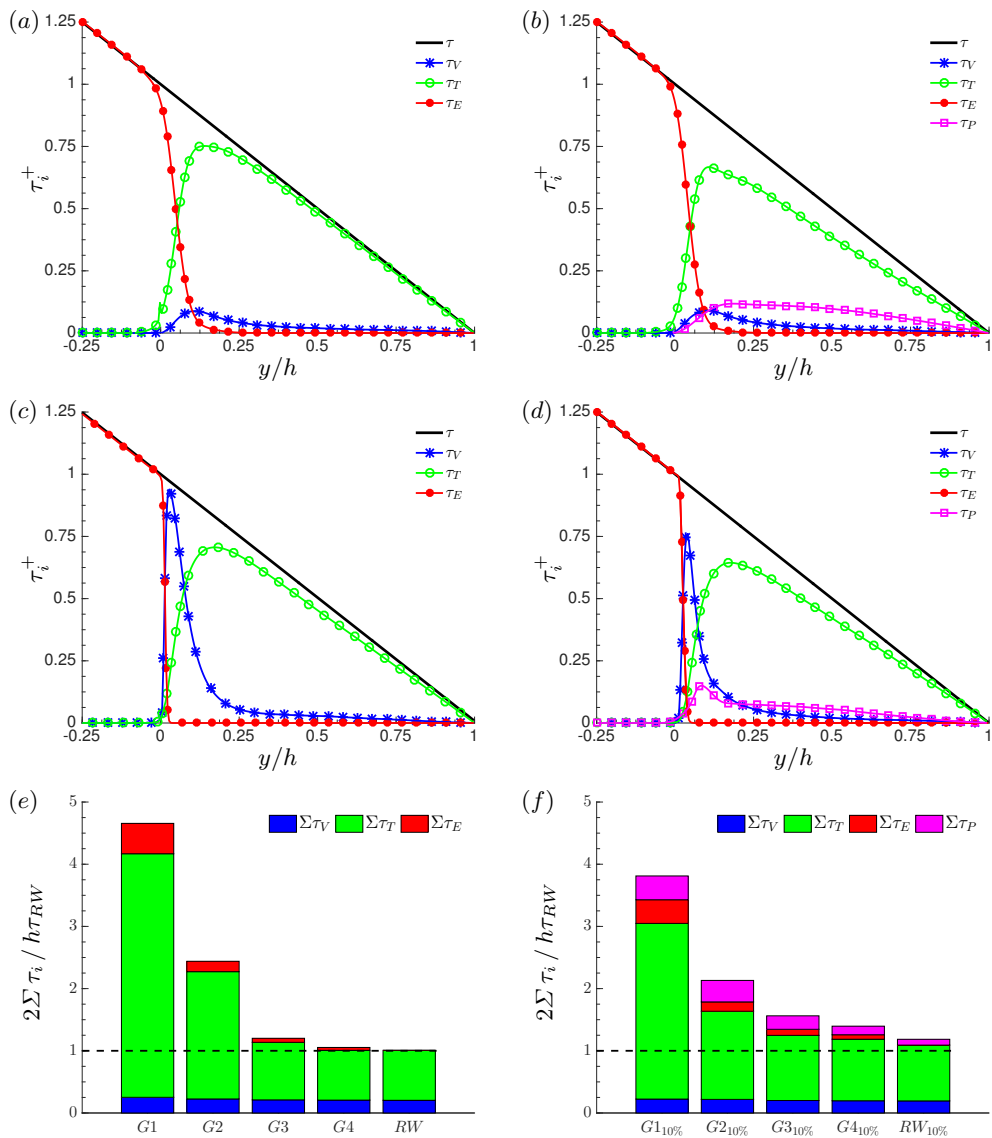

The mean velocity profiles, averaged outside of the particles are presented in figure~\ref{fig:Retau}$(a)$, together with two extra lines, pertaining the results for rigid walls \citep{Ardekani2017}. Note that the mean velocity inside the elastic walls is equal to zero. 
The results show that the mean velocity at the core of the channel increases with the wall elasticity for both particulate and single-phase cases, whereas it decreases close to the interface. Similar to the flow over rigid walls \citep{Picano2015,Ardekani2017}, particles are observed to increase the core velocity, except for the cases with highly elastic walls ($G1_{10\%}$ and $G2_{10\%}$), where the core velocity is reduced in their presence. We note here that unlike the single-phase flows over rigid walls, the mean velocity reduction close to the interface should not be interpreted as a sign of drag reduction. In fact, the friction Reynolds number $Re_{\tau}$, depicted in figure~\ref{fig:Retau}$(b)$, indicates a significant drag increase with growing wall elasticity. Note that, $Re_{\tau}$ is calculated using the mean pressure gradient needed to keep $U_b$ constant. A drag reduction, with respect to the single-phase flow of the same wall elasticity, is only achieved for the particulate cases with highly elastic walls. Results, concerning the cases $G2_{5\%}$,  $G2_{20\%}$, $G1_{FW}$ and $G4_{NPWL}$ will be discussed later in the next subsections.    

To better understand the results, concerning the drag at the wall, we perform a momentum budget analysis. In particulate turbulent flows with rigid walls, the particle-induced stress also contributes to the mean momentum transfer in the wall-normal direction, in addition to the viscous and Reynolds shear stresses. Based on the formulation proposed by \cite{Zhang2010,Picano2015} for particulate flows, and taking the hyper-elastic contribution into account \citep{Rosti20183}, we write the mean momentum balance in the channel as:
\begin{equation}
\label{eq:bx}  
\rho u^2_{\tau} \left( 1 - \frac{y}{h} \right ) \, =  \, \mu (1-\Phi) \frac{\mathrm{d} U}{\mathrm{d} y} \, - \, \rho \left[ \, \Phi \langle u^\prime_p v^\prime_p \rangle + (1-\Phi) \langle u^\prime v^\prime \rangle \right] \, + \, \langle \xi G B_{xy}  \rangle \, + \, \Phi \langle \sigma^p_{xy} \rangle \, \, \, , 
\end{equation}
where the first term on the right hand side is the viscous shear stress, denoted $\tau_V$, the second and the third terms together are the turbulent Reynolds shear stress of the combined phases, $\tau_T$, the fourth term is the contribution of the hyper-elastic wall, $\tau_E$ and the fifth term is the particle-induced stress $\tau_P$. $\sigma^p_{xy}$ in the equation above indicates the general stress in the particle phase, normal to the streamwise plane and pointing in the wall-normal direction, $\Phi$ is the mean local particle volume fraction as a function of the wall-normal distance and $u^\prime$ and $v^\prime$ are the velocity fluctuations in the streamwise and wall-normal directions with the subscript ${}_p$ denoting the particle phase. $\rho$ and $\mu$ are the density and viscosity, assumed equal in the fluid and the elastic layer. Note that in the absence of particles ($\Phi=0$) and the elastic layer ($\xi=0$), this equation reduces to the classic momentum balance for single-phase turbulent channel flow \citep{Pope2001}. The momentum transfer pertaining to each term, normalized by $\rho u^2_{\tau}$, is given in figure~\ref{fig:budgetp}$(a)-(d)$ for the cases $G1$, $G1_{10\%}$, $G4$ and $G4_{10\%}$, respectively. $\tau_P$ is computed by subtracting the three other stresses from the the total stress. Close to the interface, the share of $\tau_V$ is reduced significantly with increasing wall elasticity, as the mean velocity is shown (figure~\ref{fig:Retau}$(a)$) to decrease in this region. The hyper-elastic stress, $\tau_E$, instead takes over the momentum transfer, growing linearly within the elastic layer, where all other terms in equation~\ref{eq:bx} tend to zero. Note that, this linear line can be extrapolated to calculate the exact wall drag at $y=0$, obtaining a drag which is consistent with the ones reported above based on the mean pressure gradient. Far from the interface region, $\tau_T$ as expected plays the main role in momentum transfer. The share of $\tau_P$ for the particulate cases is observed to increase with the wall elasticity, while the peak close to the interface disappears. 

Integrating equation (\ref{eq:bx}) from $0$ to $h$ and substituting $\rho u^2_{\tau}$ with the total shear stress at the wall, $\tau_w$, leads to the contribution of each stress to $\tau_w$, which can be written as:
 \begin{equation}
\label{eq:FlInt}  
\tau_w \, =  \, \frac{2}{h} \left[ \Sigma \tau_{V} + \Sigma \tau_{T} + \Sigma \tau_{E} + \Sigma \tau_{P} \right] \, \, .
\end{equation}
Each contribution times $2/h$, normalized by the wall shear stress of the single-phase flow over rigid walls ($\tau_{RW}$), is given in figure~\ref{fig:budgetp}$(e)$ and \ref{fig:budgetp}$(f)$ for the single-phase and the particulate cases with different wall elasticities. $RW$ and $RW_{10\%}$ in this figure are the cases with rigid walls from the study of \cite{Ardekani2017}, included here for a better comparison. The contribution of the viscous shear stress to the total drag is slightly varying among the different cases, however its relative contribution is noticeably reduced with increasing wall elasticity. $\Sigma \tau_{P}$ is monotonically increased with decreasing $G^*$. A turbulence activity enhancement, with respect to $RW$, can be observed for all cases; however, we note the non-monotonic effect of the particles on  $\Sigma \tau_{T}$ and $\Sigma \tau_{E}$ when comparing the results with the single-phase flows of the same wall elasticity: turbulence attenuation for highly elastic walls, while increasing the turbulence activity in the cases with less wall elasticity.  
 
\subsection{Turbulence modulation}\label{subsec:turb}
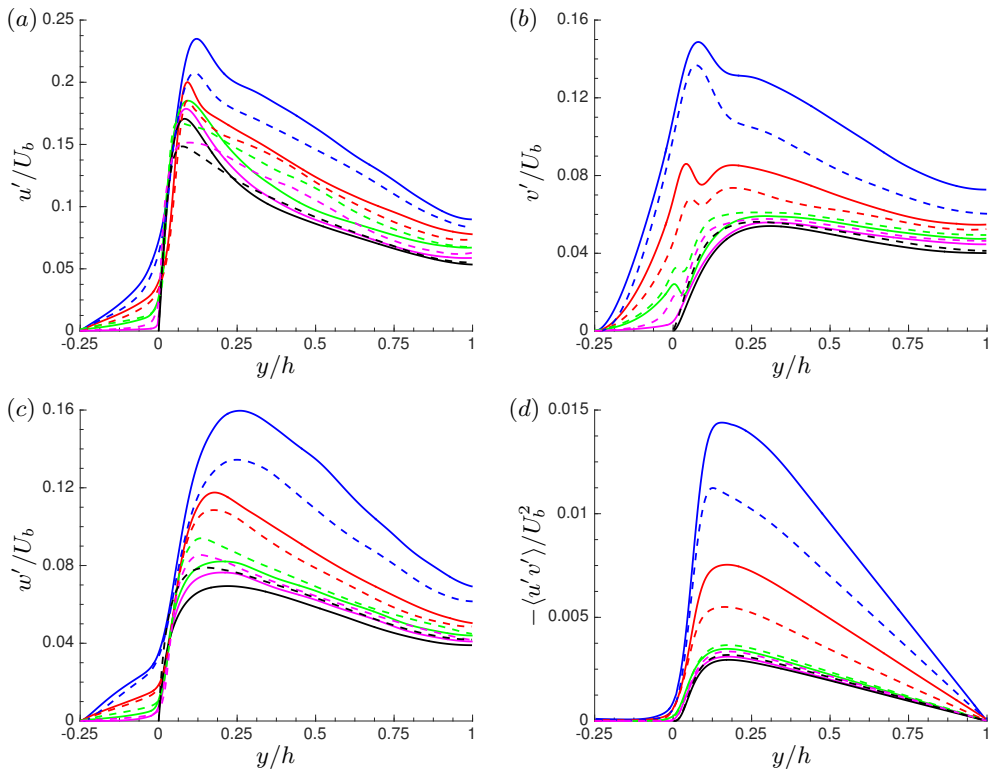
\begin{figure}
  \centering
   \includegraphics[width=0.496\textwidth]{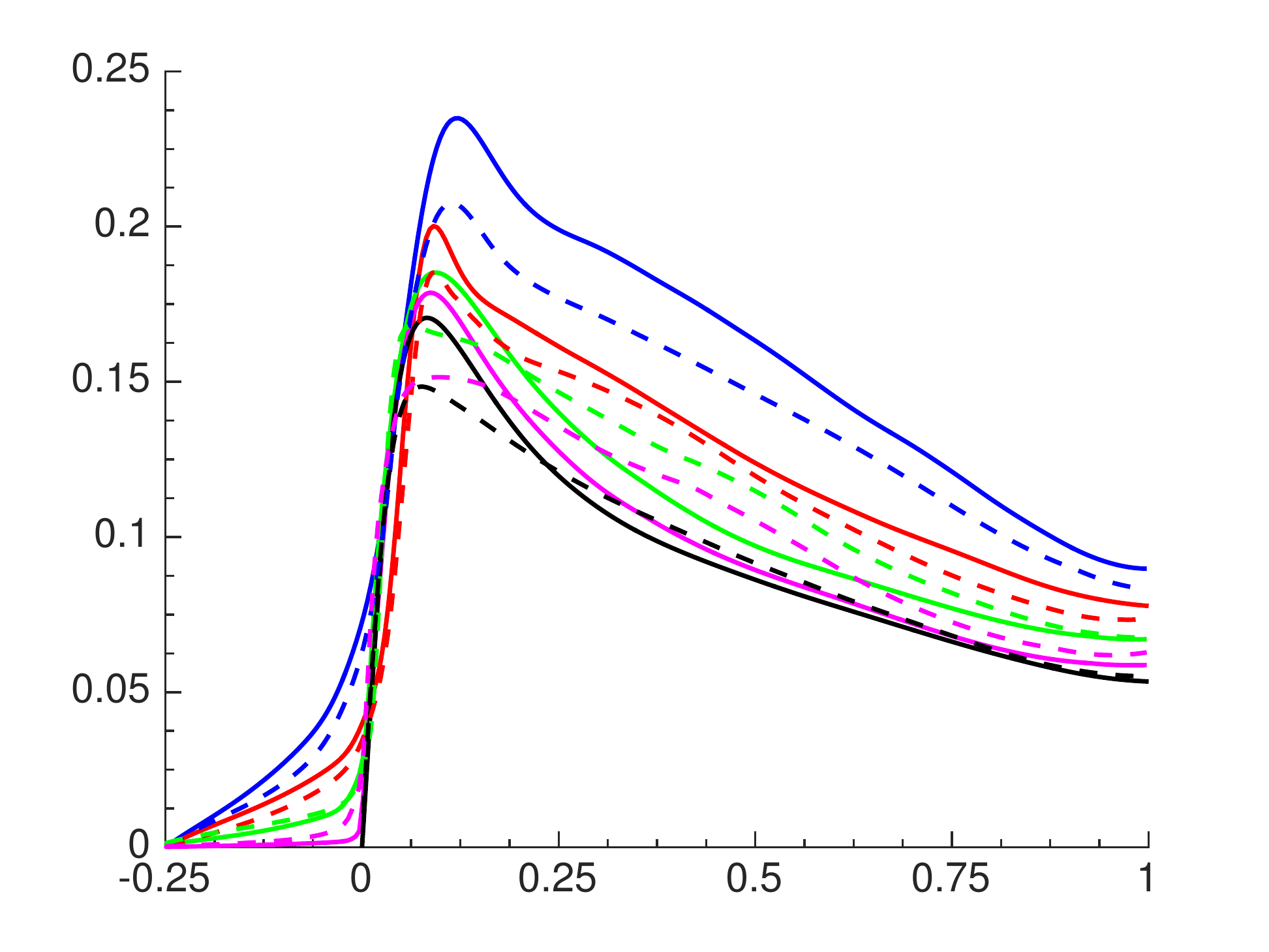}
   \includegraphics[width=0.496\textwidth]{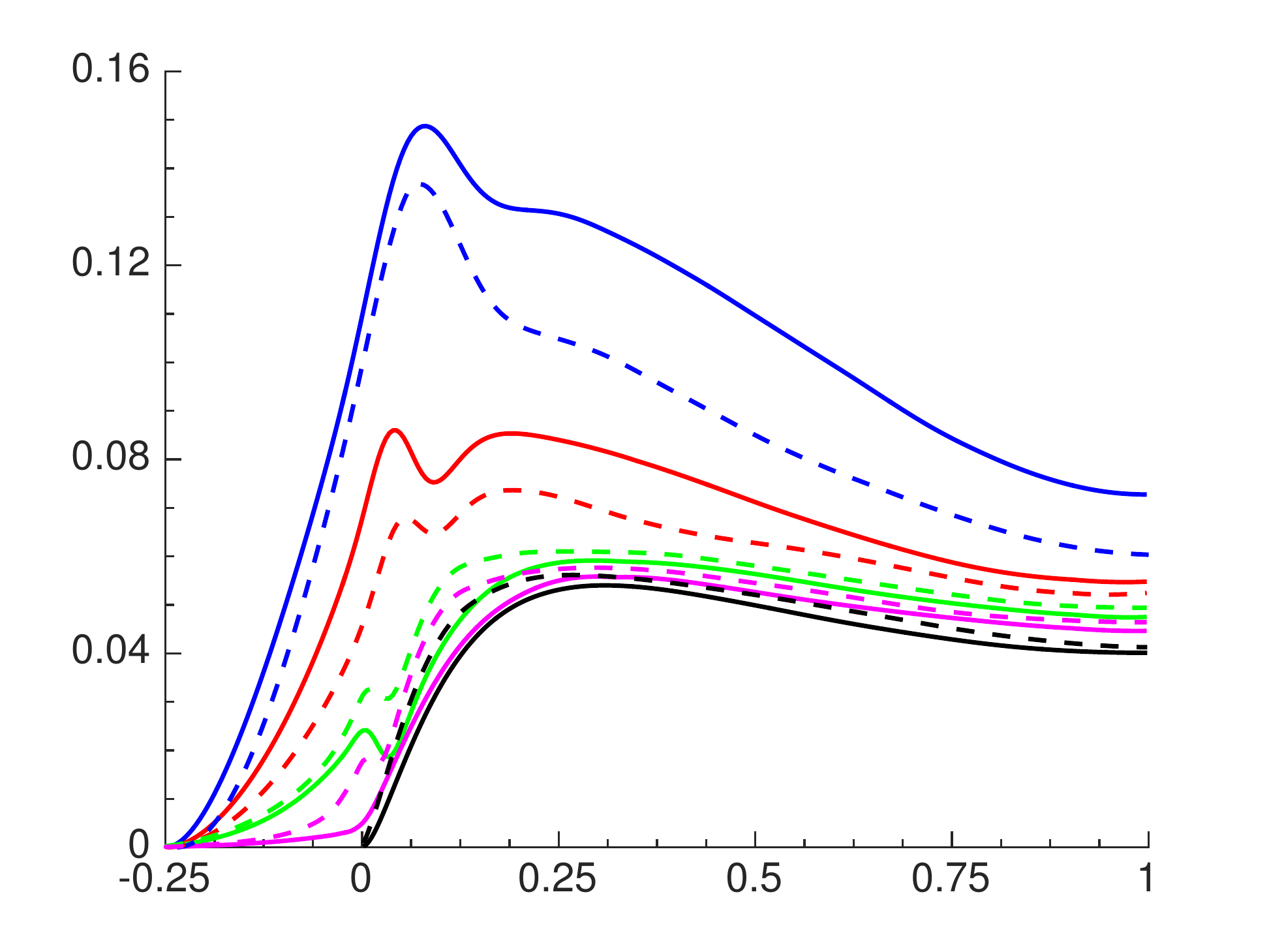} 
   \put(-385,64){\rotatebox{90}{$u^\prime / U_b$}}
   \put(-194,64){\rotatebox{90}{$v^\prime / U_b$}}   
   \put(-99,0){{$y / h$}}
   \put(-293,0){{$y / h$}}     
   \put(-387,130){\footnotesize $(a)$}
   \put(-198,130){\footnotesize $(b)$} \\ 
   \includegraphics[width=0.496\textwidth]{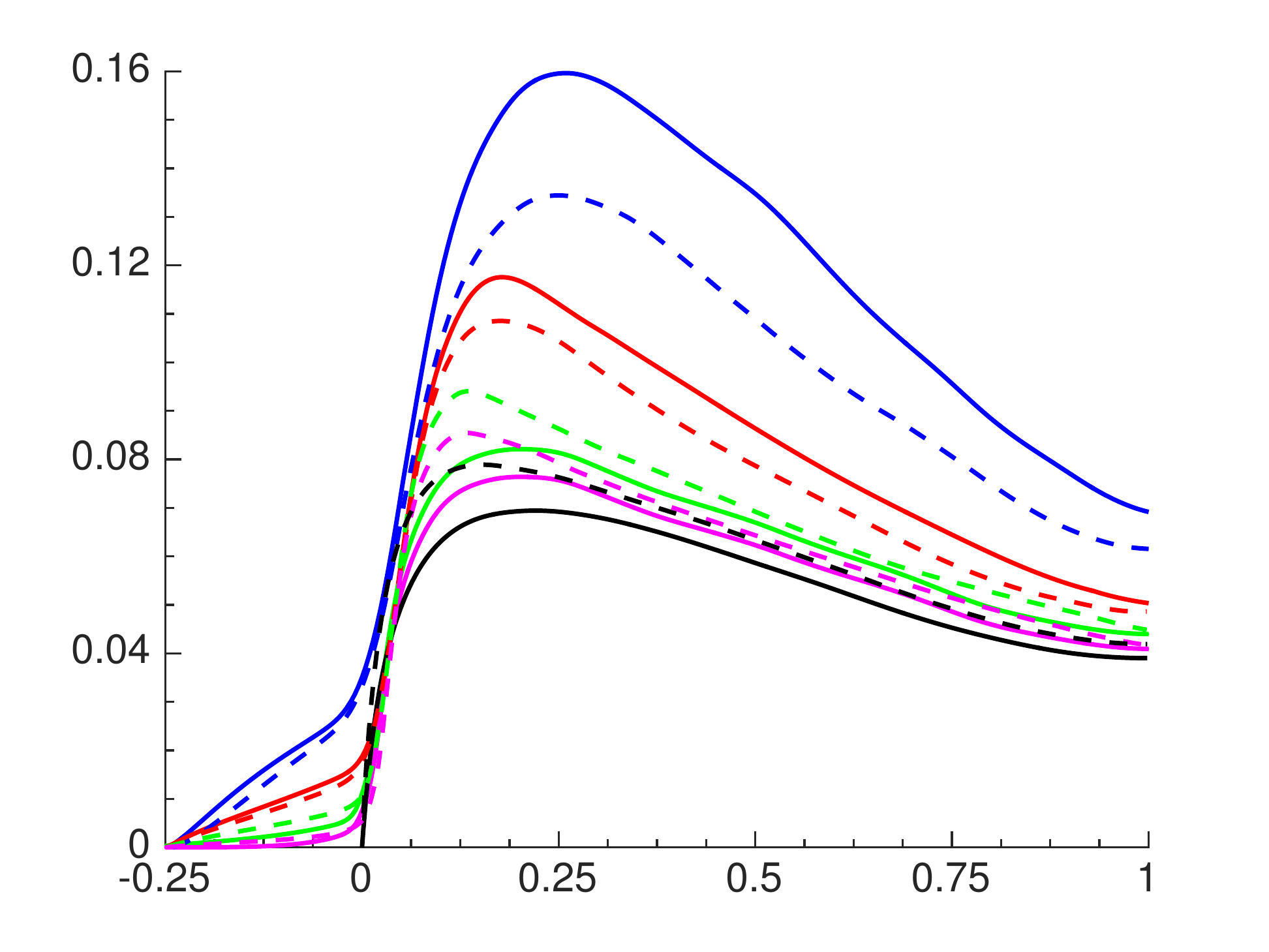}
   \includegraphics[width=0.496\textwidth]{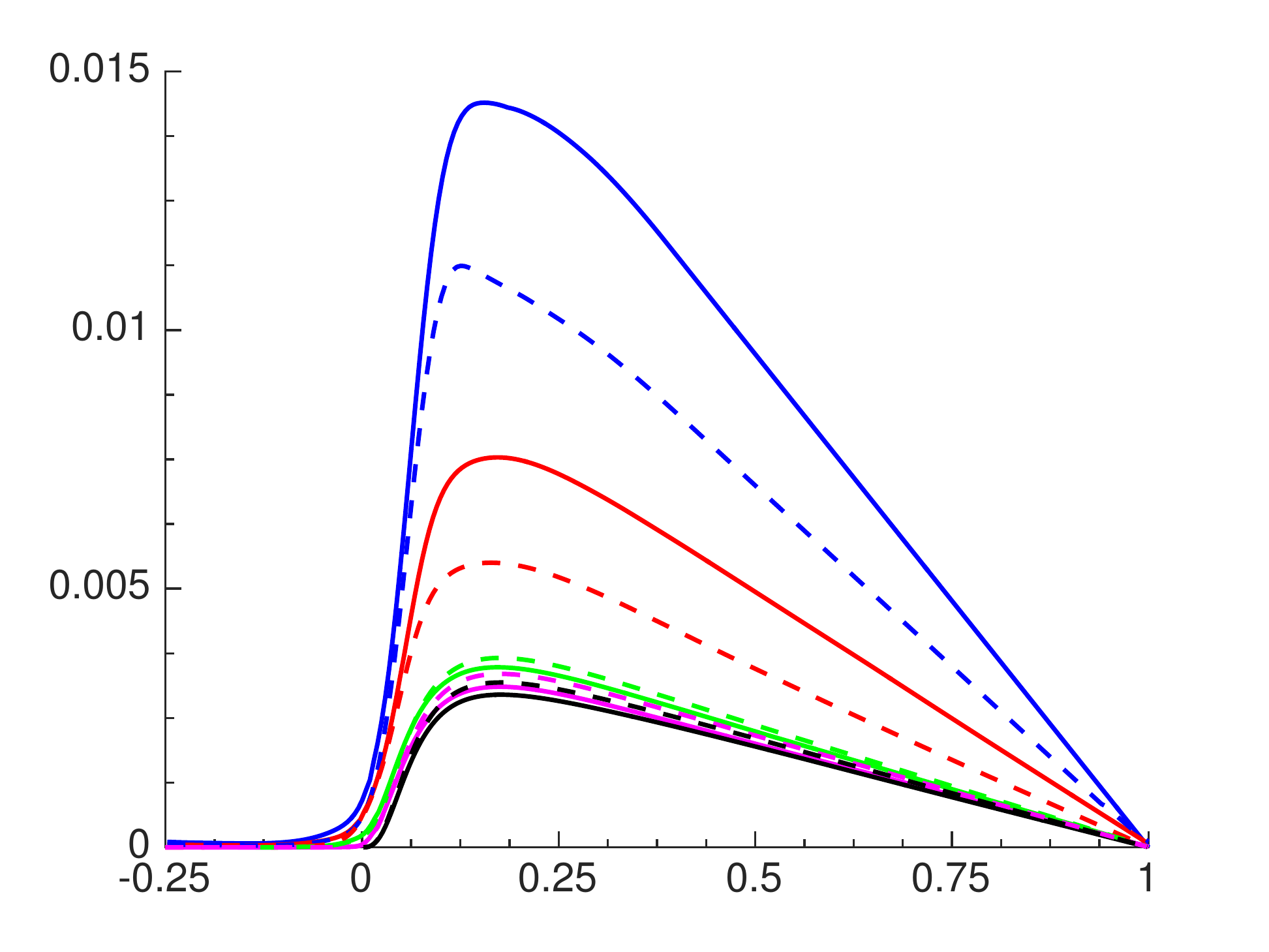}
   \put(-385,64){\rotatebox{90}{$w^\prime / U_b$}}
   \put(-194,51){\rotatebox{90}{$-\langle u^\prime v^\prime \rangle / U^2_b$}}    
   \put(-99,0){{$y / h$}}
   \put(-293,0){{$y / h$}}     
   \put(-387,130){\footnotesize $(c)$}
   \put(-198,130){\footnotesize $(d)$} \\
  \caption{Root-mean-square velocity fluctuations and Reynolds shear stress, scaled in outer units: $(a)$ streamwise $u^\prime$; $(b)$ wall-normal $v^\prime$; $(c)$ spanwise $w^\prime$ and $(d)$ Reynolds shear stress. The colour scheme is the same as figure~\ref{fig:Retau}$(a)$. }
\label{fig:rms}
\end{figure}
In this section we take a closer look at the turbulence modulation, caused by the presence of the particles and the deformable walls. We start with the velocity fluctuations and the Reynolds shear stress profiles, depicted in figure~\ref{fig:rms}. Velocity fluctuations are strongly affected by the elastic layer, lingering inside as the wall elasticity increases.  Despite the strong velocity fluctuations inside the highly elastic walls, the Reynolds shear stress (see figure~\ref{fig:rms}$(d)$) is observed to be zero within these layers, indicating a decorrelation between $u^\prime$ and $v^\prime$. The Reynolds shear stress increases in the particulate cases, except in the presence of highly elastic walls where a significant attenuation is instead observed (also shown in figure~\ref{fig:budgetp}$(f)$).  The wall-normal velocity fluctuation $v^\prime$ is the most affected by the wall elasticity, due to the weakening of the wall-blocking and wall-induced viscous effects \citep{Perot1995a,Perot1995b}. This is also observed in turbulent boundary-layer flows over rough \citep{Krogstadt1999} and porous walls \citep{Rosti2018}. $v^\prime$ goes through a secondary peak in the vicinity of the interface, associated with the oscillatory interface movement. As the wall elasticity increases, this secondary peak moves farther from the interface, overcomes the classical turbulent peak and becomes its maximum value. The peak of velocity fluctuations in the streamwise direction $u^\prime$ grows considerably less than the other two directions with increasing wall elasticity, which can be attributed to the reduction of the presence of streaky structures close to the interface \citep{Breugem2006,Rosti20171,Rosti2018}. The effect of particles on the velocity fluctuations is similar to their effect on the Reynolds shear stress, except for the peaks of $u^\prime$ which are always reduced with respect to their single-phase counterpart cases. 

\begin{figure}
  \centering
   \includegraphics[width=0.495\textwidth]{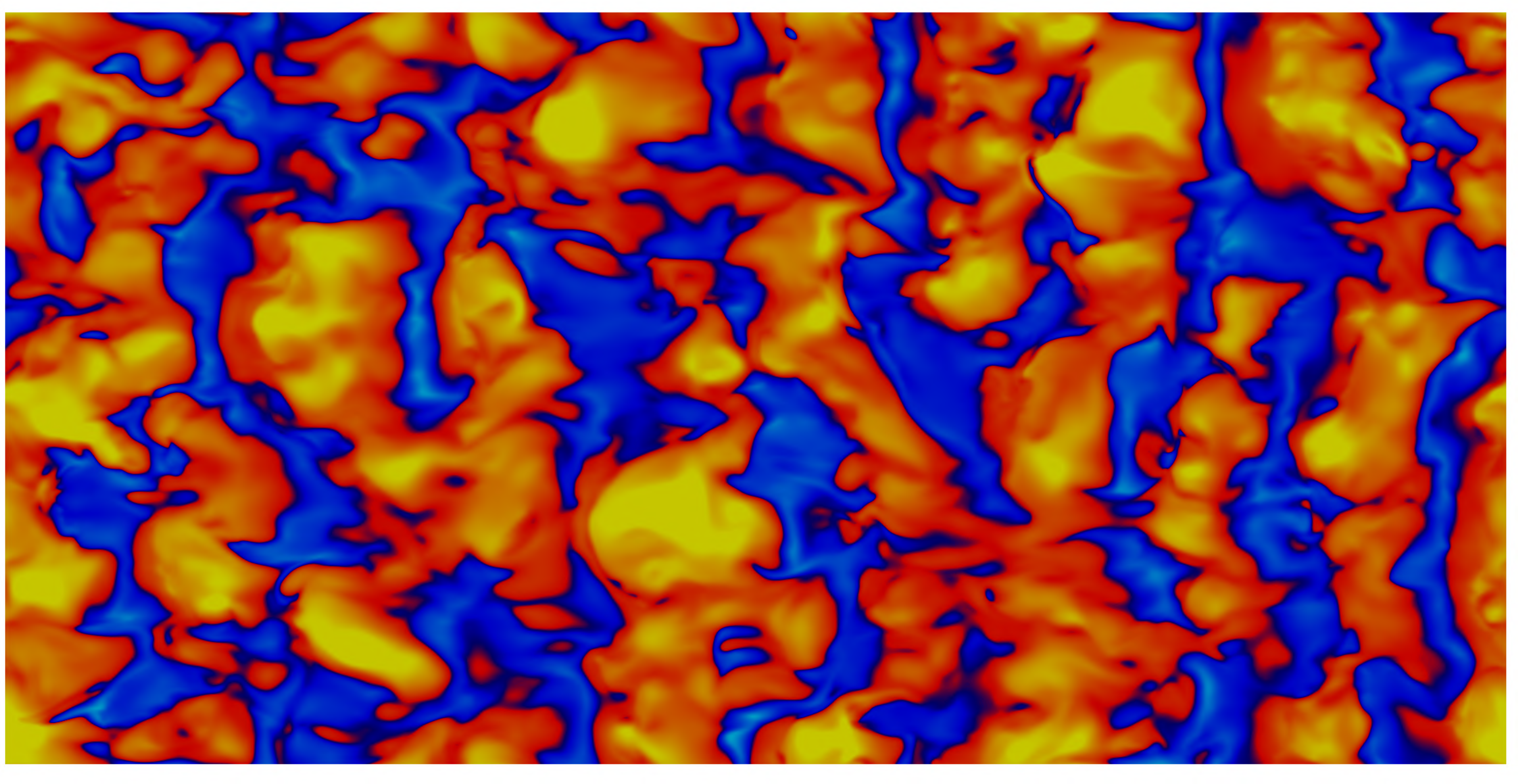}
   \includegraphics[width=0.495\textwidth]{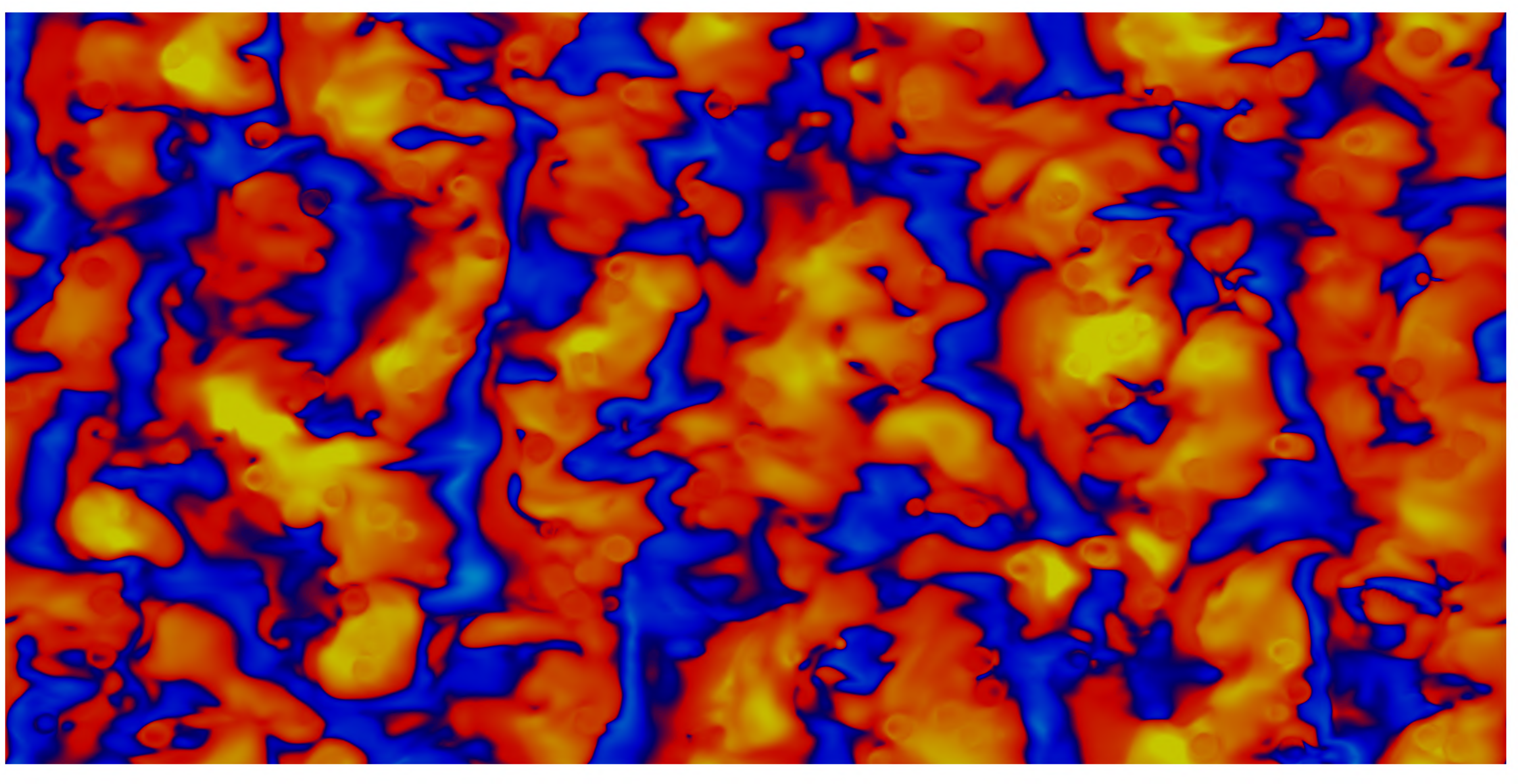}
   \put(-392,47){{$z$}}    \\
   \includegraphics[width=0.495\textwidth]{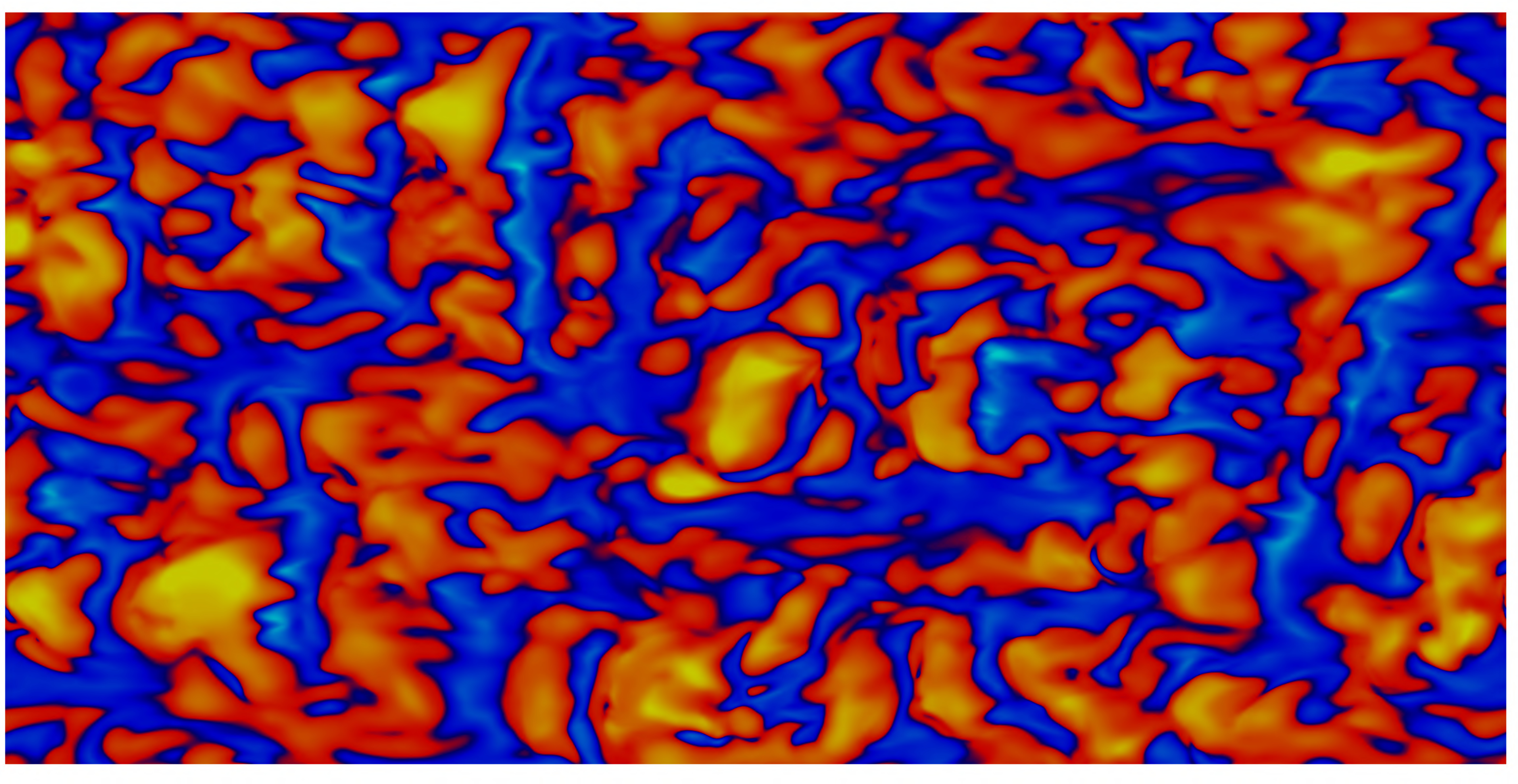}
   \includegraphics[width=0.495\textwidth]{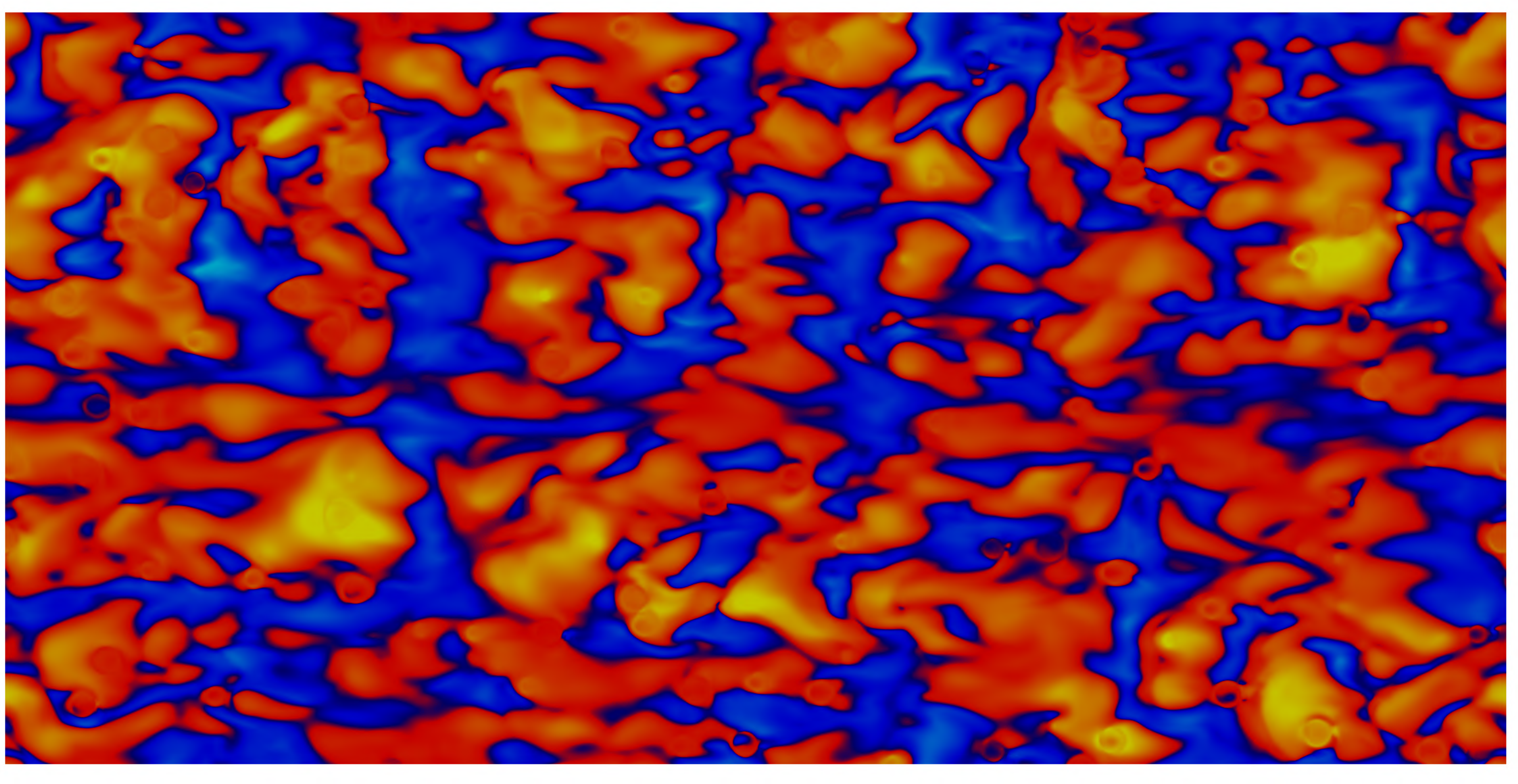}
   \put(-392,47){{$z$}}    \\
   \includegraphics[width=0.495\textwidth]{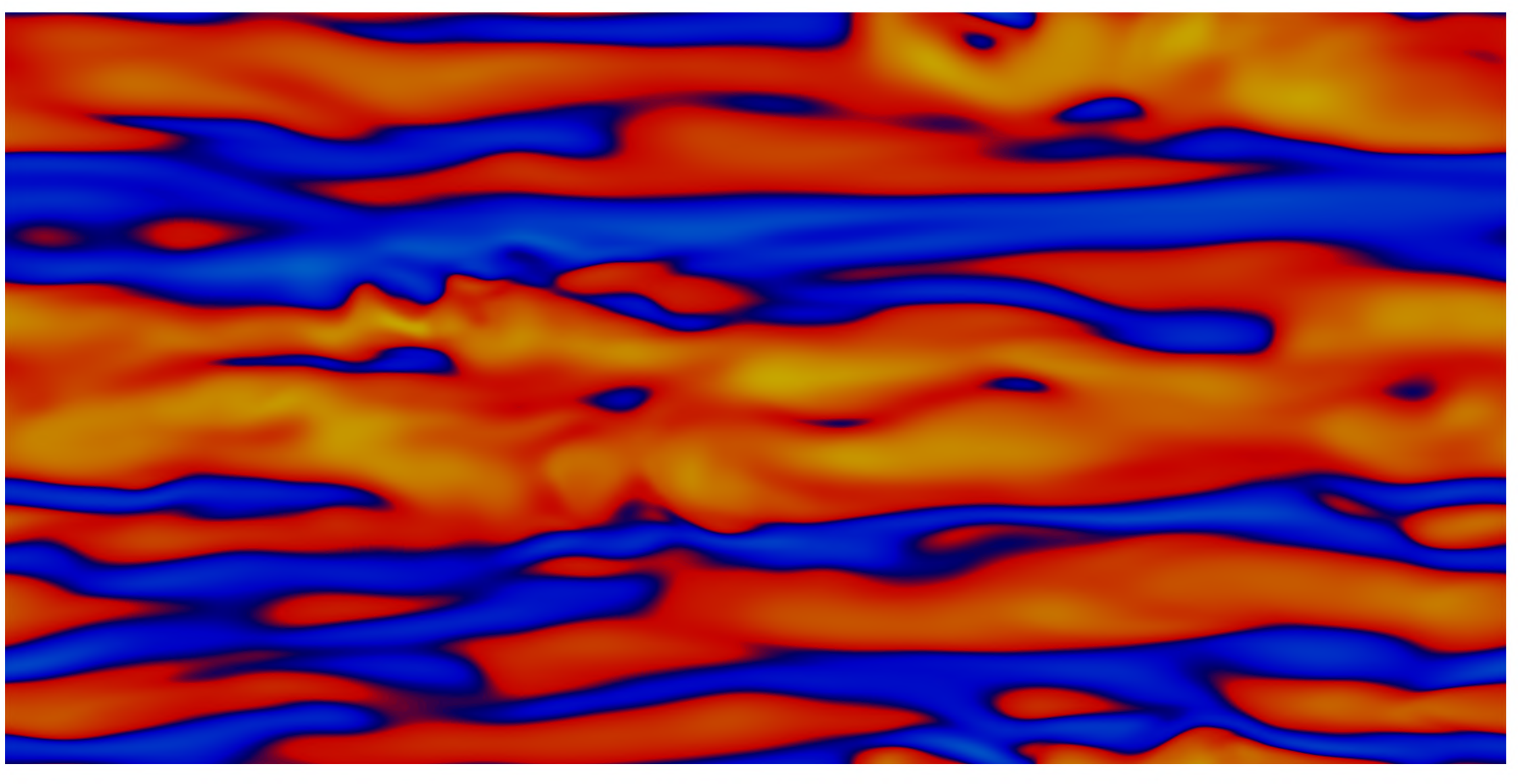}
   \includegraphics[width=0.495\textwidth]{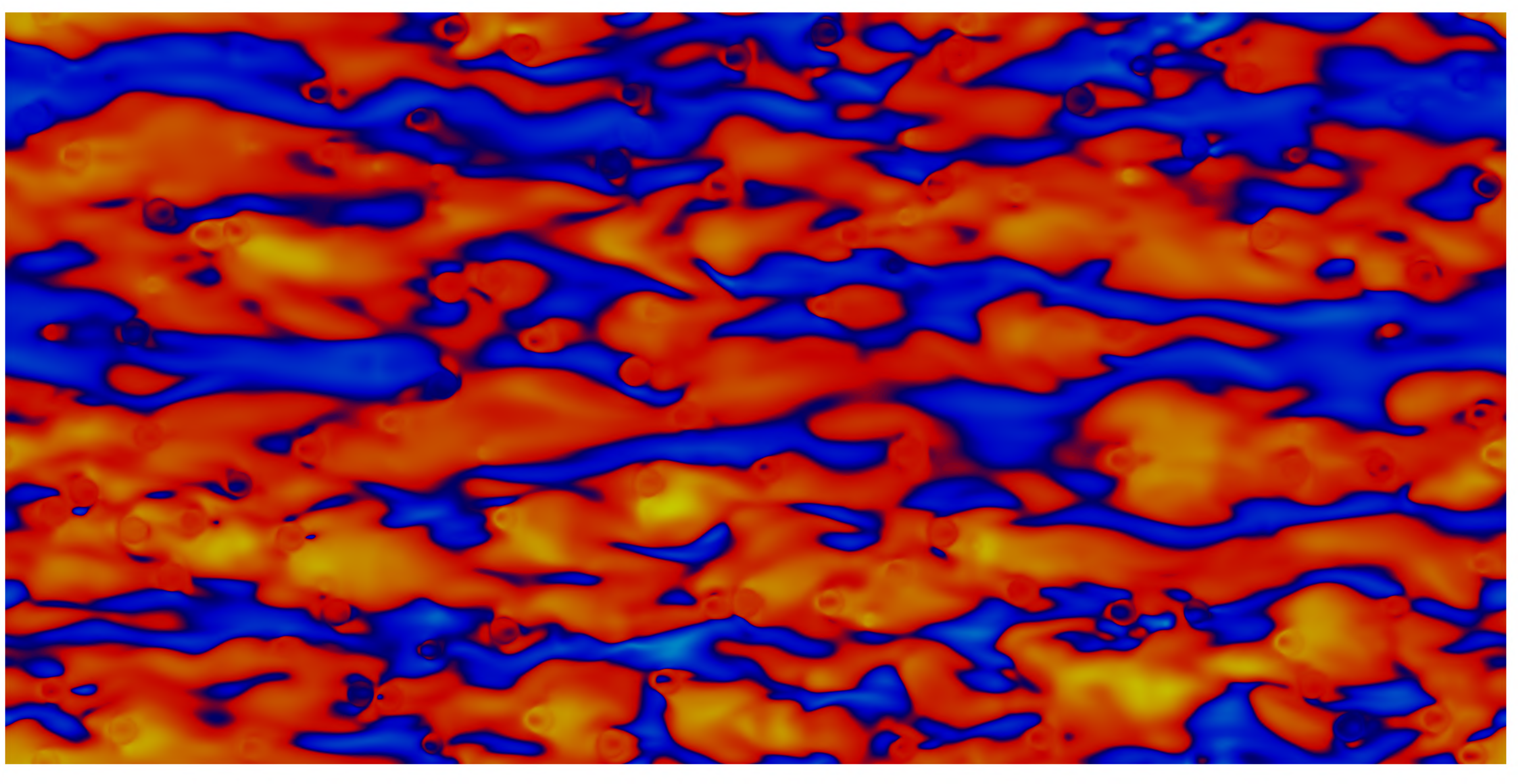}
   \put(-392,47){{$z$}}    \\   
   \includegraphics[width=0.495\textwidth]{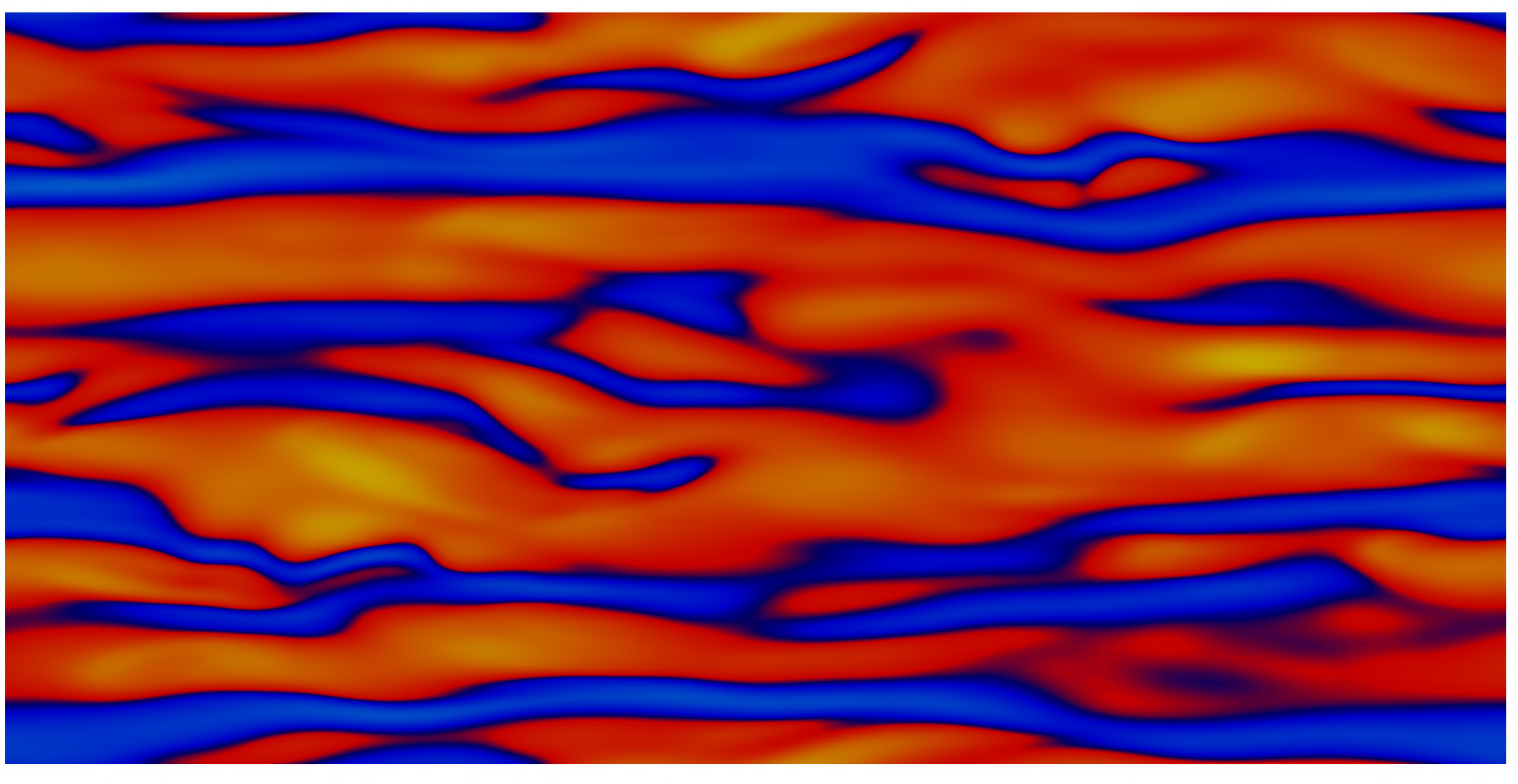}
   \includegraphics[width=0.495\textwidth]{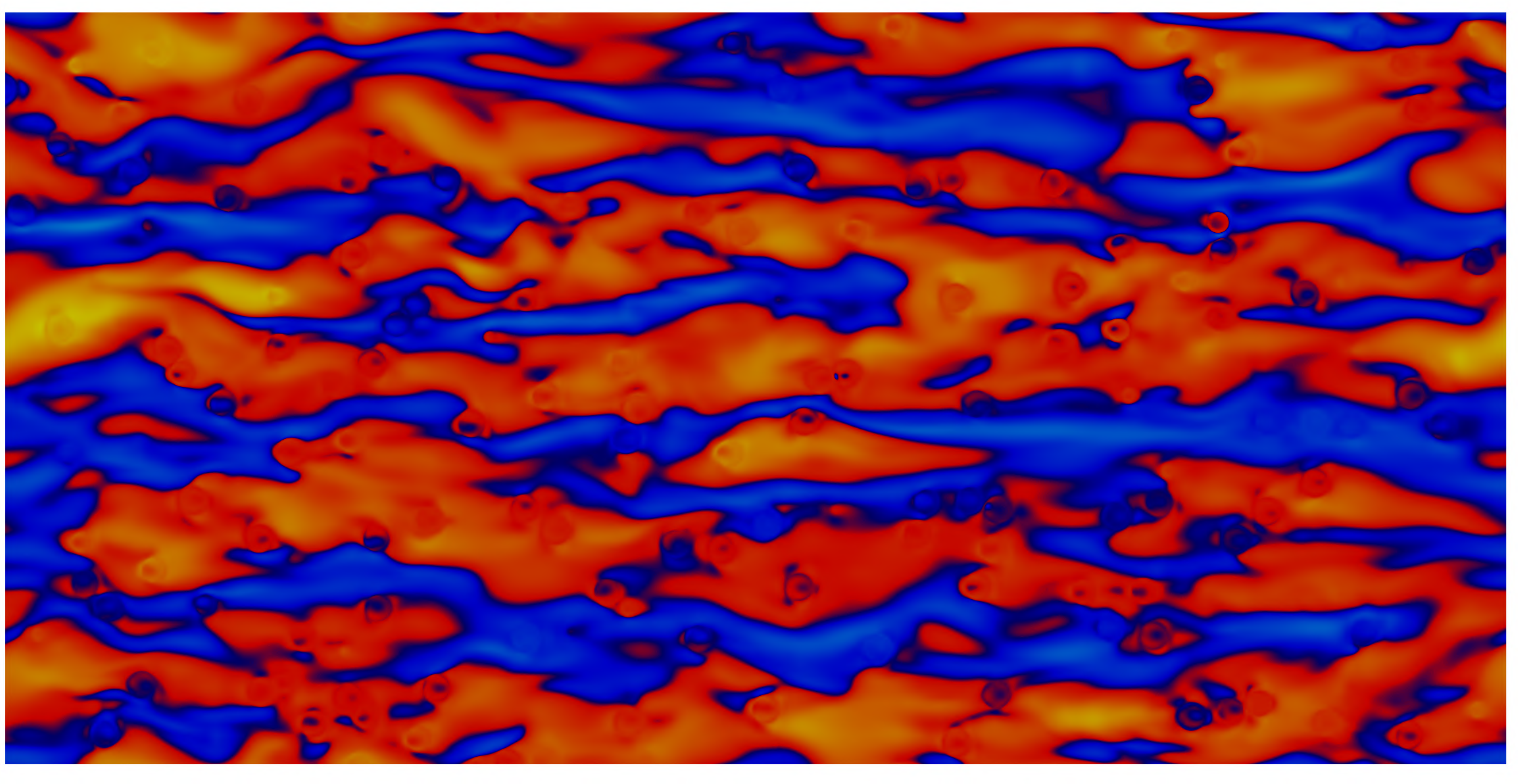}
   \put(-392,47){{$z$}}    
   \put(-110,-7){{$x \longrightarrow$}}
   \put(-298,-7){{$x \longrightarrow$}}\\       
  \caption{Instantaneous contours of the streamwise velocity fluctuations $u^\prime$ in the wall-parallel plane $x-z$ at $y/h=0.1$. The colour scale goes from $-0.5 U_b$ (blue) to $0.5 U_b$ (yellow). The left column indicates the single-phase flow cases ($G1$ to $G4$) and the right column, the particulate ones ($G1_{10\%}$ to $G4_{10\%}$), with $G^*$ increasing from top to bottom.}
\label{fig:2dsnap}
\end{figure}
We take a closer look at the turbulent flow near the interface region ($y/h=0.1$) in figure~\ref{fig:2dsnap}, where instantaneous snapshots of the streamwise velocity fluctuations $u^\prime$ are visualized in the wall-parallel plane $x-z$. The so-called high- and low-speed streaks, characteristic of wall-bounded turbulence, can be observed near the less elastic walls. As the wall elasticity increases (from bottom to top), these streamwise streaky structures appear less elongated and more fragmented, broken into smaller pieces due to the interface movement. Further increasing the wall elasticity results in the formation of relatively large structures, elongated in the spanwise direction (top left), similar to the findings of \cite{Rosti20171}. These spanwise structures have been observed before in the turbulent flow over porous walls \citep{Breugem2006,Rosti2015,Rosti2018}, rough surfaces \citep{Raupach1991,Jimenez2001} and plant canopies \citep{Finnigan2000}. \cite{Breugem2006} attributed these elongated structures to a Kelvin-Helmholtz type instability that is triggered by the inflection point in the mean velocity profiles (Rayleigh's criterion \citep{Drazin2004}), creating roller-type spanwise vortices. The snapshots of $u^\prime$ in the particulate cases are shown in the right column of figure~\ref{fig:2dsnap}. Particles can be observed to break both the streamwise and the spanwise structures into less elongated and more fragmented pieces, with the flow being rich of small-scale features. 

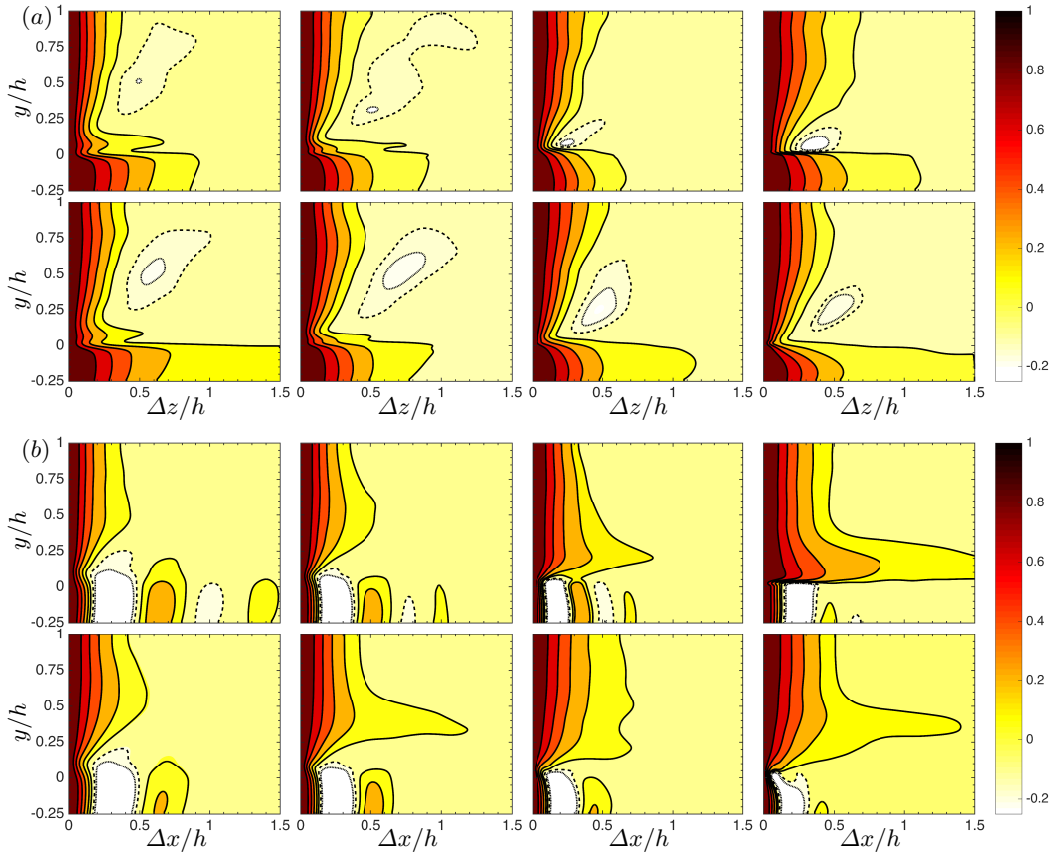
\begin{figure}
  \centering
   \includegraphics[width=1\textwidth]{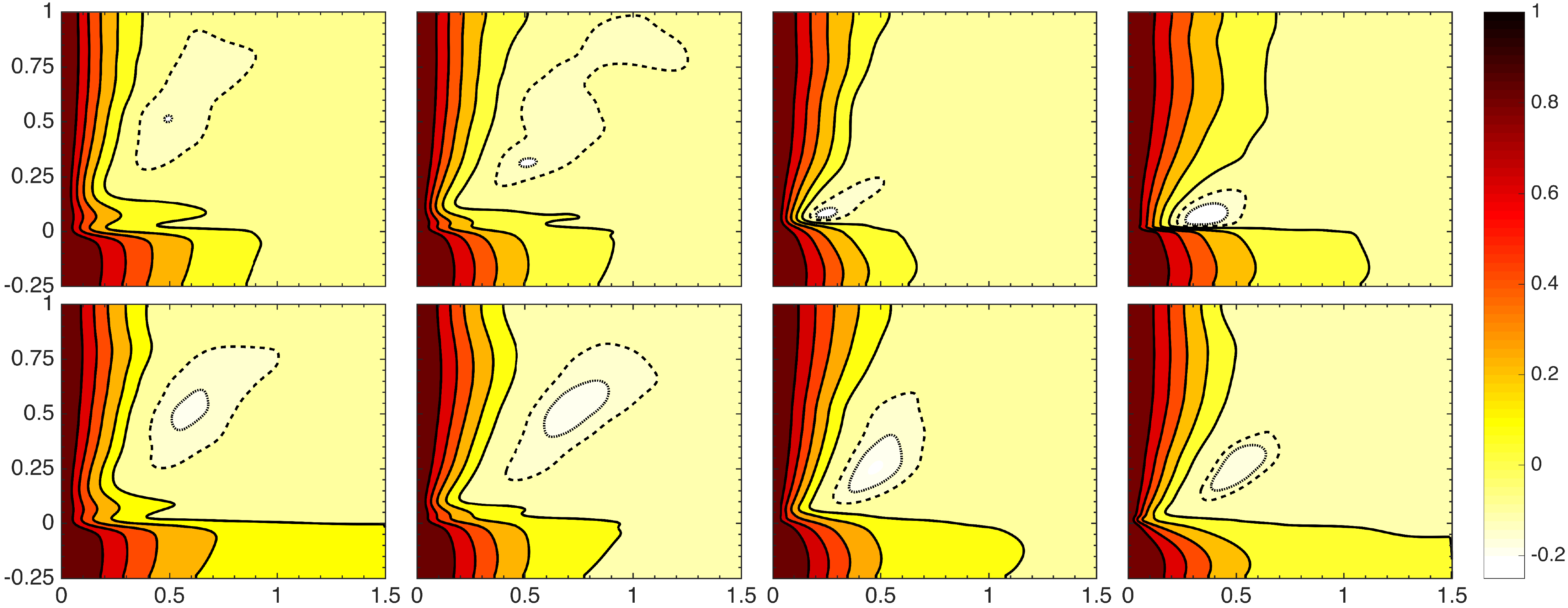}
   \put(-387,141){\footnotesize $(a)$}   
   \put(-392,34){\rotatebox{90}{$y / h$}}
    \put(-392,106){\rotatebox{90}{$y / h$}}   
   \put(-79,-5){{$\Delta z / h$}}
   \put(-166,-5){{$\Delta z / h$}}   
   \put(-253,-5){{$\Delta z / h$}}    
   \put(-340,-5){{$\Delta z / h$}}        \\ [5pt]
   \includegraphics[width=1\textwidth]{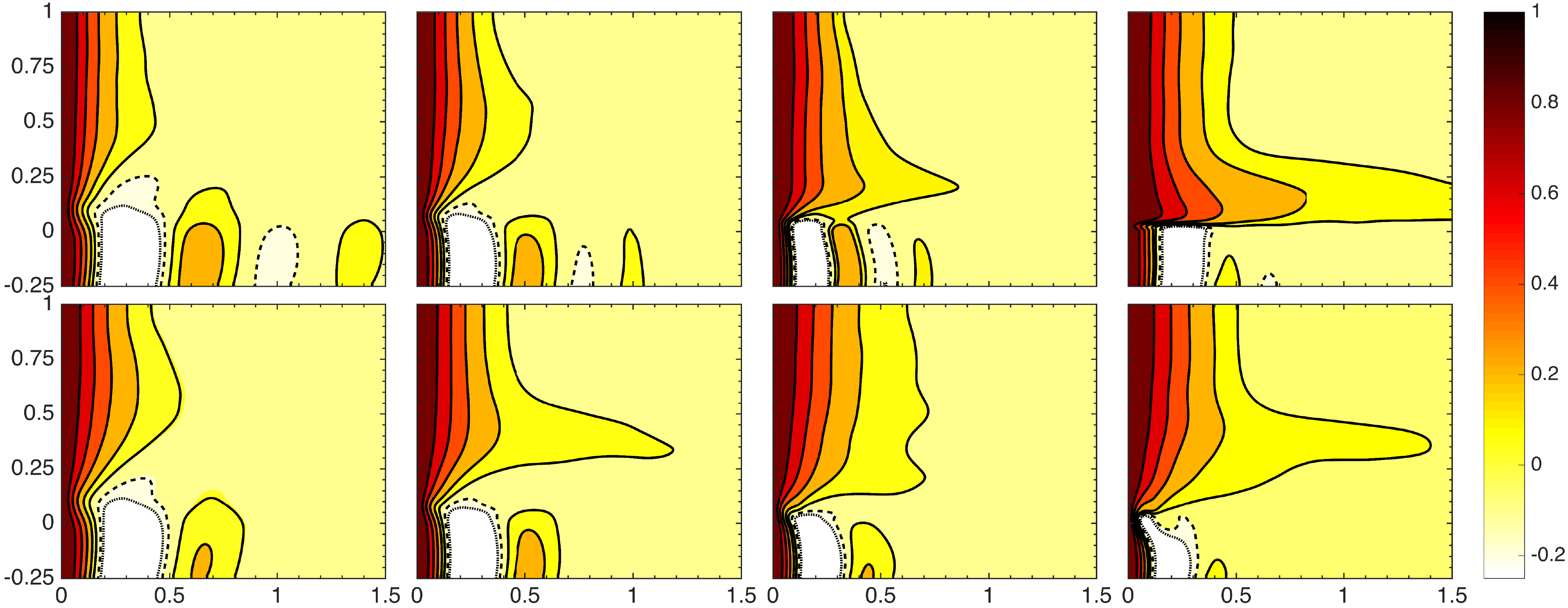}
   \put(-387,141){\footnotesize $(b)$}    
   \put(-392,34){\rotatebox{90}{$y / h$}}
    \put(-392,106){\rotatebox{90}{$y / h$}}   
   \put(-79,-5){{$\Delta x / h$}}
   \put(-166,-5){{$\Delta x / h$}}   
   \put(-253,-5){{$\Delta x / h$}}    
   \put(-340,-5){{$\Delta x / h$}}   \\   
  \caption{Line and colour contours of the one-dimensional autocorrelation of $(a)$ the streamwise velocity fluctuations as a function of the spanwise spacing ($R^z_{uu} (y,\Delta z)$) and $(b)$ the wall-normal velocity fluctuations as a function of the streamwise spacing for different $y/h$ ($R^x_{vv} (y,\Delta x)$). The solid lines correspond to positive values, ranging form $0$ to $0.8$ with a step of $0.2$ between two neighbouring lines, whereas dashed and dotted lines indicate the isolines at $-0.1$ and $-0.2$, respectively. Results, pertaining the single-phase flow (cases $G1$ to $G4$), are shown in the first row of each panel, while the autocorrelations for the particulate cases (cases $G1_{10\%}$ to $G4_{10\%}$) are depicted in the second row. $G^*$ increases from left to right.}
\label{fig:rspu}
\end{figure}
To gain a better understanding of the turbulence structures and in order to quantify the observations in figure~\ref{fig:2dsnap}, we compute the two-point spatial correlation of the streamwise and the wall-normal velocity fluctuations as a function of the spanwise and streamwise spacing, respectively as:
\begin{eqnarray}
\label{eq:UT}  
R^z_{uu} (y,\Delta z) \, &=&  \, \frac{\langle u^\prime (x,y,z,t) \, u^\prime (x,y,z+\Delta z,t) \rangle}{{u^\prime}^2} \, \, , \\ [8pt] 
R^x_{vv} (y,\Delta x) \, &= & \, \frac{\langle v^\prime (x,y,z,t) \, v^\prime (x+\Delta x,y,z,t) \rangle}{{v^\prime}^2} \, \, .
\end{eqnarray}
Line and colour contours of $R^z_{uu} (y,\Delta z)$ and $R^x_{vv} (y,\Delta x)$ are presented in figure~\ref{fig:rspu}$(a)$ and \ref{fig:rspu}$(b)$ for the different studied cases. It is well known that the autocorrelation of the streamwise velocity fluctuations along the spanwise direction in a single-phase turbulent flow exhibits a negative local minimum value in the near wall region at around $\Delta z^+ \approx 50-60$ \citep{Pope2001}, which indicates half of the spacing between two neighboring  streaks. A similar spacing is obtained in figure~\ref{fig:rspu}$(a)$ for the cases with less elastic walls $G3$ and $G4$ (top right), as they go though a local minimum at $\Delta Z^+\approx55$ and $\Delta Z^+\approx58$, respectively. The autocorrelations plot show the disruption of the high- and low-speed streaks, when increasing the wall elasticity, and their shift away from the interface region. The absence of high- and low-speed streaks, close to the highly elastic walls, can be attributed to i) a strong reduction in the mean shear  (see figure~\ref{fig:Retau}$(a)$), an important ingredient in the streaks formation mechanism \citep{Lee1990} and ii) strong wall-normal velocity fluctuations preventing the development of these elongated structures. A positive and correlated region for $R^z_{uu}$ can be observed instead, at $y/h\approx0.1$ and $0.07$ for the highly elastic cases $G1$ and $G2$: an indication of the elongated structures, forming in the spanwise direction (also shown in figure~\ref{fig:2dsnap}) with an approximate length of $0.7-0.8h$. $R^z_{uu}$ for the particulate cases are given in the second row of figure~\ref{fig:rspu}$(a)$. The presence of particles shortens the spanwise length of the structures in the cases $G1_{10\%}$ and $G2_{10\%}$, having an effect similar to the wall elasticity on disrupting and shifting away the high- and low-speed streaks towards the channel centre. The disruption of the near wall streaks was reported before in the turbulent channel flows of finite-size spherical particles \citep{Picano2015,Ardekani2018AR,Wang2018add,Costa2018}.  

Next, we show the autocorrelations of the wall-normal velocity fluctuations along the streamwise direction $R^x_{vv} (y,\Delta x)$ in figure~\ref{fig:rspu}$(b)$. Results clearly reveal that the long quasi-streamwise vortices, associated with the presence of streaks, vanish when increasing the wall elasticity and are substituted, instead, with roller-type vortices in the spanwise direction. The active range of these vortical structures grows with the wall elasticity, reaching to a size of approximately $140^+$ in the most elastic case $G1$. Particles are responsible for reducing the height of these vortices in the highly elastic cases, whereas they shorten and disrupt the quasi-streamwise vortices in the less deformable cases. Indeed, breaking the large spanwise vortices near highly elastic walls results in less wall deformation and thus, an overall attenuation of the turbulent activity (see also figure~\ref{fig:snapshots}$(b)$); however, it should be also noted that breaking the quasi-streamwise vortices does not necessarily decrease the turbulent activity. Indeed, it has been shown before that spherical particles with the same size and volume fraction as here induce an increase of the turbulent activity near the rigid walls \citep{Picano2015,Ardekani2017}, similarly to what observed in the results of this study for the less elastic cases (see figure \ref{fig:rms}$(d)$). 

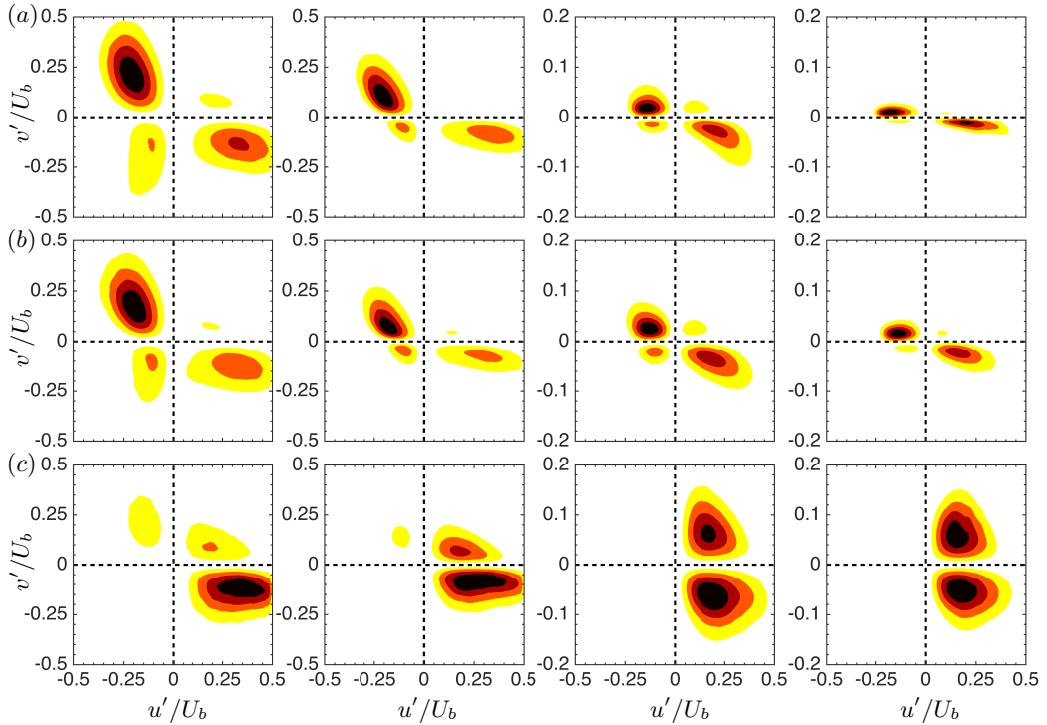
\begin{figure}
  \centering
   \includegraphics[width=1\textwidth]{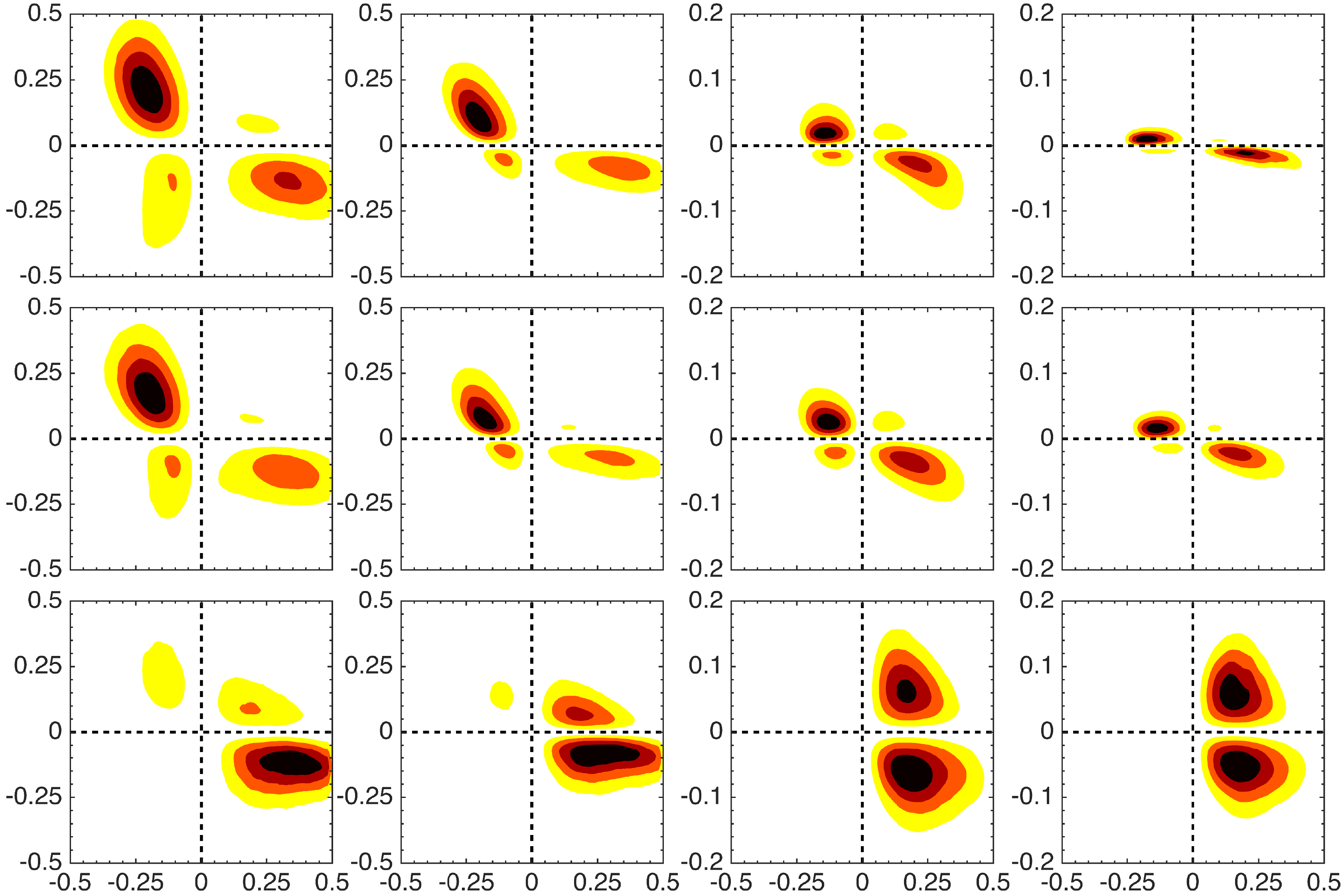}
   \put(-389,250){\footnotesize $(a)$}   
   \put(-389,166){\footnotesize $(b)$}  
   \put(-389,82){\footnotesize $(c)$}        
   \put(-388,37){\rotatebox{90}{$v^\prime / U_b$}}      
   \put(-388,121){\rotatebox{90}{$v^\prime / U_b$}}  
   \put(-388,205){\rotatebox{90}{$v^\prime / U_b$}}  
   \put(-53,-10){{$u^\prime / U_b$}}
   \put(-147,-10){{$u^\prime / U_b$}}     
   \put(-242,-10){{$u^\prime / U_b$}}
   \put(-336,-10){{$u^\prime / U_b$}}  \\                
  \caption{Contours of the weighted Reynolds shear stress at $y/h = 0.05$, given by multiplying the absolute value of the Reynolds shear stress by the joint probability density of its occurrence in the $u^\prime-v^\prime$ plane for: $(a)$ the single-phase flow cases $G1$ to $G4$, $(b)$ the particulate cases $G1_{10\%}$ to $G4_{10\%}$ and $(c)$ spherical shells 5\% larger than the particles in $G1_{10\%}$ to $G4_{10\%}$. $G^*$ increases from left to right and the colours go from yellow to black (maximum)}.
\label{fig:jpdf}
\end{figure}
So far we have presented a clear picture of the coherent turbulence structures near elastic walls also in the presence of finite-size particles; however, the way these structures contribute to the enhancement or attenuation of the Reynolds shear stress is yet to be understood. To this purpose, we compute the weighted Reynolds shear stress, given by multiplying the absolute value of Reynolds shear stress by the joint probability density of its occurrence in the $u^\prime-v^\prime$ plane \citep{Zhou1999}. This analysis indicates the turbulent events that contribute the most to the Reynolds shear stress, dividing them into four quadrants of the $u^\prime-v^\prime$ plane, denoted $Q_1$ to $Q_4$.
$Q_2$ (ejections: $u^\prime<0$, $v^\prime>0$) and $Q_4$ (sweeps: $u^\prime>0$, $v^\prime<0$) events result in turbulence production, while $Q_1$ and $Q_3$ are responsible for damping. Contours of the weighted Reynolds shear stress are depicted in figure~\ref{fig:jpdf} for a wall-normal plane, slightly above the elastic wall, $y/h = 0.05$. The panel $(a)$ shows the results for the single-phase cases with elasticity decreasing from left to right. The major contribution to the turbulence production near the highly elastic walls comes from the ejection events with significantly large values of $v^\prime$ and small negative $u^\prime$, whereas the contribution of sweeps comes from the events with an opposite combination: small negative $v^\prime$ and a relatively large positive $u^\prime$. This observation reveals the mechanism by which the spanwise roller-type vortices, forming close to the highly elastic walls, contribute to the turbulence production: these vortices bring down the high momentum flow towards the elastic wall with a small decelerated wall-normal velocity due to the decreased wall-blocking effect of the elastic layer, thus deforming the interface in the streamwise and the wall-normal directions. The interface then releases the elastic stress by pumping the decelerated flow with a strong wall-normal velocity towards the centre of the channel, contributing to the turbulence production and regeneration of new spanwise vortical structures. The low magnitudes of $u^\prime$ in the ejection events is due to the blocking effect of the upstream deformed interface. This is consistent with the shape of the deformed wall, previously shown in figure~\ref{fig:snapshots}$(a)$, where the interface descends with a larger slope, compared to its ascending. As the wall elasticity decreases, the distribution of the strong ejection events over the $u^\prime-v^\prime$ plane appears to become less elongated in the vertical direction ($v^\prime$ axis), while stretching along the $u^\prime$ axis. The relative contribution of sweep events increases when lowering the wall-elasticity, although with small magnitudes of $u^\prime v^\prime$. The distribution for the least elastic case (top right) is similar to what found in turbulent flows in the vicinity of rigid walls \citep{Kim1987}, where the sweep events are slightly more powerful than the ejections. The results of the particulate cases are depicted in figure~\ref{fig:jpdf}$(b)$. The shape of the distributions are similar to the ones of the single-phase flow cases, except for an increase ($G3_{10\%}$ and $G4_{10\%}$) or a reduction ($G1_{10\%}$ and $G2_{10\%}$) in the magnitude of $v^\prime$. To have a better understanding of the particles role, we show the same analysis in figure~\ref{fig:jpdf}$(c)$, sampling only the fluid inside a spherical shells which is $5\%$ larger than the particles (at the same wall-normal plane $y/h=0.05$). For the cases with highly elastic walls, the particles approach the interface region, trapped inside the strong sweep events with a relatively large streamwise velocity. The asymmetry in the magnitude of the wall-normal fluctuations (large magnitudes for positive $v^\prime$) causes the particles to move away from the wall considerably faster than when approaching it. Therefore, it is more likely to sample particles which are approaching the wall, surrounded by $Q_4$ events. It should be noted here than the particles are scarce in the vicinity of highly elastic walls as we will discuss later in \S\ref{subsec:particles}, and thus, the strong sweep events found around the particles do not contribute significantly to the turbulent production. Indeed, particles reduce the active range of the large spanwise vortices, causing smaller wall deformation and in turn a turbulence attenuation. In addition to sweeps, stronger $Q_1$ events appear in the flow surrounding the particles, when the wall elasticity decreases. This is due to the particles with large streamwise velocity, bouncing back from the less deformable walls. This mechanism increases the wall-normal velocity fluctuations close to the wall and therefore enhances the turbulent activity.

\subsection{Particle dynamics}\label{subsec:particles}
\begin{figure}
  \centering
   \includegraphics[width=0.495\textwidth]{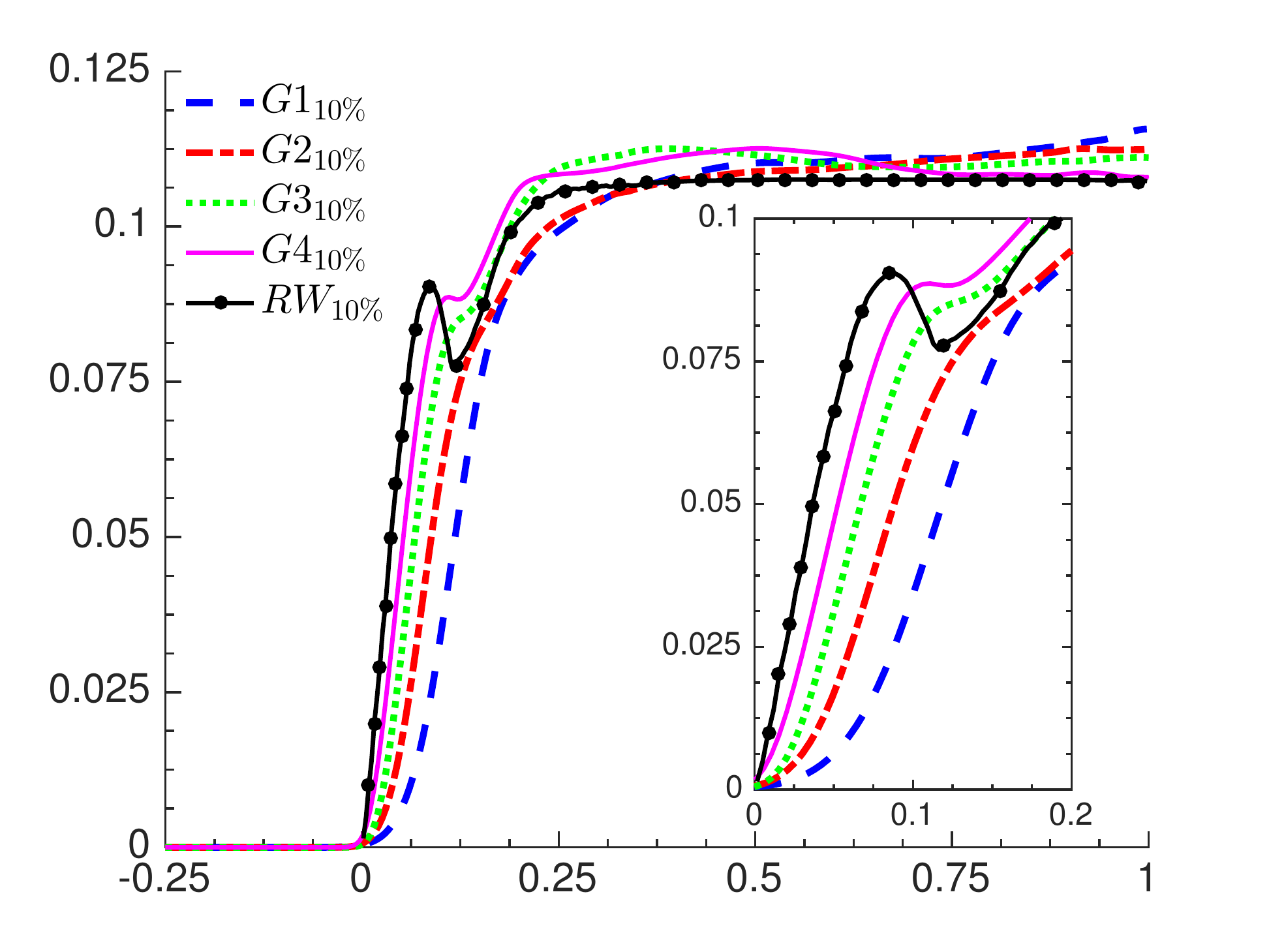}
   \includegraphics[width=0.495\textwidth]{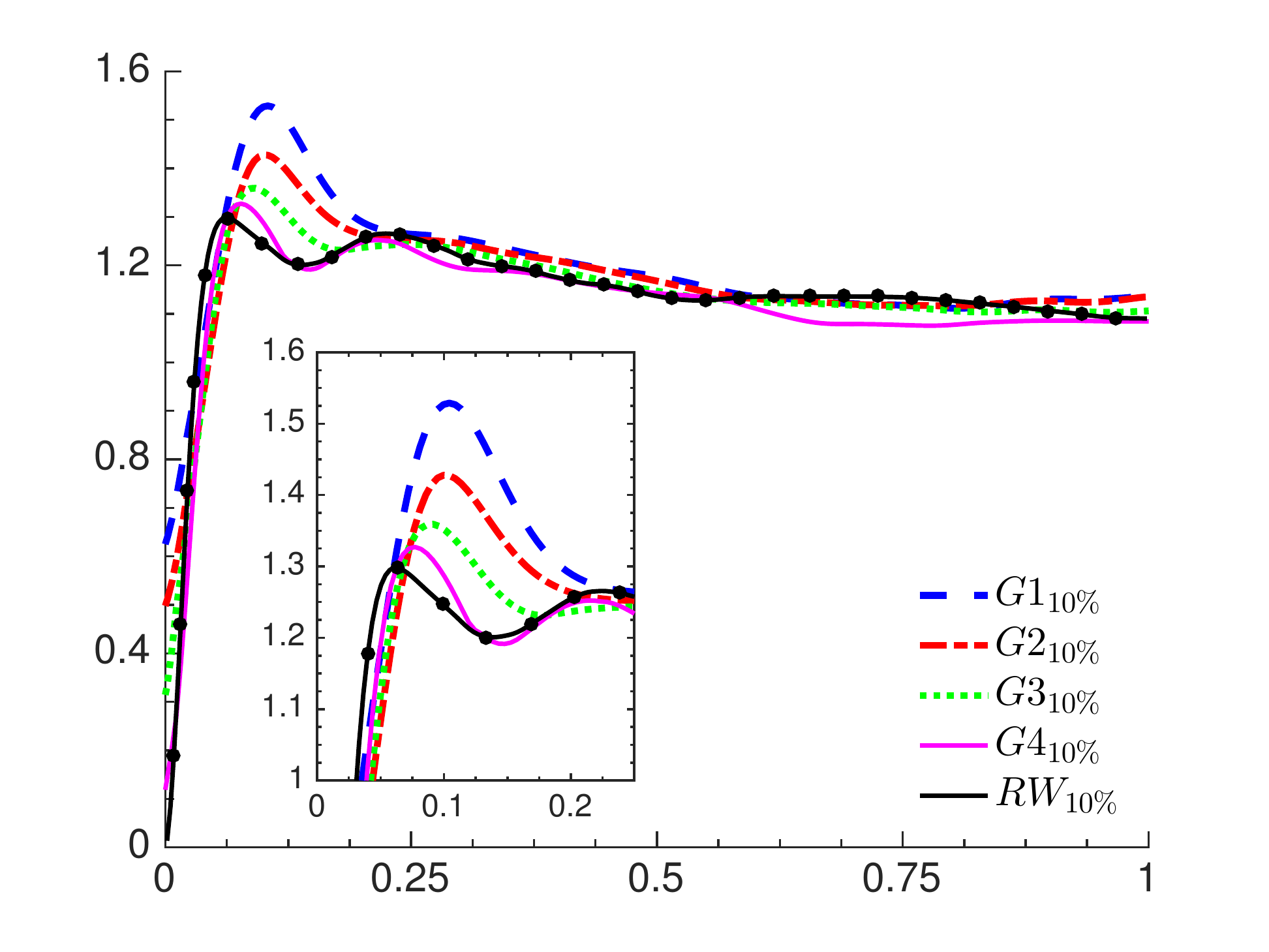} 
   \put(-99,0){{$y / h$}}
   \put(-293,0){{$y / h$}}      
   \put(-383,70){\rotatebox{90}{$\Phi$}}
   \put(-192,59){\rotatebox{90}{$K^\prime_f / K^\prime_p$}}      
   \put(-388,130){\footnotesize $(a)$}
   \put(-194,130){\footnotesize $(b)$} \\ 
   \includegraphics[width=0.495\textwidth]{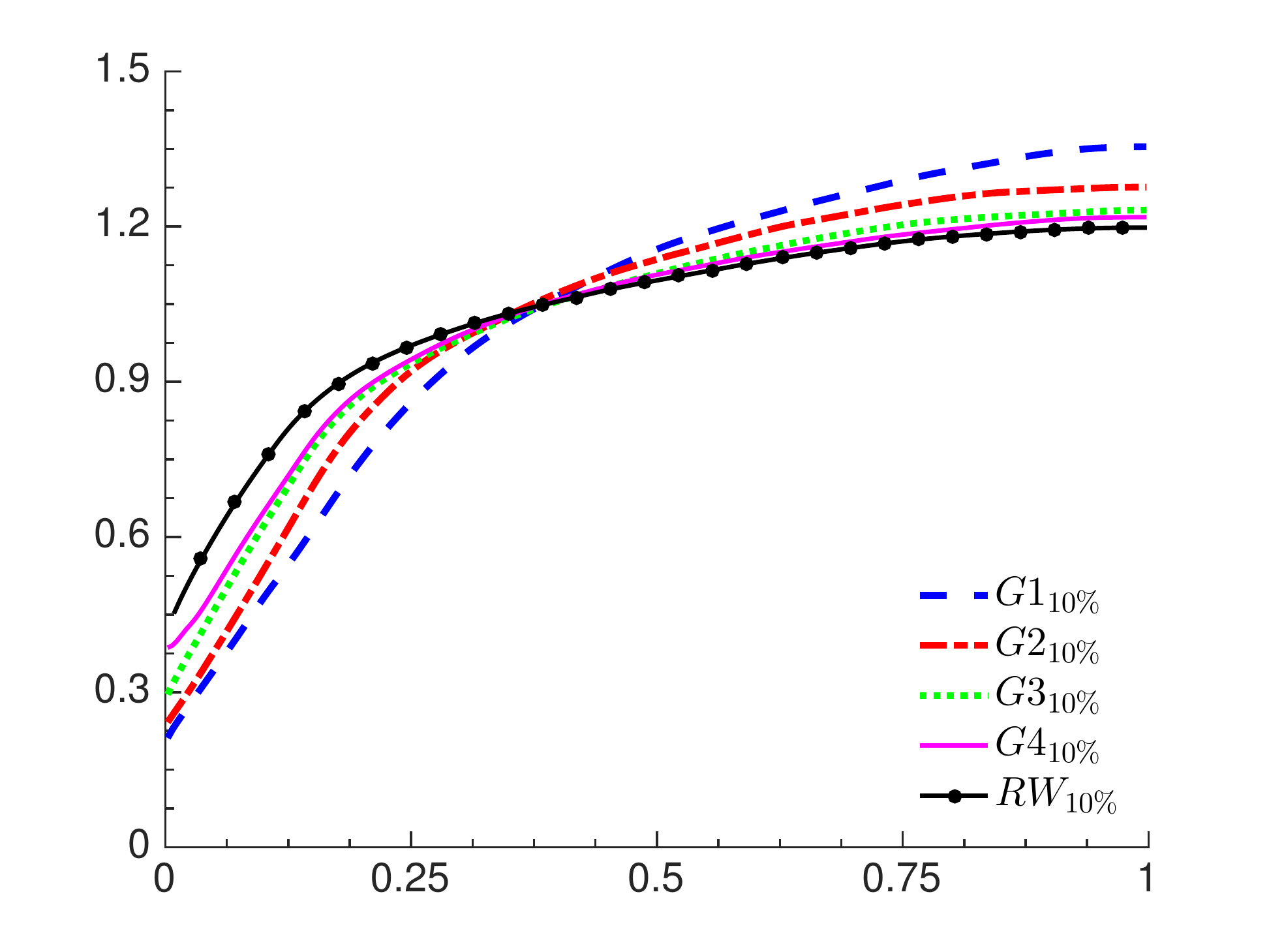}
   \includegraphics[width=0.495\textwidth]{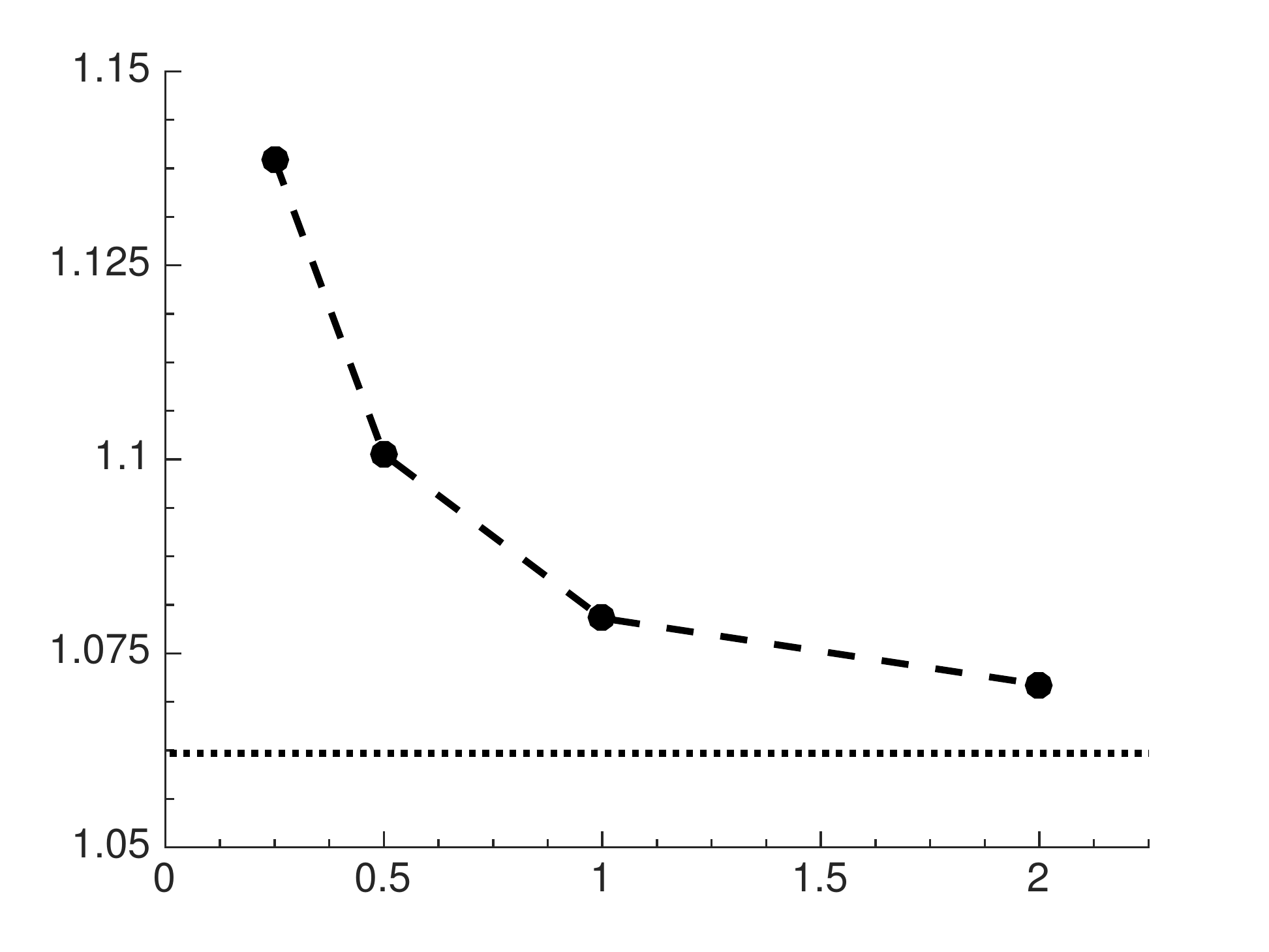}
   \put(-383,63){\rotatebox{90}{$U_p / U_b$}}
   \put(-192,63){\rotatebox{90}{$\overline{U}_p / U_b$}}     
   \put(-97,0){{$G^*$}}
   \put(-293,0){{$y / h$}}         
   \put(-388,130){\footnotesize $(c)$}
   \put(-194,130){\footnotesize $(d)$} \\
  \caption{Profiles of particle-phase averaged data versus $y/h$: $(a)$ mean local volume fraction, $(b)$ the ratio between the turbulent kinetic energy of the fluid and of the particle phase and $(c)$ mean particle velocity profiles, normalized by the bulk velocity $U_b$. $RW_{10\%}$ refers to the case with rigid walls from the study of \cite{Ardekani2017}. $(d)$ Displays the overall average of particle velocity versus the modulus of transverse elasticity $G^*$. The average of particle velocity for $RW_{10\%}$ is indicated with a dotted line.}
\label{fig:parts}
\end{figure}
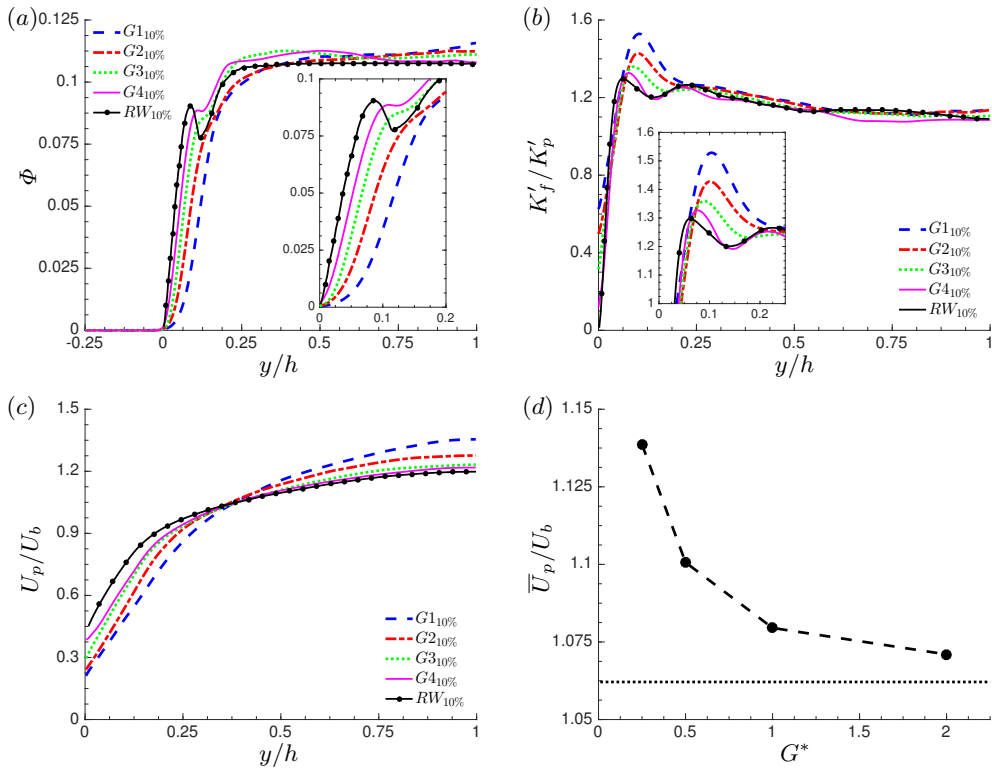
The mean local volume fraction $\Phi$ is depicted in figure~\ref{fig:parts}$(a)$. Spherical particles in the turbulent flow with rigid walls (case $RW_{10\%}$) display a local maximum at a distance slightly larger than one particle radius from the wall. \cite{Picano2015} attributed this local maximum to the formation of a particle layer at the wall, due to the wall-particle interactions that stabilize the particle position. \cite{Costa2016} further explained that the presence of this particle layer always results in drag increase, with respect to the single-phase flow. Interestingly, the local maximum for $\Phi$ is observed to reduce and move farther away from the wall when increasing the wall elasticity. The migration of the particles from the interface region can be explained by the presence of the strong ejection events with large $v^\prime$ in the flow in the case of highly elastic walls (see figure~\ref{fig:jpdf}). Indeed, the strong asymmetry in the magnitudes of the wall-normal velocity fluctuations, which favour the positive $v^\prime$, pushes the inertial particles towards the channel centre. However, as the wall elasticity decreases, a more symmetric distribution of $v^\prime$ allows the particles to form a layer close to the interface, similarly to the rigid wall cases. This wall-layer of particles contributes to the turbulent production by increasing the cross-flow velocity fluctuations, as shown in previous subsection. Next, we display in figure~\ref{fig:parts}$(b)$ the ratio between the turbulent kinetic energy of the fluid and of the particle phase, $K^\prime_f / K^\prime_p = ( {u^\prime}^2 + {v^\prime}^2 + {w^\prime}^2 ) / ( {u^\prime_p}^2 + {v^\prime_p}^2 + {w^\prime_p}^2 )$. For the particle statistics presented in this figure the rigid body motion of the particles (angular velocities) is taken into account. Particles are observed to fluctuate less than the fluid at the same wall-normal distance, except for a tiny region close to the wall ($y/h \lessapprox 0.05$), where the ratio is below $1$ for all the cases. The high particle fluctuation level in this region suggests that this is the cause of the near-wall enhancement of the fluid fluctuations, observed in the cases with less elastic walls. The ratio at $y=0$ increases with the wall elasticity, reaching $0.6$ in the most elastic case, where particles are scarce in this region. Particle-wall/particle collisions are observed to play an important role in increasing the fluctuations near the wall. Indeed, the particles in this region have a mean velocity larger than the surrounding fluid, due to their size and the fact that they can have a relative tangential motion at the wall. Therefore any collision between the particles can result in strong velocity fluctuations. Figure~\ref{fig:parts}$(c)$ depicts the mean particle velocity profiles, normalized by the bulk velocity $U_b$. Comparing the profiles with the mean flows in figure~\ref{fig:Retau}$(a)$, shows that the particle and fluid phases flow with the same mean velocity throughout the channel, except for a small layer close to the wall with the width of approximately one particle diameter $D$, where particles experience a larger mean velocity. The slip velocity at $y=0$ is observed to increase as the wall elasticity decreases. The particle velocity is averaged in figure~\ref{fig:parts}$(d)$ versus the modulus of transverse elasticity $G^*$.  It can be observed in this figure that increasing the wall elasticity results in a faster transport of particles as the averaged velocity increases from approximately $106\%$ of $U_b$ in the case of rigid walls (shown with a dotted horizontal line) to $114\%$ in the case with the most elastic walls. This increase in the averaged velocity is due to the migration of the particles away from the interface region close to the elastic walls. 

\begin{figure}
  \centering
   \includegraphics[width=0.496\textwidth]{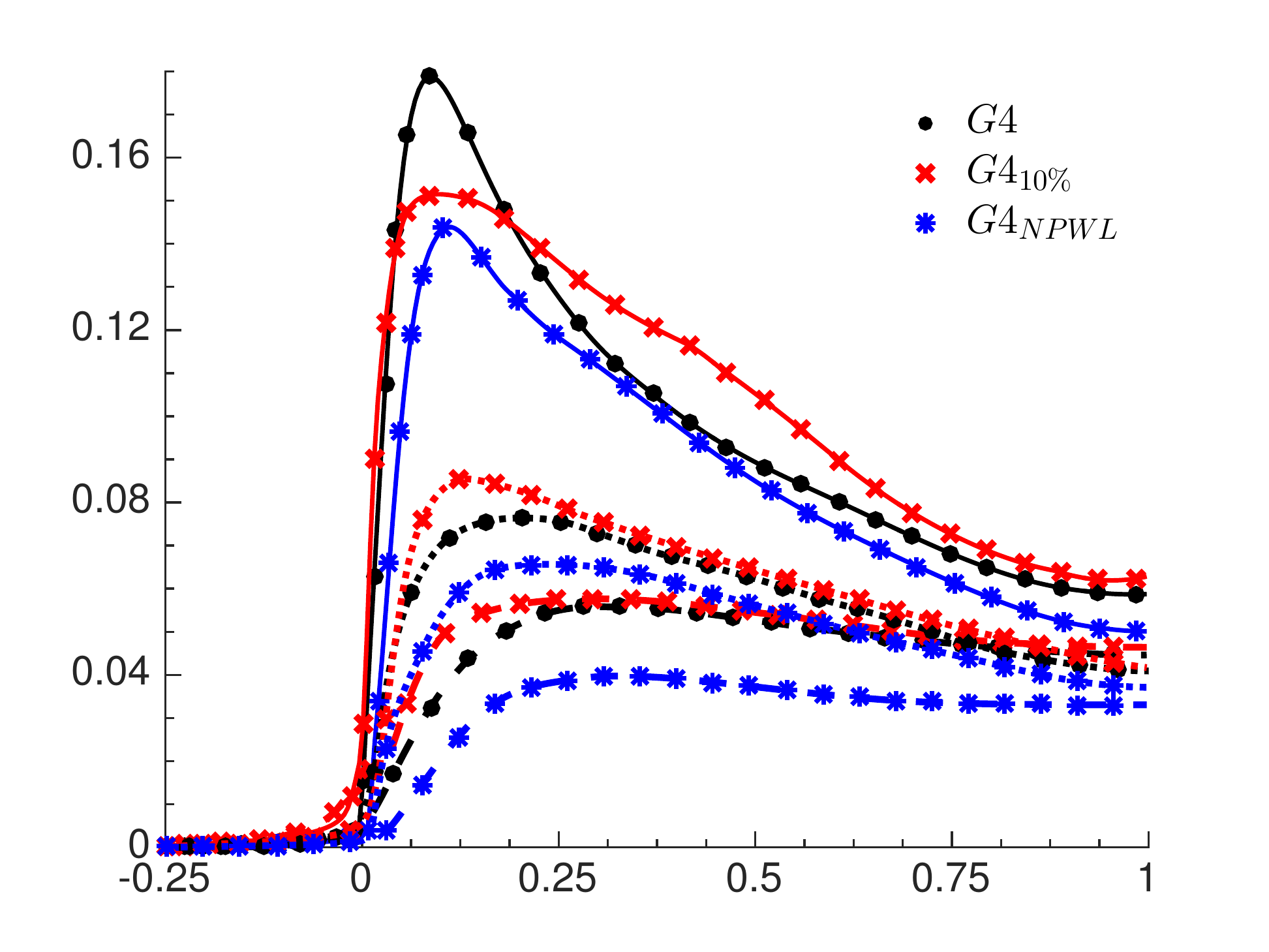}
   \includegraphics[width=0.496\textwidth]{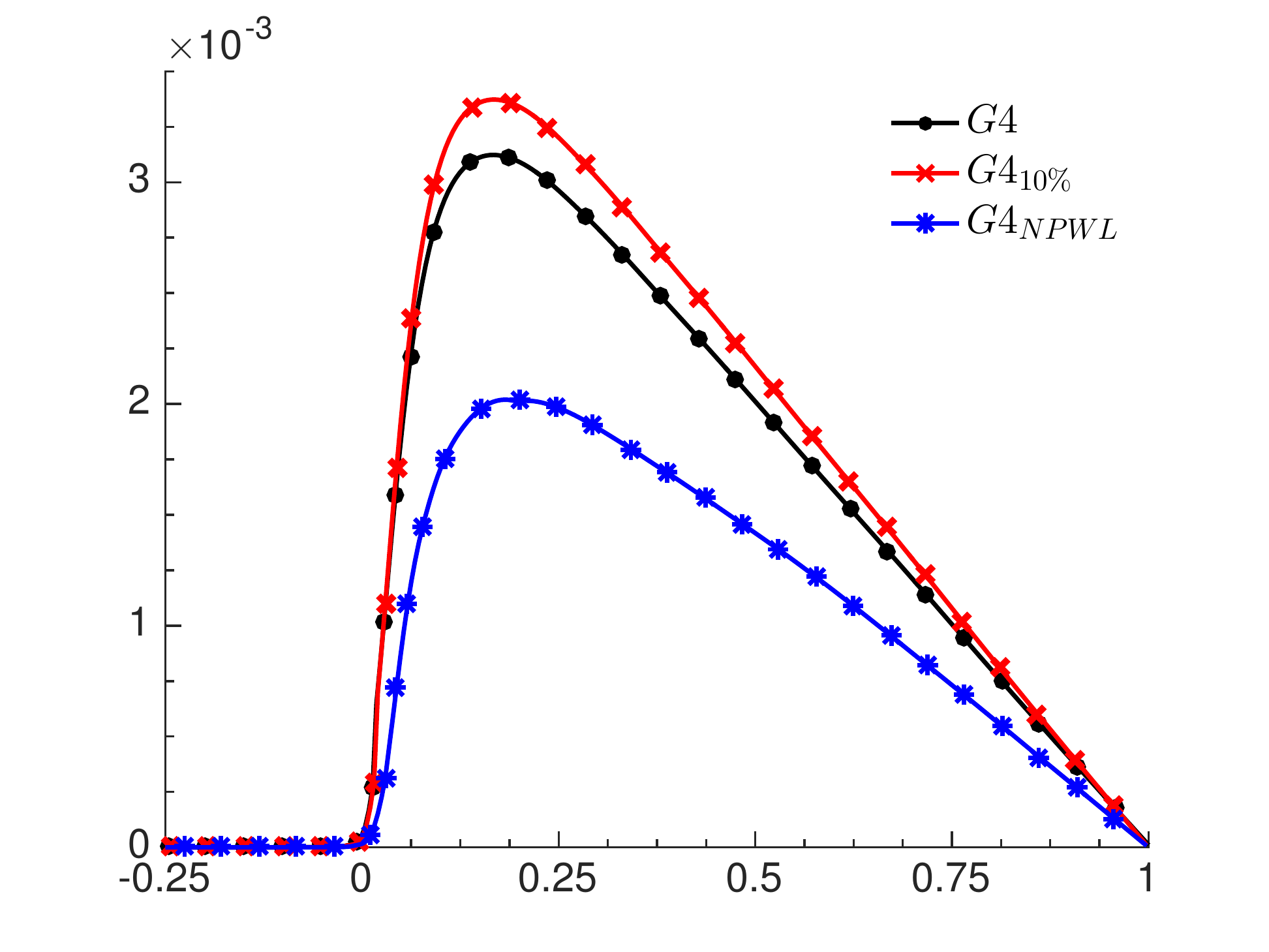} 
   \put(-186,51){\rotatebox{90}{$-\langle u^\prime v^\prime \rangle / U^2_b$}}
   \put(-385,46){\rotatebox{90}{$u^\prime $, $v^\prime $, $w^\prime \, / U_b$ }}   
   \put(-99,0){{$y / h$}}
   \put(-293,0){{$y / h$}}     
   \put(-387,130){\footnotesize $(a)$}
   \put(-194,130){\footnotesize $(b)$} \\ 
  \caption{$(a)$ Root-mean-square velocity fluctuations $u^{\prime}$, $v^{\prime}$ and $w^{\prime}$ in the streamwise, ---~, wall-normal, -~-~-~, and spanwise, ...~, directions and $(b)$ Reynolds shear stress, scaled in outer units for the cases $G4$, $G4_{10\%}$ and $G4_{NPWL}$. }
\label{fig:npwl}
\end{figure}
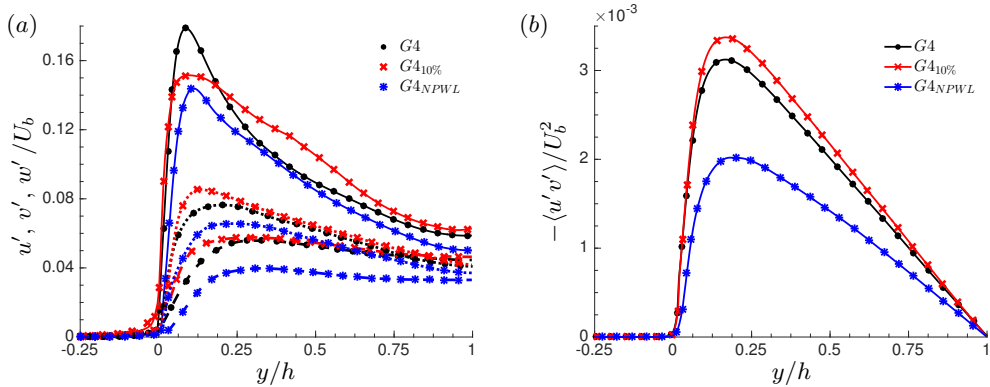
To show the importance of the particle wall-layer on the enhancement of the velocity fluctuations in this region, we perform a numerical experiment where we remove this layer. In this simulation, denoted $G4_{NPWL}$ (no particle wall-layer), particles bounce back towards the core region of the channel before approaching the interfaces, following a collision with two virtual walls, located at a distance $h/10$ from the two real interfaces ($y/h=0.1$). The elasticity of the wall is set to $G^*=2$, i.e.,~the most rigid elastic wall case previously considered where the effect of the particle wall-layer is more pronounced. The velocity fluctuations and the Reynolds shear stress, obtained for this additional case is compared with the results for $G4$ and $G4_{10\%}$ in figure~\ref{fig:npwl}. The increase in the cross-flow fluctuations previously observed for the case $G4_{10\%}$ disappears by removing the particle wall-layer ($G4_{NPWL}$), and as a consequence, a strong turbulence attenuation (see figure~\ref{fig:npwl}$(b)$) is obtained, being $v^\prime$ the most reduced velocity fluctuation component. With particles not contributing to the turbulent production in the near-wall region, the observed attenuation of the turbulence activity can be associated with the increased effective viscosity of the suspension \citep{Picano2015,Ardekani2017}. The effective viscosity of a rigid particle suspension is always higher than that of the single-phase flow, which cause a reduction in the turbulent activity; however, this effect is usually compensated by the formation of a particle wall-layer and its contribution to the turbulence production, thus resulting in an overall enhancement of the turbulent activity ($G4$). On the other hand, a strong turbulence attenuation and thus a drag reduction ($Re_\tau \approx 167$) even with respect to the single-phase flow over rigid walls ($Re_\tau \approx 180$) can be obtained by removing the particle wall-layer (see figure~\ref{fig:Retau}$(b)$).

\subsection{Effect of the volume fraction}\label{subsec:volume}
\begin{figure}
  \centering
   \includegraphics[trim={3.9cm 0 6.5cm 0},clip,width=0.496\textwidth]{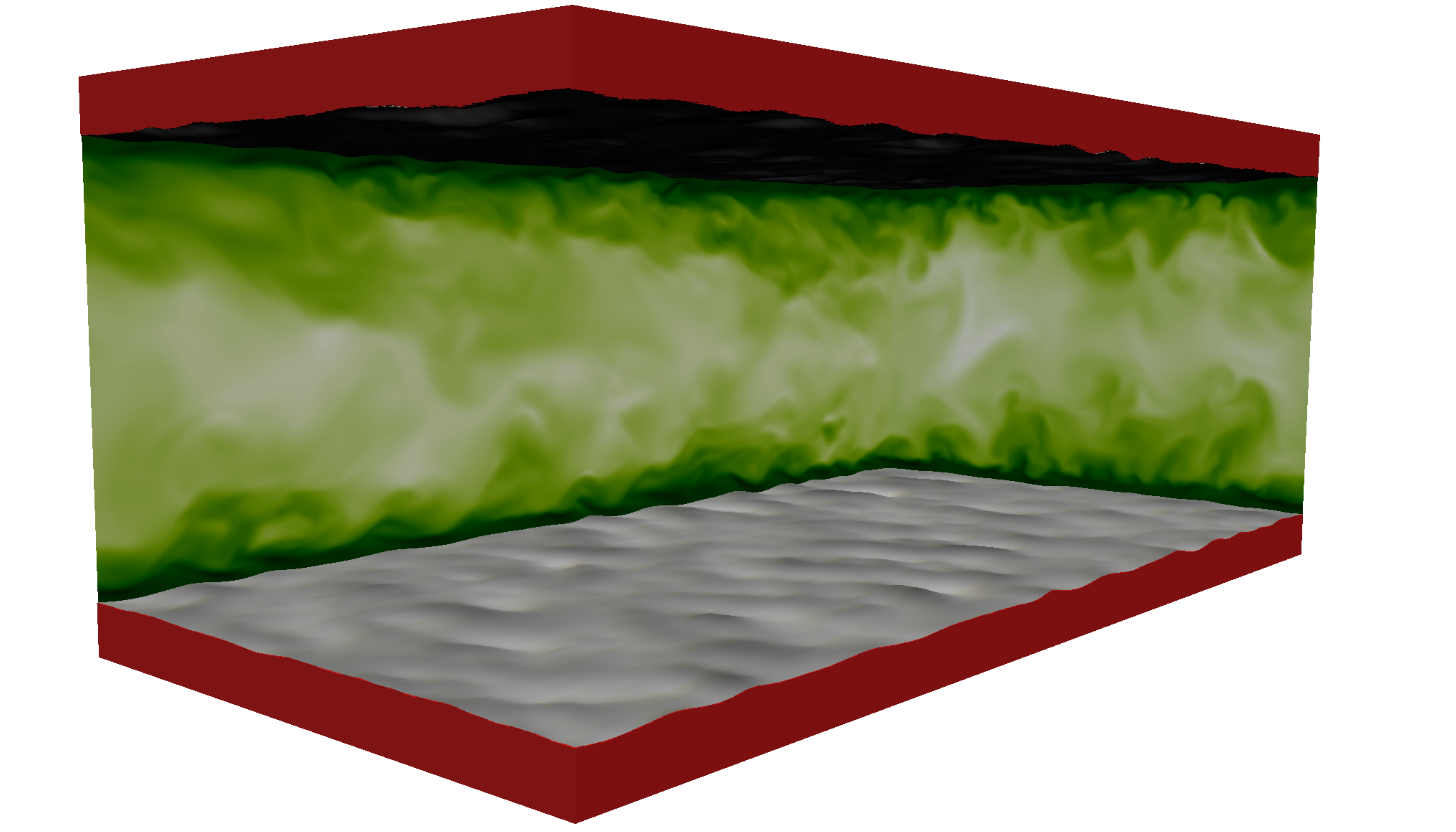}
   \includegraphics[trim={3.9cm 0 6.5cm 0},clip,width=0.496\textwidth]{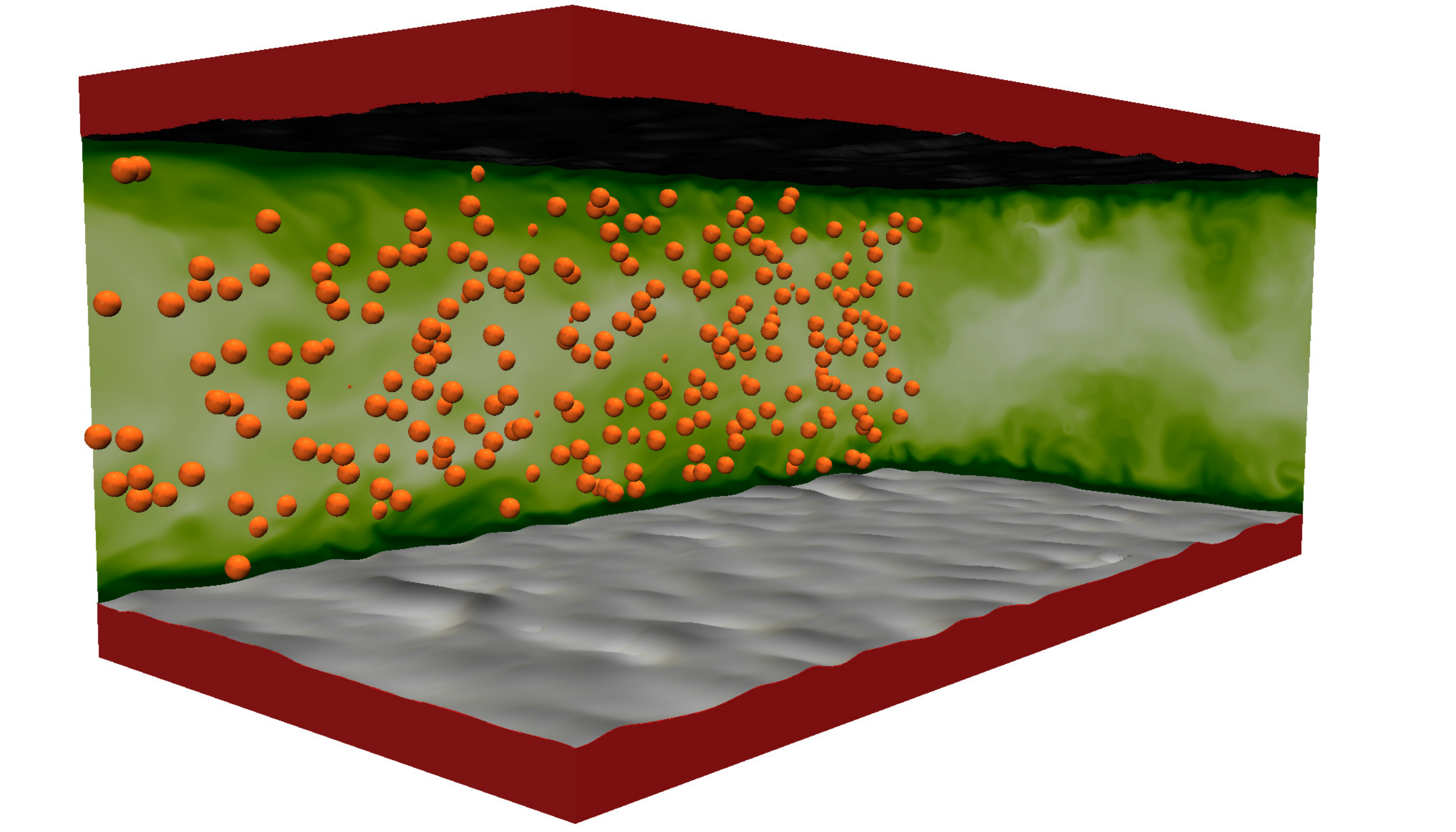} 
   \put(-387,125){\footnotesize $(a)$}
   \put(-194,125){\footnotesize $(b)$} \\ 
   \includegraphics[trim={3.9cm 0 6.5cm 0},clip,width=0.496\textwidth]{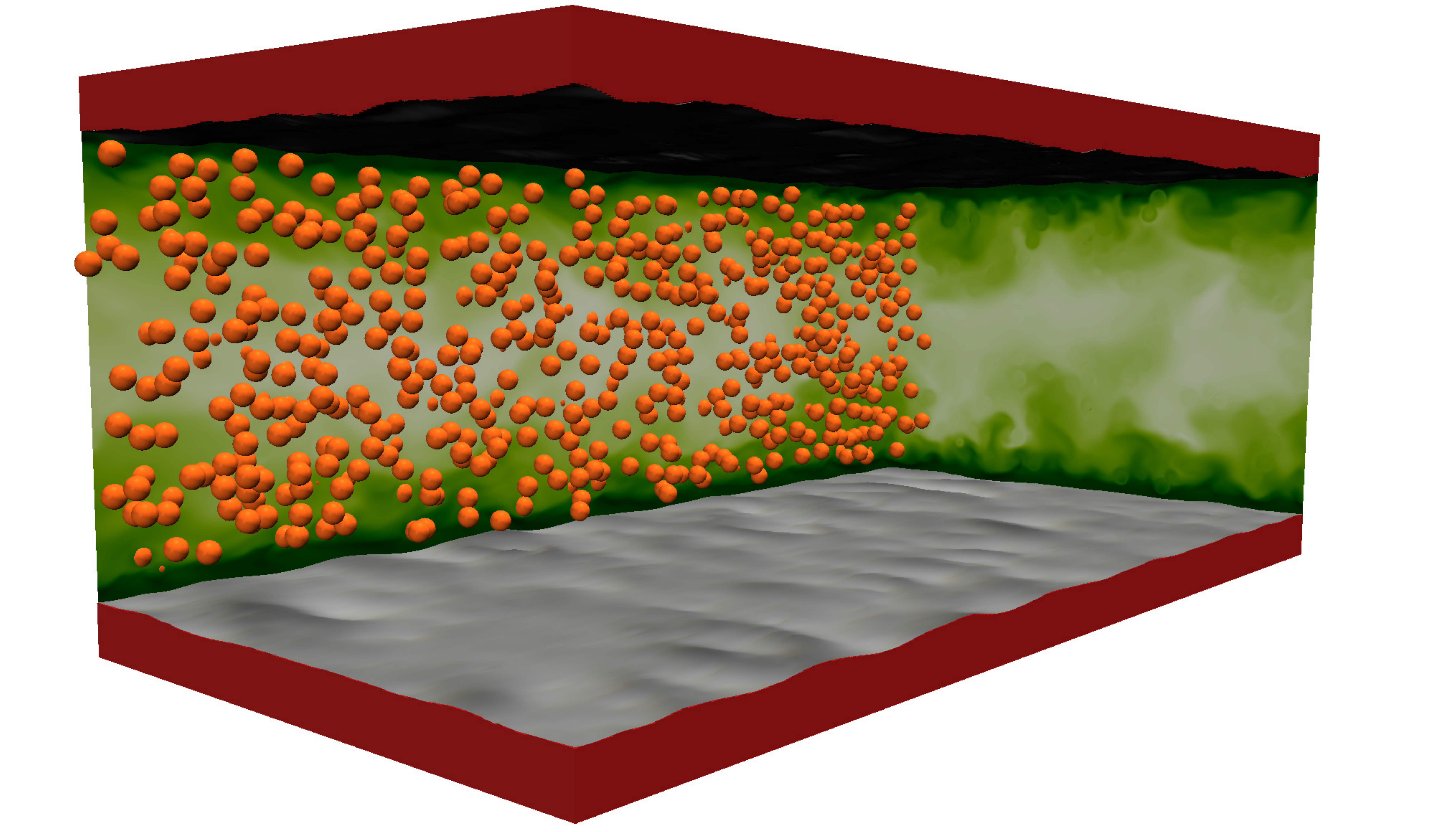}
   \includegraphics[trim={3.9cm 0 6.5cm 0},clip,width=0.496\textwidth]{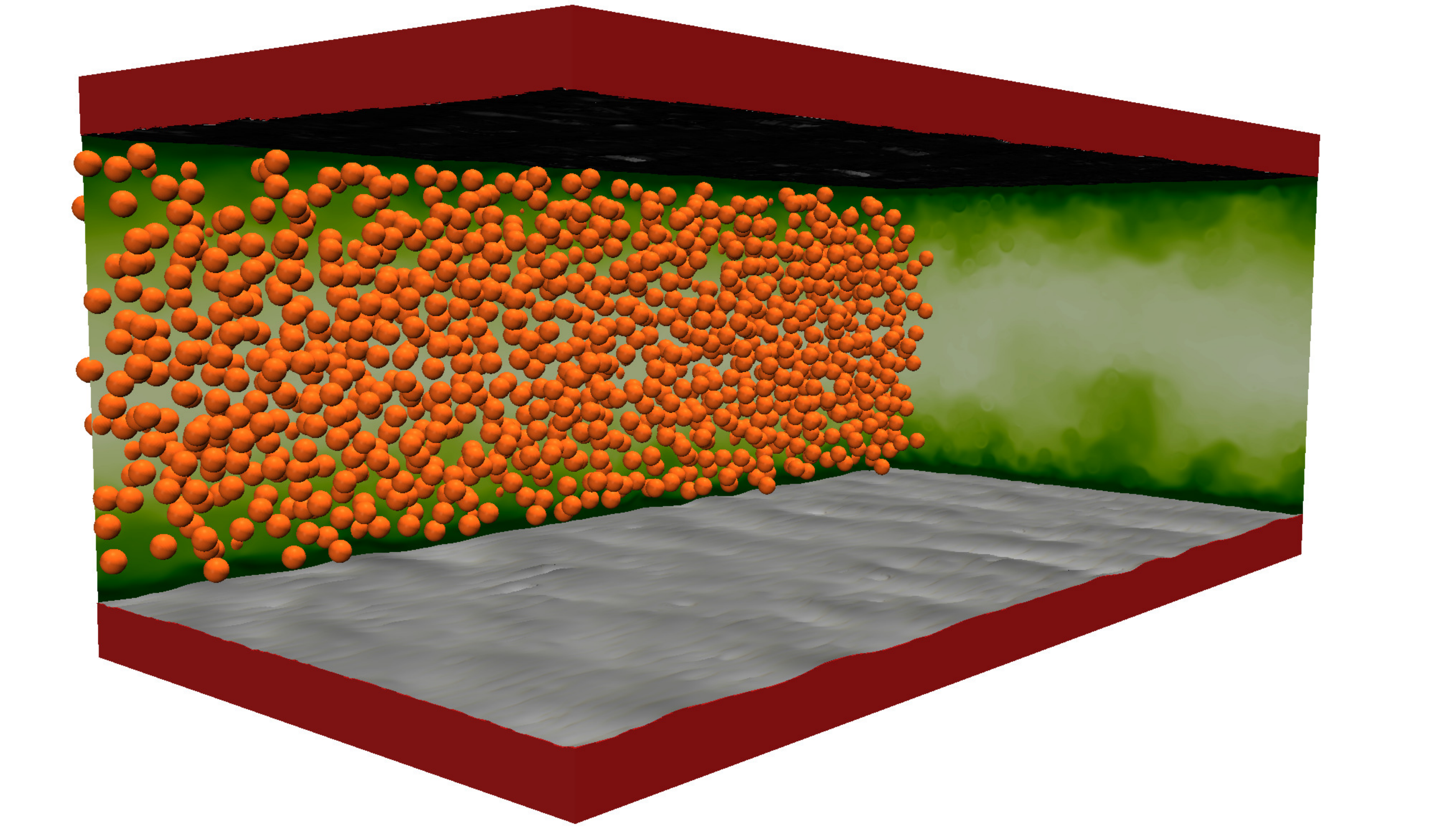}
   \put(-387,125){\footnotesize $(c)$}
   \put(-194,125){\footnotesize $(d)$} \\
  \caption{Instantaneous snapshots of the streamwise velocity $u$ on planes $x-y$ and $y-z$ for $G^*=0.5$ at different volume fractions $(a)$ $\phi=0\%$ (case $G2$), $(b)$ $\phi=5\%$ (case $G2_{5\%}$), (c) $\phi = 10\%$ (case $G2_{10\%}$) and $(d)$ $\phi = 20\%$ (case $G2_{20\%}$). For clarity, only the particles lying within the selected $x-y$ plane are displayed. The colour scale for the streamwise velocity ranges from $0$ (dark green) to $1.5u/U_b$ (white). The elastic walls are represented by the isosurfaces of $\xi=0.5$, coloured by the wall-normal distance, ranging from $-0.15h$ (white) to $0.15h$ (black).}
\label{fig:snapshots_vol}
\end{figure}
\cite{Picano2015} investigated dense suspensions of spherical particles in a turbulent channel flow with rigid walls up to a volume fraction equal to $20\%$. Their study revealed that the overall drag increase is mainly due to the enhancement of the turbulence activity up to $\phi=10\%$ and to the particle-induced stress at higher concentrations ($\phi=20\%$), where the turbulence is instead attenuated. In this section we investigate the effect of the particle volume fraction in the same channel but in a channel with elastic walls. The case $G2$ with elasticity $G^*=0.5$ is simulated with two additional volume fractions, $\phi=5\%$ and $20\%$, to compare the change in drag with respect to the cases with rigid walls. 

Snapshots of the flow and particles are displayed in figure~\ref{fig:snapshots_vol}, where the instantaneous streamwise velocity $u$ is depicted on $x-y$ and $y-z$ planes for the cases with $G^*=0.5$ and $\phi=0$, $5\%$, $10\%$ and $20\%$. For clarity, just a fraction of the particles (those lying within the visualized $x-y$ plane) are displayed. The wall deformation and the turbulent activity appear to decrease when increasing the particle volume fraction, especially in the case with $\phi=20\%$. Note also the absence of a particle wall-layer close to the interface, even at highest volume fraction $\phi=20\%$ considered here.

\begin{figure}
  \centering
   \includegraphics[width=0.496\textwidth]{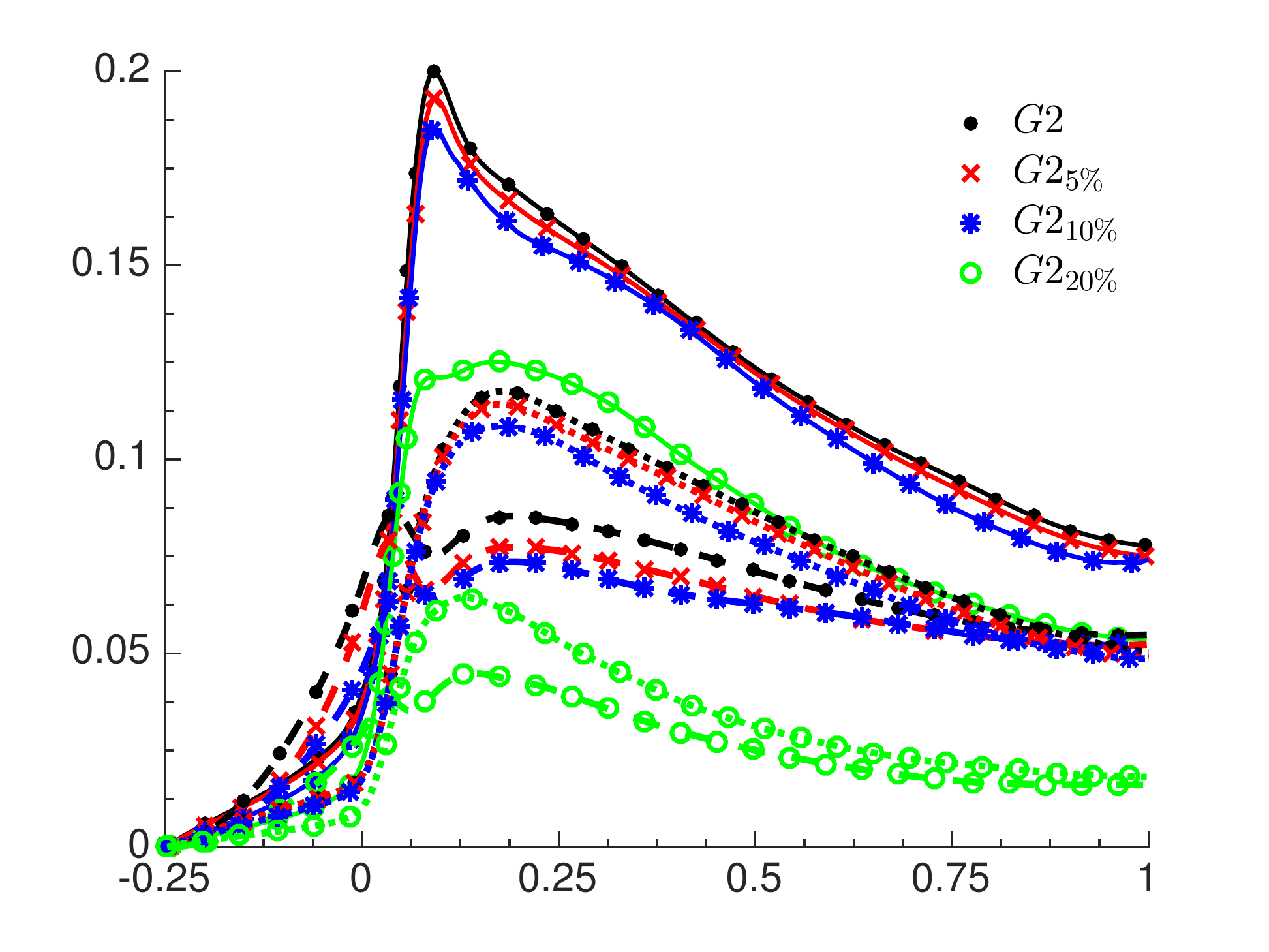}
   \includegraphics[width=0.496\textwidth]{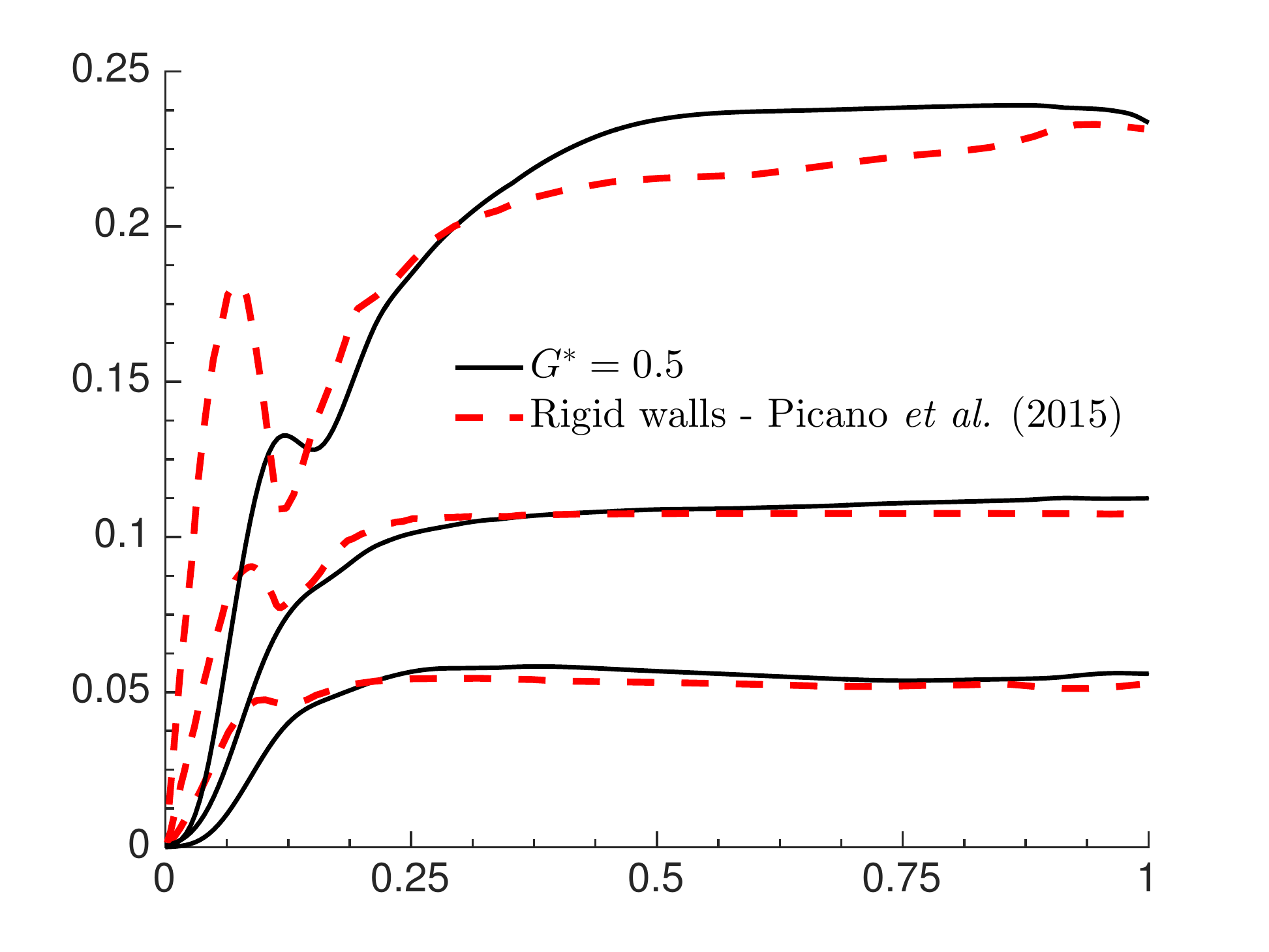} 
   \put(-385,46){\rotatebox{90}{$u^\prime $, $v^\prime $ \& $w^\prime \, / U_b$ }}   
   \put(-188,70){\rotatebox{90}{$\Phi$}}      
   \put(-99,0){{$y / h$}}
   \put(-293,0){{$y / h$}}      
   \put(-387,130){\footnotesize $(a)$}
   \put(-194,130){\footnotesize $(b)$} \\ 
   \includegraphics[width=0.496\textwidth]{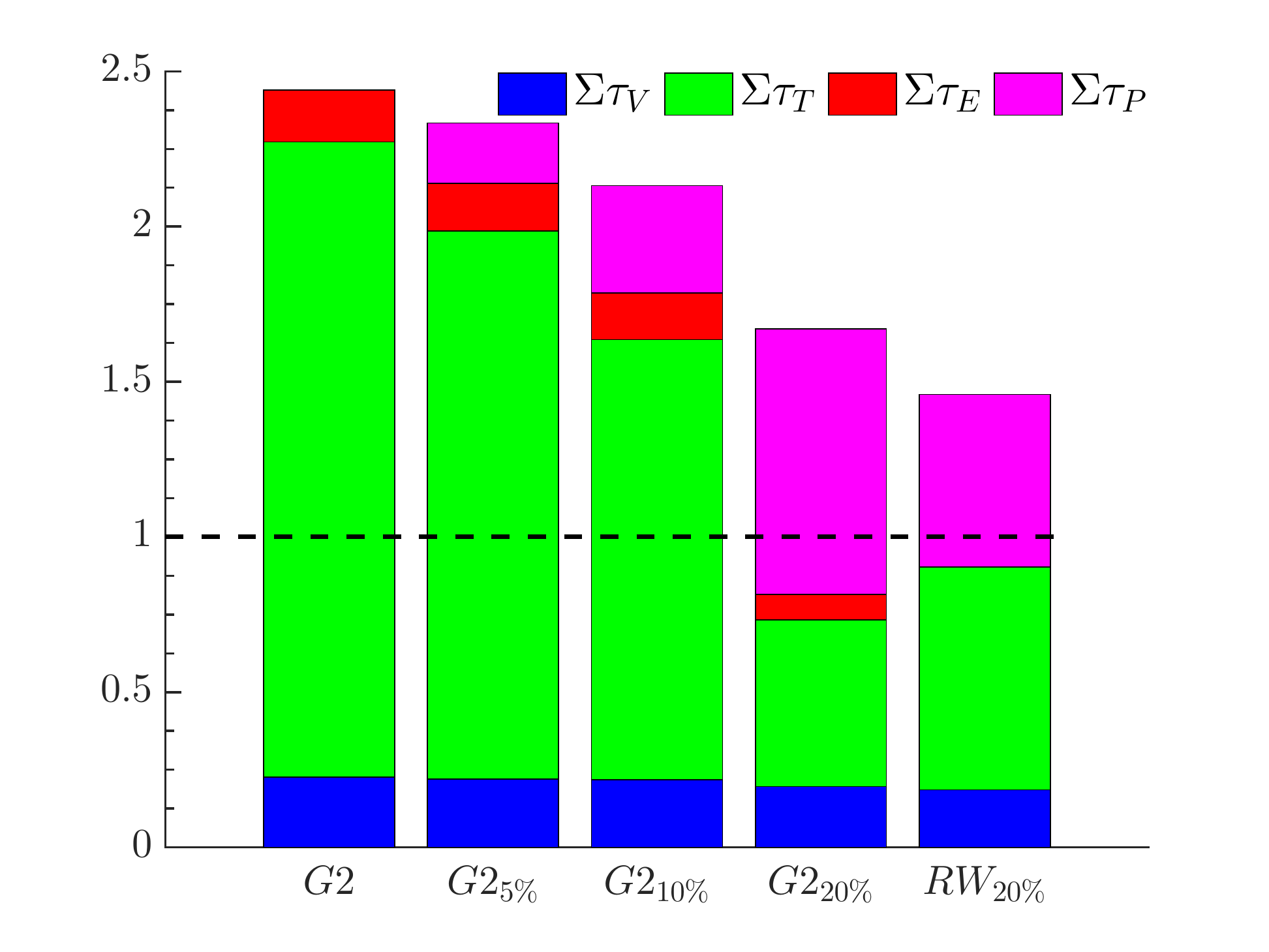}    
   \includegraphics[width=0.496\textwidth]{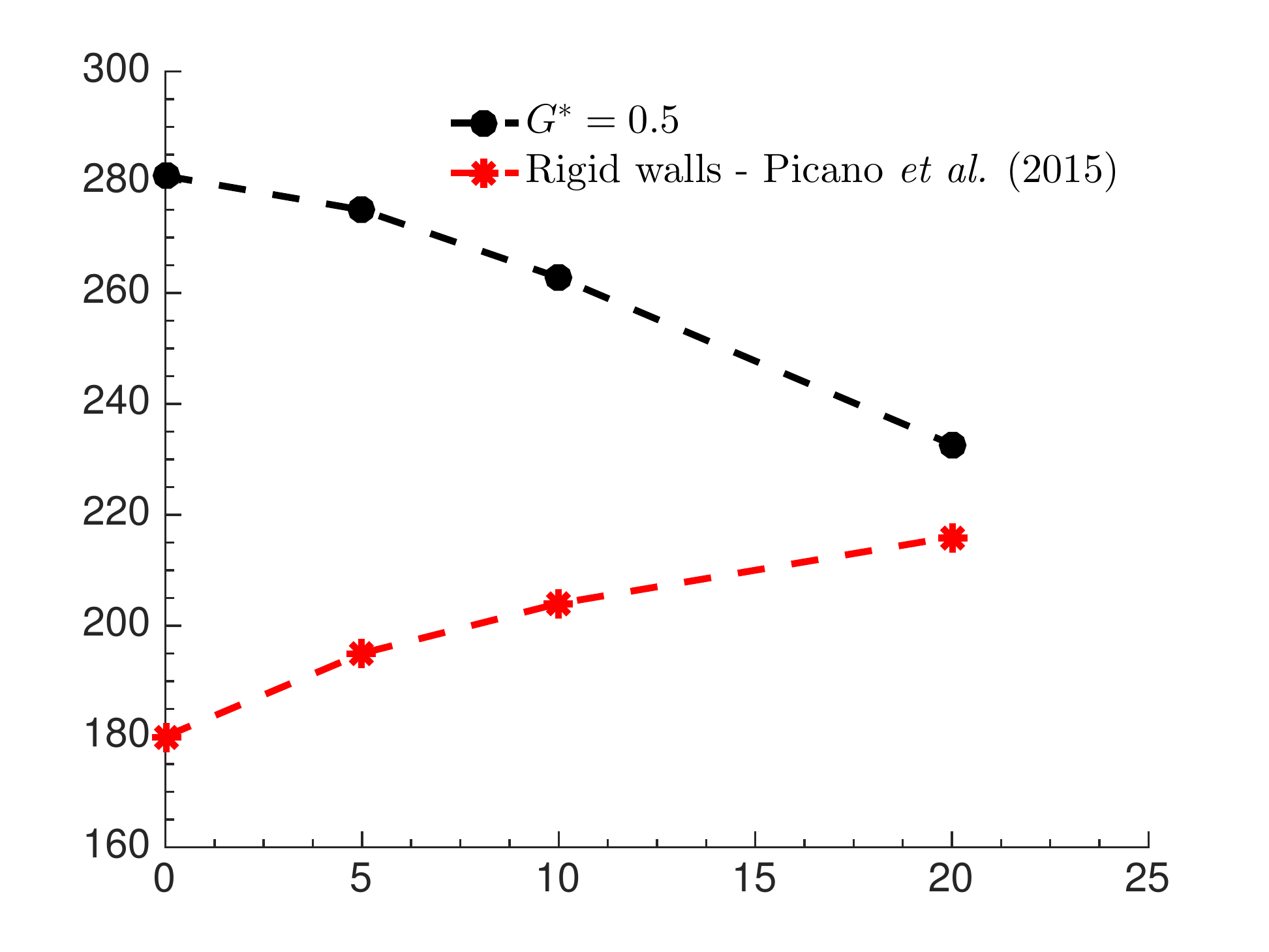}
   \put(-385,50){\rotatebox{90}{$2 \Sigma \,\tau_{i} \, / \, h \tau_{RW}$}}   
   \put(-101,0){{$\phi (\%)$}}
   \put(-188,68){\rotatebox{90}{$Re_{\tau}$}}   
   \put(-387,130){\footnotesize $(c)$}
   \put(-194,130){\footnotesize $(d)$} \\    
  \caption{$(a)$ Root-mean-square velocity fluctuations $u^{\prime}$, $v^{\prime}$ and $w^{\prime}$ in the streamwise, ---~, wall-normal, -~-~-~, and spanwise, ...~, directions. $(b)$ Mean local volume fraction profiles against the results for rigid walls \citep{Picano2015}. $(c)$ The total contribution of each stress to the drag at the interface, normalized by the drag of the single-phase flow with rigid walls, $\tau_{RW}$ (the dashed line with $Re_\tau=180$). The case $RW_{20\%}$ (rigid walls at $\phi=20\%$) is extracted from the reported data in \cite{Picano2015} $(d)$ The friction Reynolds number $Re_{\tau}$, versus the volume fraction $\phi$.}
\label{fig:phi}
\end{figure}
The root-mean-square velocity fluctuations $u^\prime$, $v^\prime$ and $w^\prime$ are depicted in figure~\ref{fig:phi}$(a)$, versus the wall-normal distance $y/h$. The results show a progressive reduction in the velocity fluctuations with incresing volume fractions. The decrease with respect to $G2$ is moderate for $G2_{5\%}$ and $G2_{10\%}$, while for a volume fraction of $20\%$ a significant attenuation of all the velocity fluctuations is evident, especially in the core of the channel where the turbulent flow is almost relaminarized (see also figure~\ref{fig:snapshots_vol}$(d)$). Differently from the results with rigid walls \citep{Picano2015}, the particles are observed here to attenuate the turbulence activity for all the volume fractions, with a reduction for the case at $\phi=20\%$ which is more than that with rigid walls. We explain this by plotting the mean local volume fraction profiles in figure~\ref{fig:phi}$(b)$ against the results for rigid walls \citep{Picano2015}. Indeed, the particle wall-layer disappears for $\phi \leq 10\%$ and the remaining near-wall peak of the profile for the case $G2_{20\%}$ is significantly reduced and displaced farther away from the wall ($y/h\approx0.1$). The particles migration away from the interface region causes a stronger attenuation of turbulence than with the rigid walls at $\phi=20\%$. 

Following equation~\ref{eq:FlInt}, the contribution of each shear stress (viscous, $\tau_V$, Reynolds, $\tau_T$, hyper-elastic wall, $\tau_E$ and the particle-induced $\tau_P$), normalized by the drag of the single-phase flow over rigid walls ($\tau_{RW}$), is given in figure~\ref{fig:phi}$(c)$ for different volume fractions at $G^*=0.5$. $RW_{20\%}$ in this figure denotes the case with $\phi=20\%$ taken from \cite{Picano2015}. In spite of a reduced turbulent activity in $G2_{20\%}$, with respect to $RW_{20\%}$, $\tau_P$ compensates for the loss of Reynolds shear stress, thus resulting in an overall higher drag for the elastic wall case. The friction Reynold number $Re_\tau$ is finally compared with the results for rigid walls in figure~\ref{fig:phi}$(d)$, where the effect of wall-elasticity is observed to reduce significantly when the volume fraction of the particles is increased.

\subsection{Roughness \& elasticity}\label{subsec:rough}
The single-phase turbulent flow obtained in this study for highly elastic walls shows similarities to what has been reported in the literature for turbulent flows over porous walls \citep{Breugem2006,Rosti2015,Samanta2015,Rosti2018}, rough surfaces \citep{Raupach1991,Jimenez2001} and plant canopies \citep{Finnigan2000}. In this section, we distinguish the main effects of wall elasticity on the near wall turbulence from the modulation caused by the roughness. 
To this purpose, an additional simulation is performed where an instantaneous configuration of the deformed interface, obtained in the case with the largest wall elasticity ($G1$), is chosen and frozen in time ($G^*\rightarrow\infty$). A volume penalization IBM \citep{Kajishima2001,Breugem2013,Ardekani2018pipe} is employed to impose the n-s/n-p boundary conditions on the surfaces, with the local solid (referring to the wall) volume fraction at each grid cell being known from the field of $\xi$ in $G1$.

\begin{figure}
  \centering
   \includegraphics[width=0.496\textwidth]{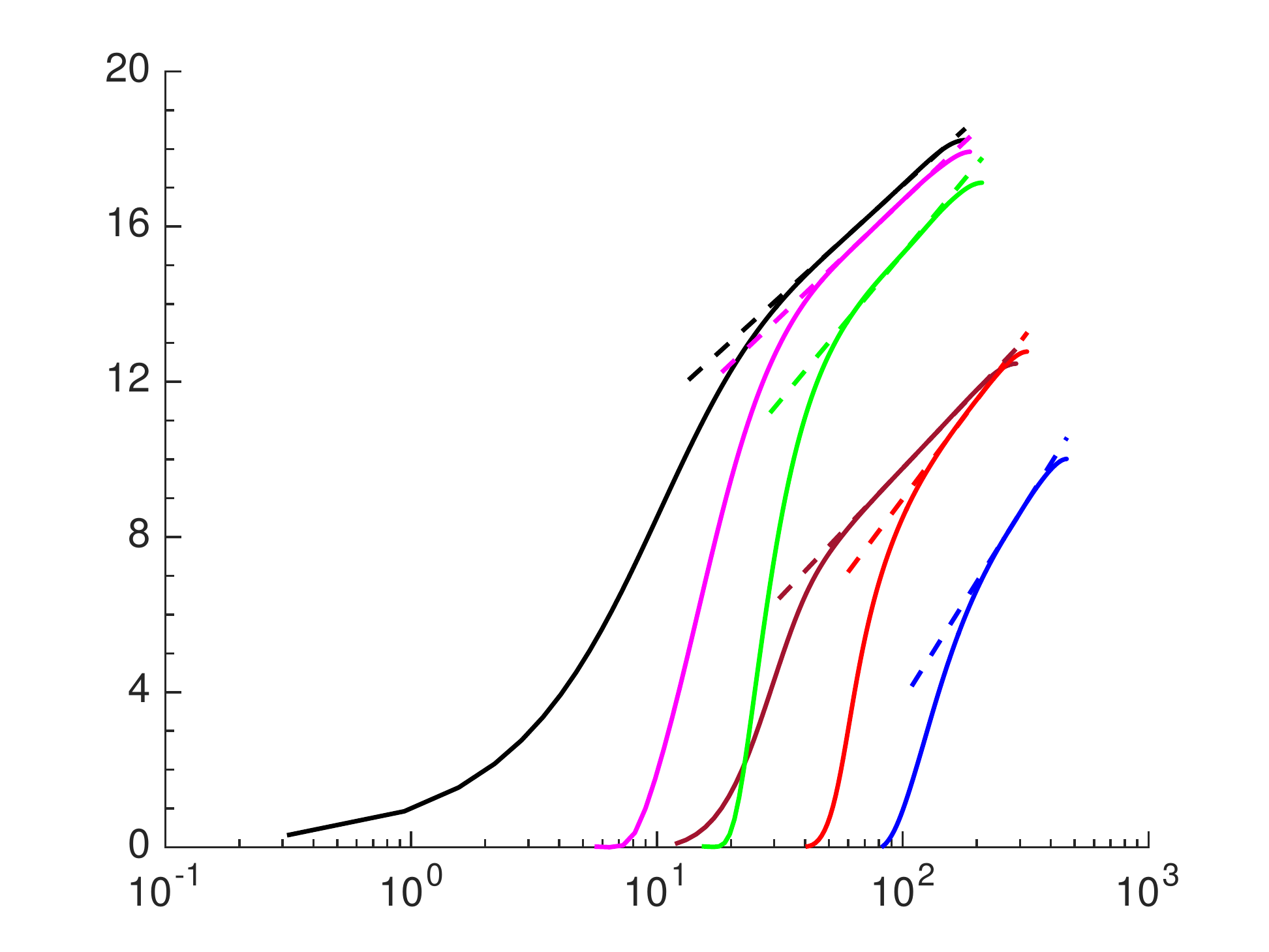}
   \includegraphics[width=0.496\textwidth]{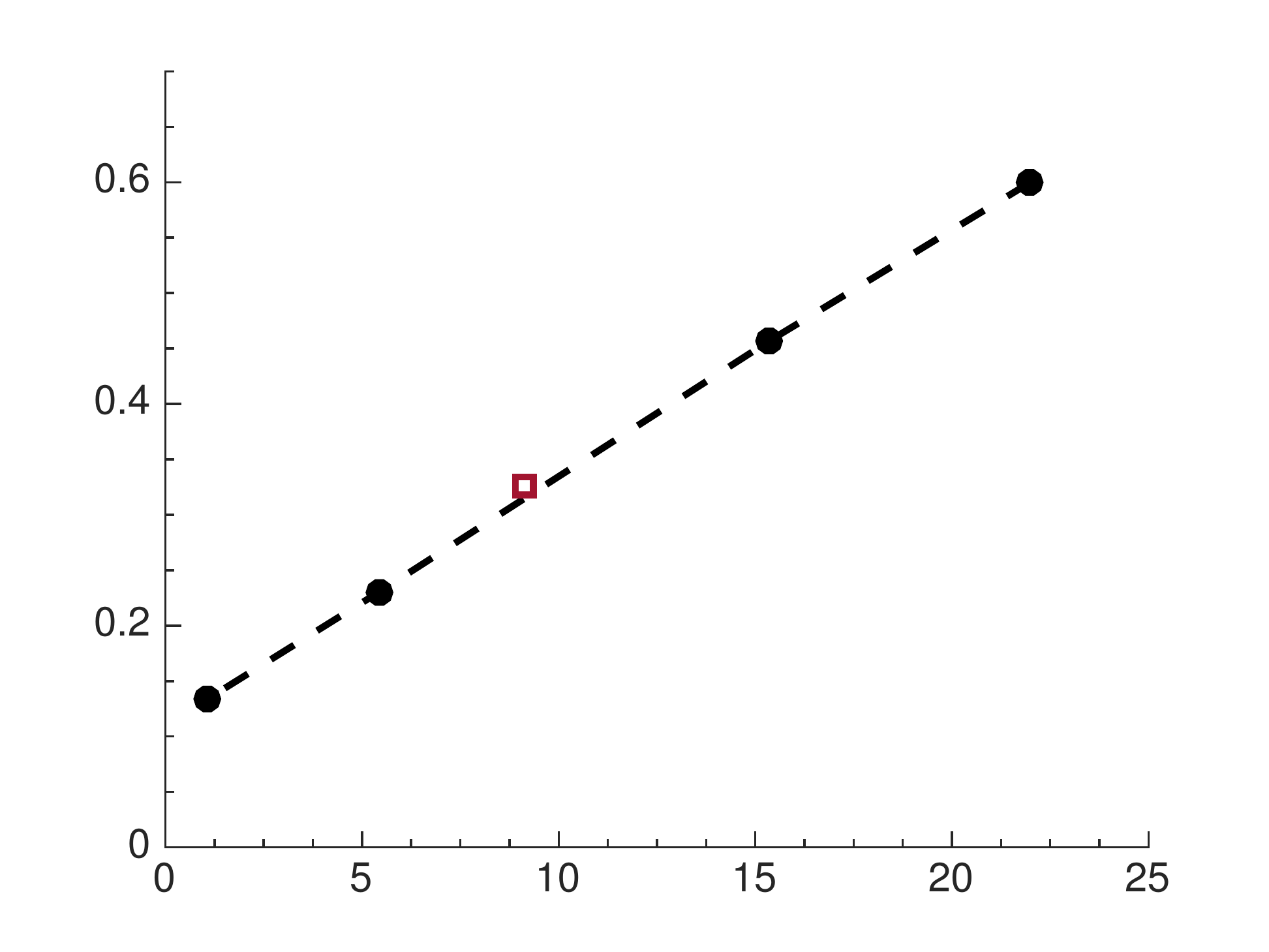} 
   \put(-301,-3){{$(y + d)^+$}}
   \put(-100,-3){{$\Delta U^+$}}         
   \put(-383,70){\rotatebox{90}{$U^+$ }}   
   \put(-190,49){\rotatebox{90}{$v^\prime(0) / U_b |\Delta k|$}}    
   \put(-387,130){\footnotesize $(a)$}
   \put(-194,130){\footnotesize $(b)$} \\ 
  \caption{$(a)$ Mean velocity profiles $U^+$, scaled in inner units, versus ${(y+d)}^+$, where $d$ is a shift of the origin. The blue, red, green and magenta solid lines are used for the cases $G1$ to $G4$, respectively, while the profile for $G1_{FW}$ is indicated with a brown solid line. $(b)$ Root-mean-square of the wall-normal velocity fluctuation at $y=0$, divided by $|\Delta k| U_b$, versus the velocity shift $\Delta U^+$. The data for $G1_{FW}$ is indicated with a brown square.}
\label{fig:scale}
\end{figure}
\begin{figure}
  \centering
   \includegraphics[trim={2.65cm 0.75cm 2.65cm 0},clip,width=0.496\textwidth]{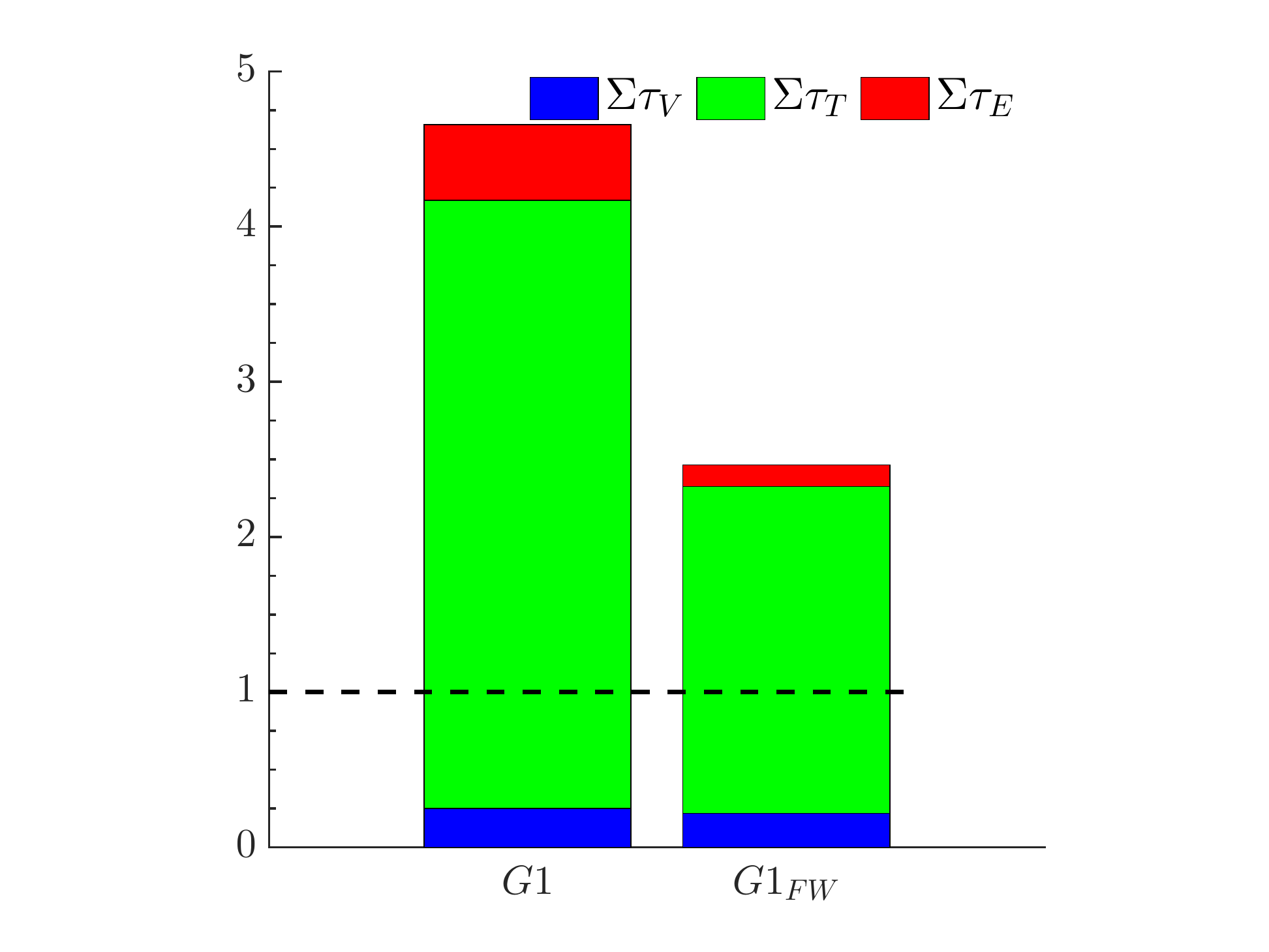}
   \includegraphics[width=0.496\textwidth]{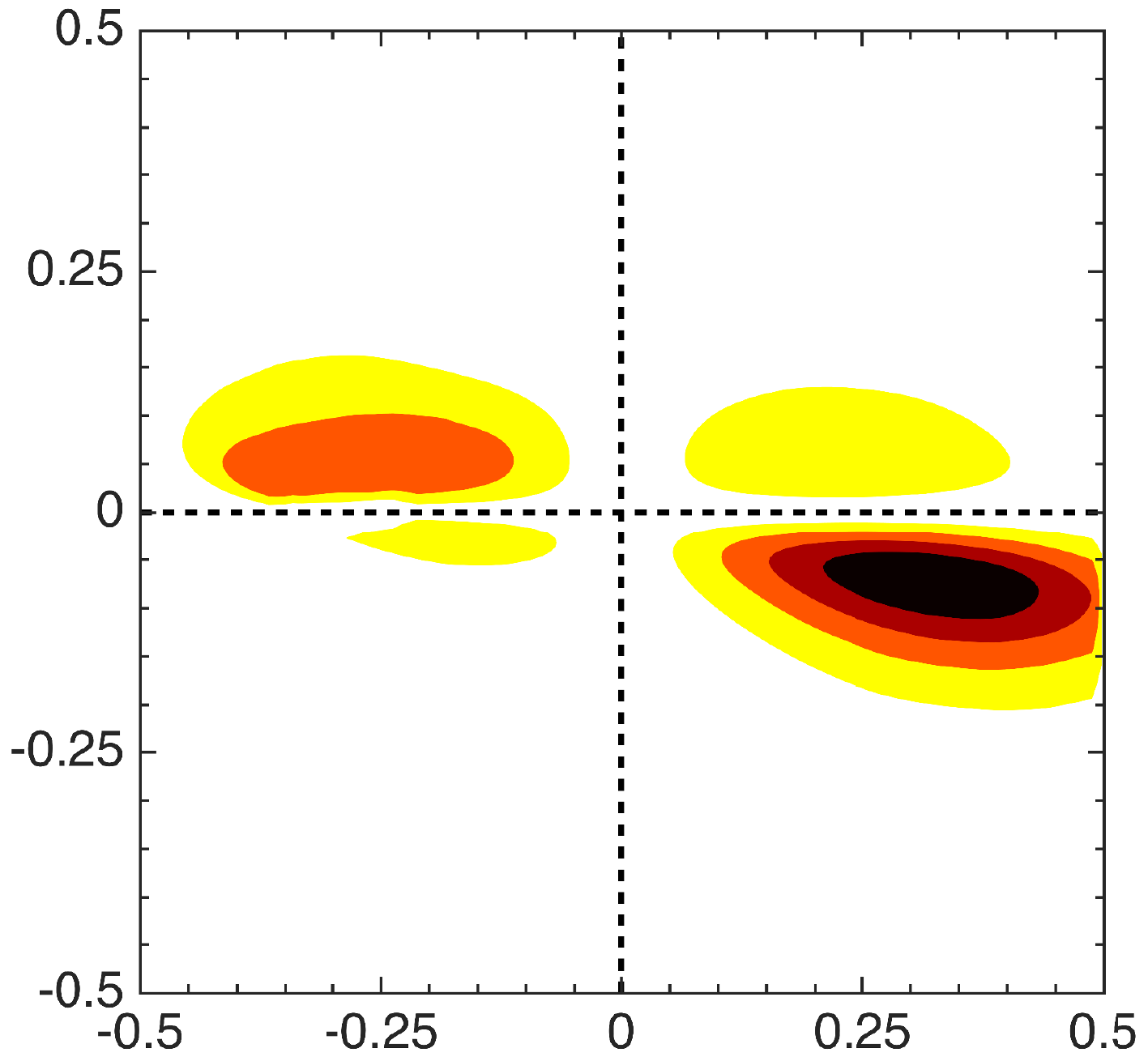} 
   \put(-385,67){\rotatebox{90}{$2 \Sigma \,\tau_{i} \, / \, h \tau_{RW}$}}   
   \put(-99,-9){{$u^\prime/U_b$}}
   \put(-190,82){\rotatebox{90}{$v^\prime/U_b$}}     
   \put(-387,162){\footnotesize $(a)$}
   \put(-194,162){\footnotesize $(b)$} \\
   \includegraphics[width=0.493\textwidth]{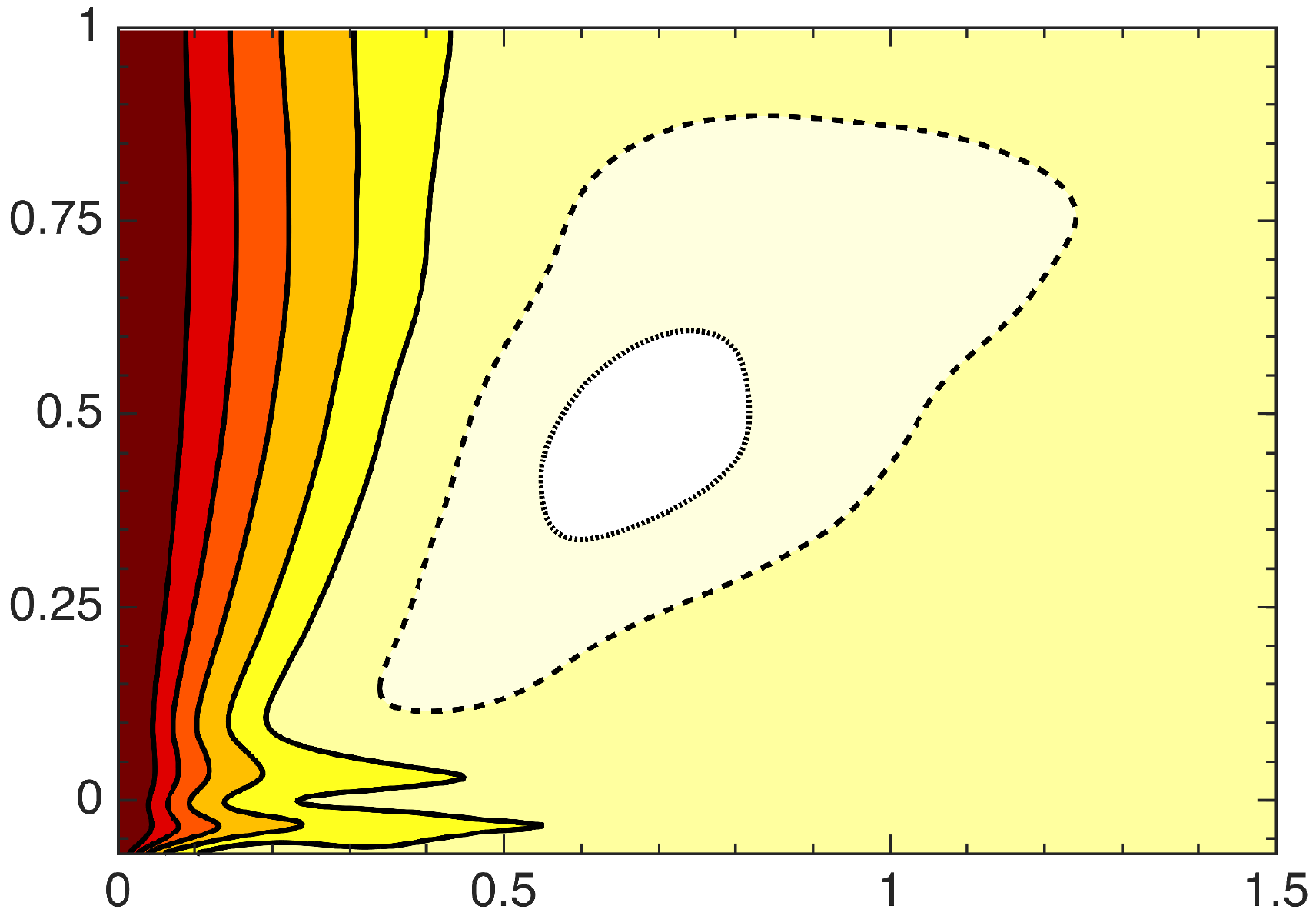}
   \includegraphics[width=0.493\textwidth]{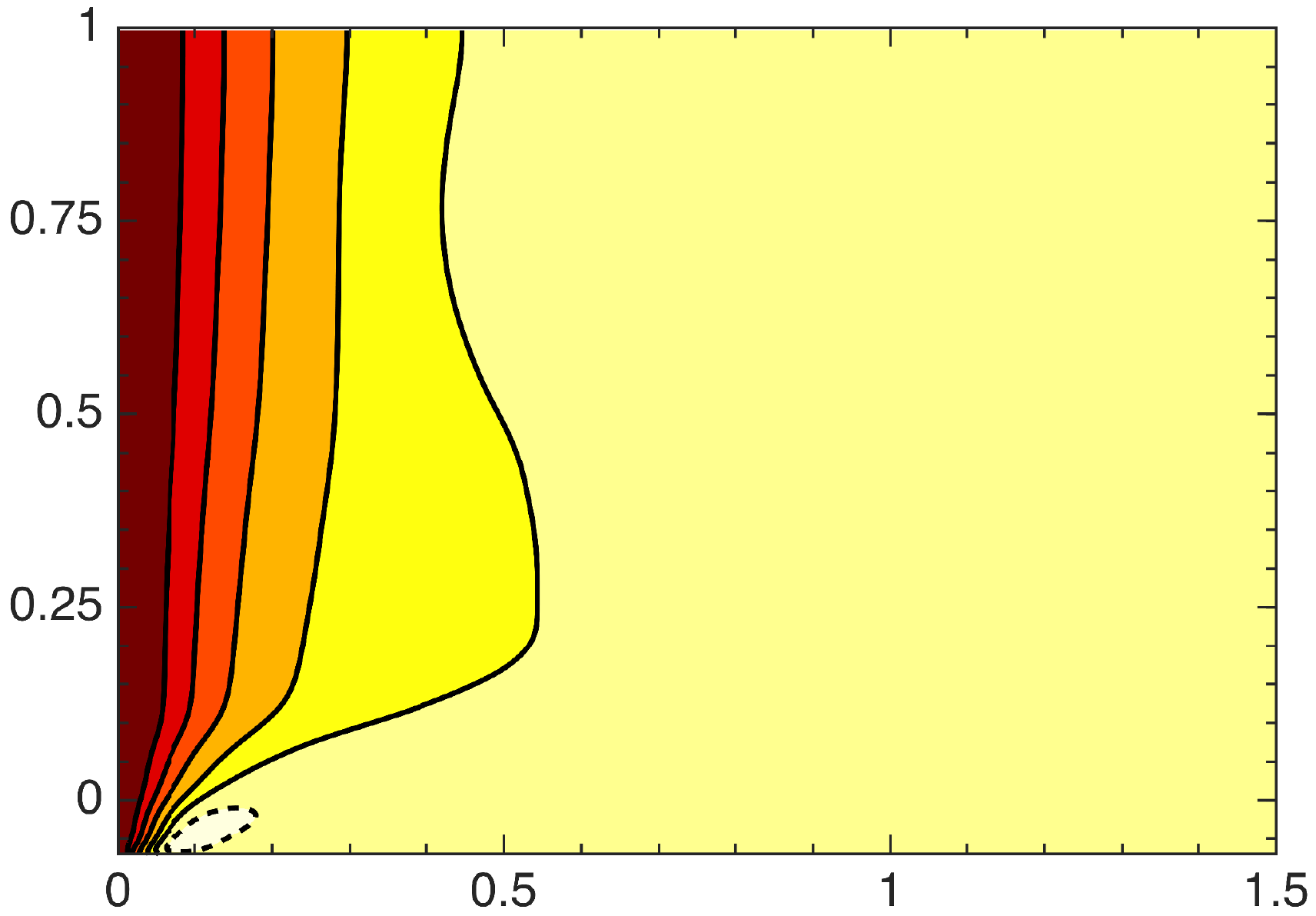}
   \put(-388,65){\rotatebox{90}{$y / h$}}   
   \put(-195,65){\rotatebox{90}{$y / h$}}   
   \put(-99,-6){{$\Delta x / h$}}   
   \put(-291,-6){{$\Delta z / h$}}      
   \put(-387,127){\footnotesize $(c)$}
   \put(-194,127){\footnotesize $(d)$} \\
  \caption{$(a)$ The contribution of each stress to the total drag, normalized by the drag of the single-phase flow with rigid walls, $\tau_{RW}$ (the dashed line with $Re_\tau=180$). $(b)$ Contours of the weighted Reynolds shear stress for $G1_{FW}$ at $y/h = 0.05$, given by multiplying the absolute value of the Reynolds shear stress with the joint probability density of its occurrence in the $u^\prime-v^\prime$ plane. $(c)$ One-dimensional autocorrelation of $u^\prime$ as a function of the spanwise spacing ($R^z_{uu} (y,\Delta z)$) and $(d)$ One-dimensional autocorrelation of $v^\prime$ as a function of the streamwise spacing for different $y/h$ ($R^x_{vv} (y,\Delta x)$). The colour scheme in $(c)$ and $(d)$ is the same reported for figure~\ref{fig:rspu}.}
\label{fig:rough_turb}
\end{figure}
We start by comparing the mean velocity of the single-phase elastic cases with the results obtained for $G1_{FW}$. The mean velocity profiles, scaled in inner units, are depicted in figure~\ref{fig:scale}$(a)$, versus ${(y+d)}^+$. $d$ is a shift of the origin (zero-plane) \citep{Jackson1981,Breugem2006,Suga2010,Rosti20171}, following the modified log-law:  
\begin{equation}
\label{eq:loglaw}  
U^+ = \frac{1}{k+\Delta k} \log {\left( y + d \right)}^+ \,+ \, B \, - \, \Delta U^+  \, . 
\end{equation}
$k+\Delta k$ and $B$ in the equation above are the modified von K\'arm\'an and the additive constants ($k=0.4$ and $B=5.5$) and $\Delta U^+$ is the velocity shift. $d$ is calculated similarly to \cite{Breugem2006,Rosti20171} by iterating between several values of $d$ until a region of constant ${(y+d)}^+ \mathrm{d} U^+ / \mathrm{d} y^+ = 1/(k+\Delta k)$ is achieved. The mean flow profiles show a significant reduction in the length of the logarithmic layer with increasing wall elasticity, while no shrinkage is observed for $G1_{FW}$. The values of the fitting parameters, reported in table~\ref{tab:rough}, indicate a downward shift of the inertial range and an increase of its slope with increasing wall elasticities. The mean flow profile for $G1_{FW}$ shows a more pronounced downward shifting than the change in slope of the logarithmic layer. \cite{Rosti20171} found a linear relation between the wall-normal velocity fluctuations at $y=0$, divided by $\Delta k$ and the velocity shift $\Delta U^+$ by modifying the correlation proposed by \cite{Orlandi2008} for turbulent channel flows over rough walls. This linear correlation is shown in figure~\ref{fig:scale}$(b)$ for the single-phase cases, and interestingly the correlation is also valid for the rough case, depicted with a brown square in this figure.

\begin{table}
  \begin{center}
\def~{\hphantom{0}}
  \begin{tabular}{lccccccc}
  Case & $d/h$ & $k+\Delta k$  & $\Delta U^+$ & $v^\prime(0) / U_b$ \\[5pt]
      $G1$  & $0.211$ & $0.23$ &  $22.01$ & $0.108$   \\
      $G2$  & $0.143$ & $0.27$ &  $15.37$ & $0.064$ \\
      $G3$  & $0.075$ & $0.31$ &  $5.42$   & $0.023$ \\
      $G4$  & $0.028$ & $0.38$ &  $1.05$   & $0.004$ \\
      $G1_{FW}$  & $0.041$ & $0.35$ & $9.12$ & $0.021$ \\
  \end{tabular}
  \caption{Summary of the log-law fitting parameters with $d$ the origin shift, $k+\Delta k$ the modified von K\'arm\'an constant and $\Delta U^+$ the shift of the mean velocity profile in inner scales. $v^\prime(0)$ denotes the wall-normal velocity fluctuation at $y=0$.} 
 \label{tab:rough}
 \end{center}
\end{table}
The contribution of different stresses is compared between $G1$ and $G1_{FW}$ in figure~\ref{fig:rough_turb}$(a)$. It should be noted that $\tau_E$ for $G1_{FW}$ does not indicate the elastic stress anymore, but it describes the shear stress in the solid roughness elements that blocks the streamwise velocity. This is computed based on the difference between the total drag (calculated by the pressure gradient) and the summation of the viscous and Reynolds shear stresses inside the channel. This value is indeed significantly smaller than the contribution of the elastic stress in $G1$. The total drag is reduced from $4.65 \tau_{RW}$ to $2.42  \tau_{RW}$ in $G1_{FW}$, mostly due to the less turbulent activity in the case with the rough walls. To understand the cause of this reduction, we show the contours of the weighted Reynolds shear stress (previously discussed in figure~\ref{fig:jpdf}) for $G1_{FW}$ at $y/h = 0.05$. Interestingly, the strong and dominant ejection events, previously observed for highly elastic walls, are absent near the rough walls. In fact, this further proves that these energetic events are indeed associated with the elastic potential energy of the deformable walls. Sweep events can be observed to be the dominant contribution to the turbulent production near the rough walls, similarly to turbulent flows over permeable walls \citep{Breugem2006,Suga2010} or plant canopies \citep{Zhu2007}. Finally, the line and colour contours of $R^z_{uu} (y,\Delta z)$ and $R^x_{vv} (y,\Delta x)$ are presented in figure~\ref{fig:rough_turb}$(c)$ and \ref{fig:rough_turb}$(d)$ for $G1_{FW}$. The high- and low-speed streaks are observed to move upwards in the case of rough walls, with a larger spacing between them. Moreover, the spanwise roller-type vortices, previously observed near highly elastic walls, are absent in this case of rough walls. Note that, the two correlated regions (just below and above $y=0$) in the contours of $R^z_{uu}$ are present due to the chosen geometry of the rough surface that has the same spanwise deformations as in $G1$.               

\section{Final remarks}\label{sec:Final_Remarks}

We have reported results from direct numerical simulations of single-phase and particulate turbulent channel flows, bounded by two incompressible hyper-elastic walls at bulk Reynolds number $5600$. Four different wall elasticities are studied, ranging from an almost rigid to a highly elastic wall. Both single-phase and particulate cases are simulated at each wall elasticity, considering a $10\%$ volume fraction of finite-size neutrally-buoyant rigid spherical particles with a diameter of $D_p=h/9$ for the suspended particles.

Our data show a significant drag increase and an enhancement of the turbulent activity with growing wall elasticity for both the single-phase and particulate cases. A drag reduction and a turbulent attenuation is obtained for the particulate cases with highly elastic walls, albeit with respect to the single-phase flow of the same wall elasticity; whereas, an opposite effect of the particles is observed on the flow with less elastic walls. 


The strong asymmetry in the magnitudes of wall-normal velocity fluctuations (favouring the positive $v^\prime$) is found to push the particles towards the channel centre.  However, as the wall elasticity decreases, a more symmetric distribution of $v^\prime$ allows the particles to form a layer close to the interface, similar to the suspension flow over rigid walls. The particle layer close to the wall is shown to contribute to increasing the wall-normal velocity fluctuations, while in the absence of this layer, smaller wall deformation and in turn a turbulence attenuation is observed. We further address the importance of the particle wall-layer in turbulence production by performing a numerical experiment, where we prevent the formation of this layer for the case with the least elastic walls. In this simulation the particles bounce back towards the core of the channel before approaching the interfaces, following a collision with the two virtual walls located at a distance $h/10$ from the two real interfaces. The results of this simulation show a strong turbulence attenuation and thus a drag reduction ($Re_\tau \approx 167$), even with respect to the single-phase flow over rigid walls ($Re_\tau \approx 180$). 

The effect of the volume fraction is further studied at a moderate wall elasticity, by increasing the particle volume fraction up to $20\%$. Migration of the particles from the interface region is found to be the cause of an increased turbulence attenuation, in comparison to the same volume fraction with rigid walls. However, the particle induced stress compensates for the loss of Reynolds shear stress, thus resulting in an overall drag increase in the case of elastic walls. The effect of the wall-elasticity on the drag is significantly reduced when the volume fraction of the particles increases.

We finally perform an extra simulation to distinguish the effect of wall elasticity on the near wall turbulence from the modulation caused by a rough wall. In this simulation, a snapshot of the deformed interface, obtained in the single-phase case with the highest wall elasticity, is fixed and frozen in time (no elasticity). The drag is strongly reduced in the rough wall case, mostly due to the less turbulence activity in the rough walls case than in the elastic one. The strong and dominant ejection events, observed in the near-wall turbulence of highly elastic walls, are indeed absent near the rough walls. Therefore we speculate that migration of spherical particles toward the channel centre is absent in the case of rough surfaces, however, a detailed analysis of this subject is required in future studies.
          
\section*{Acknowledgements}
This work was supported by the European Research Council Grant No. ERC-2013-CoG-616186, TRITOS. The authors acknowledge computer time provided by SNIC (Swedish National Infrastructure for Computing) and the support from the COST Action MP1305: Flowing matter.

\bibliographystyle{jfm}
\bibliography{Paper_Elast}

\end{document}